\documentclass[pdflatex,sn-mathphys-num]{sn-jnl}

\usepackage{graphicx}%
\usepackage{multirow}%
\usepackage{amsmath,amssymb,amsfonts}%
\usepackage{amsthm}%
\usepackage{mathrsfs}%
\usepackage[title]{appendix}%
\usepackage{xcolor}%
\usepackage{textcomp}%
\usepackage{manyfoot}%
\usepackage{booktabs}%
\usepackage{algorithm}%
\usepackage{algorithmicx}%
\usepackage{algpseudocode}%
\usepackage{listings}%
\usepackage{bm}%
\usepackage{float,subfigure,tikz}
\usepackage{enumitem}
\usepackage{etoolbox}

\theoremstyle{thmstyleone}%
\newtheorem{lemma}{Lemma}[section]%
\newtheorem{corollary}{Corollary}[section]%
\newtheorem{theorem}{Theorem}[section]
\theoremstyle{thmstyletwo}%
\newtheorem{remark}{Remark}[section]%

\theoremstyle{thmstylethree}%
\newtheorem{definition}{Definition}[section]%

\def\bi{{\bf i}}
\makeatletter
\patchcmd{\@maketitle}{\artauthors}{\centering{\artauthors}}{}{}
\makeatother

\begin{document}

\title[Article Title]{Local Robustness of Bound States
  in the Continuum through Scattering-Matrix Eigenvector Continuation}


\author[1]{\fnm{Ya Yan} \sur{Lu}}\email{mayylu@cityu.edu.hk}

\author[2]{\fnm{Jiaxin} \sur{Zhou}}\email{jiaxzhou@cityu.edu.hk}

\affil[1]{\orgdiv{Department of Mathematics}, \orgname{City University
  of Hong Kong}, \orgaddress{\street{Kowloon}, \city{Hong Kong}, \country{China}}}

\affil[2]{\orgdiv{Liu Bie Ju Centre for Mathematical Sciences},
  \orgname{City University of Hong Kong},
  \orgaddress{\street{Kowloon}, \city{Hong Kong}, \country{China}}}


\abstract{We consider the diffraction of time-harmonic plane
  waves by a periodic structure, governed by the Helmholtz
  equation. Bound states
  in the continuum (BICs) are quasi-periodic fields that remain
  $L^{2}$-bounded over one period and occur at frequencies embedded in
  the continuous spectrum. Perturbations that break a BIC can lead to
  ultra-strong resonances, enabling various applications in photonics.
  Employing the implicit function theorem, we demonstrate how 
  a simple BIC continuously deforms into a propagating
  field as system parameters vary in a neighborhood,
  with the frequency adjusting accordingly. In this setting,   
  the incident coefficients of the field persist as an eigenvector
  of the scattering matrix with a fixed eigenvalue. By introducing a  
  mapping $\mathcal{P}$ from the parameters to these  
  coefficients, the zeros of $\mathcal{P}$ correspond precisely to BICs.
  When such a zero is isolated and the dimensions of the
  domain and codomain coincide, the BIC can be related to the mapping
  degree of $\mathcal{P}$ in a small neighborhood. This perspective clarifies
  the phase singularity  
  associated with BICs and provides a general topological
  interpretation of their local robustness with respect to the given
  parameters. Moreover, it yields a practical numerical criterion for
  detecting and verifying BICs via computation of the mapping degree of
  $\mathcal{P}$.}

\keywords{Bound states in the continuum, Helmholtz equation, Implicit
  function theorem, Scattering matrix, Mapping degree}



\maketitle

\section{Introduction}\label{sec1}
For the diffraction of time-harmonic plane waves incident upon
a periodic structure, the governing wave equation admits
a unique solution except at a discrete closed set of frequencies
\cite{nedelec91,chen91,abboud93,bao95}. The loss of uniqueness at
these frequencies is attributed to the presence of bound states in
the continuum (BICs, also referred to as trapped modes or guided
modes above the light line), which have been found in numerous
classical wave systems \cite{vonneumann29,friedrich85,evans94,anne94}.
Small perturbations of wave systems that support BICs can give
rise to ultra-strong resonances, leading to significant local field
enhancement and scattering anomalies. These phenomena enable a wide
range of applications in photonics
\cite{hsu16,kodigala17,yesilkoy19,Koshelev19,yuan20_2,hwang21,sadreev21,koshelev23}.

The existence of bound states in the continuum (BICs) was first
established in symmetric structures, introducing the concept of
symmetry-protected BICs. In such systems, symmetry-induced
decoupling enables an isolated eigenvalue of one subsystem to remain
embedded in the continuous spectrum of another
\cite{evans94,anne94,shipman07}. BICs can also arise from the
trapping of waves between two scatterers
\cite{mciver96,chesnel19,mai25}. Another
important class is Friedrich–Wintgen BICs, which emerge from
destructive interference between resonant modes
\cite{friedrich85}, with further theoretical elaboration
provided in \cite{lyapina15,yu25,zhou26}. In addition, examples and
analyses of the mechanisms underlying BIC formation in
three-dimensional structures appear in \cite{hsu13,chesnel25}.


Considerable research has also focused on the robustness of BICs. In
a symmetric structure supporting a symmetry-protected BIC, 
perturbations that preserve the symmetry merely
shift the BIC to a slightly adjusted frequency. In contrast,
for other types of BICs---or for symmetry-protected BICs subjected to
symmetry-breaking perturbations---the BIC cannot generally be recovered
by frequency tuning alone. Robustness must then be explored
through the variation of additional parameters, including
Bloch wavenumber, permittivity, or geometry. A common approach to
studying this form of robustness
involves relating BICs to topological indices. The first example of
such robust BICs was investigated in \cite{porter05} for 
two-dimensional rectangular arrays, where BICs were associated with
curve crossings. Later, BICs in photonic crystal slabs were identified
through winding numbers of resonance polarization directions in the
plane of Bloch wave vector \cite{zhen14}. While this discovery
has drawn
significant interest,
its mathematical foundation remains incomplete; some progress on the
well-definedness of these winding numbers appears in \cite{zhang25}.
Additional studies have sought to understand robustness by
establishing sufficient conditions for the parametric dependence
of BICs \cite{nazarov12,yuan17_2,yuan20_3}. Despite these advances, a
general mathematical framework describing BIC robustness remains
elusive.


If a structure supporting a BIC is slightly perturbed such that the
BIC cannot be recovered by frequency tuning alone, sharp variations in
total reflection and transmission can be observed near the original
BIC frequency. This type of scattering anomaly is known as a Fano
resonance, which has been extensively studied in
\cite{shipman12,chesnel18,chesnel20,junshanstephenhai20}. The present work
contributes a rigorous analysis of another type of scattering
anomaly: the ability to achieve arbitrary phase variations between
incident and scattered coefficients. This phenomenon was first studied
in \cite{chesnel18} through an asymptotic analysis and has also
recently been explored in \cite{liu25}.

We consider quasi-periodic, time-harmonic fields in a two-dimensional
periodic dielectric structure with a single periodic direction. Let
$\beta$ denote the Bloch wavenumber, $k$ the
frequency, and let the dielectric function depend continuously on a
vector parameter $\bm{\delta}$. We employ a variational formulation of the
scattering problem within a bounded rectangular domain, and denote by
$\bm{a}$ and $\bm{b}$ the coefficients of the incident and outgoing
fields, respectively. Assuming a simple BIC exists at
$(\beta_{*},\bm{\delta}_{*},k_{*})$, we demonstrate how it continuously deforms
into a propagating field as $\beta$ and $\bm{\delta}$ vary, with $k$ adjusted
accordingly via the implicit function theorem. For any $\theta\in[0,2\pi)$
except a finite set, we prove that there exists a unique frequency
$k(\beta,\bm{\delta})$ near $k_{*}$ such that the propagating field satisfies
$\bm{b}=e^{i\theta}\bm{a}$. Provided $\bm{a}\neq\bm{0}$, this identity also
implies that $\bm{a}$ is an eigenvector of the scattering matrix with
eigenvalue $e^{i\theta}$. This result, formalized in Theorem
\ref{thm:sec4:PiM:bic}, elucidates the phase singularity (a special
type of scattering anomaly) associated with BICs.

Building on this framework, we define $\mathcal{P}$ as the mapping from
parameters $(\beta,\bm{\delta})$ to the incident coefficients $\bm{a}$,
noting that zeros of $\mathcal{P}$ correspond precisely to
BICs. We analyze four distinct symmetry cases determined by the spatial
symmetry of the structure and show how $\mathcal{P}$ reduces to a
lower-dimensional mapping when additional symmetry is
present. Consequently,
if a BIC is isolated and the domain and codomain dimensions of $\mathcal{P}$ (or
its reduced forms) coincide, the BIC can be characterized by the
mapping degree of $\mathcal{P}$ near that point. This dimensional constraint
aligns with those
derived in \cite{abdrabou23}. A nonzero degree implies that the BIC is
robust with respect to the parameters $(\beta,\bm{\delta})$ under perturbations
that preserve the corresponding symmetry. Furthermore, when the
dielectric function is $C^{1}$ in
$\bm{\delta}$, the implicit function theorem ensures $\mathcal{P}$ is $C^{1}$ in
$(\beta,\bm{\delta})$. This regularity yields sufficient conditions for BIC
robustness via the non-vanishing of the corresponding Jacobian
determinant, recovering the conditions previously obtained through
perturbation theory in \cite{yuan17_2,yuan20_3}.

The paper is organized as follows. Section 2 introduces the problem
formulation and establishes the key notation. Section 3 presents the
variational formulation for the scattering problem in a bounded
domain. Building on
this, Section 4 employs the implicit function theorem to construct a
continuous family of propagating fields emerging from a simple
BIC. The concept
of a BIC index, which quantifies BIC robustness under parameter
variation, is introduced in Section 5. Sufficient conditions for a
nonzero index are then derived in Section 6. Numerical results
presented in Section 7 validate the theoretical analysis. We conclude
in Section 8 with a summary and outlook for future research.

\section{Problem formulation and notation}\label{sec2}
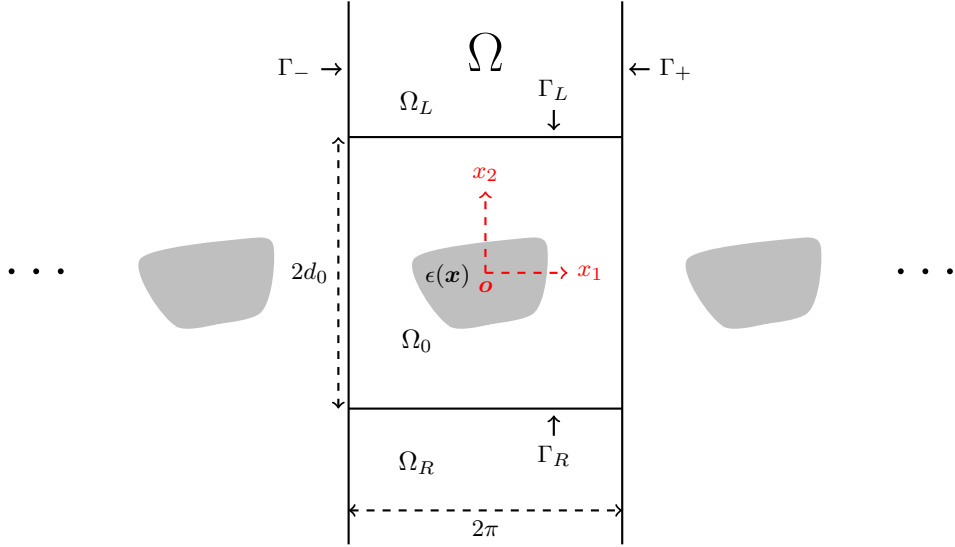
\begin{figure}[htb]
  \centering
  \begin{tikzpicture}[scale=0.9]
    \draw [thick](2,4) -- (2,-4);
      \draw [thick](-2,4) -- (-2,-4);    
      \fill [lightgray] plot [smooth] coordinates {(0.6,-0.6) (-0.5,-0.8)
        (-1.05,0.2) (0.45,0.5) (0.9,0.4) (0.7,-0.6) (-0.55,-0.8)};
      \fill [lightgray] plot [smooth] coordinates {(-3.4,-0.6) (-4.5,-0.8)
        (-5.05,0.2) (-3.55,0.5) (-3.1,0.4) (-3.3,-0.6) (-4.55,-0.8)};
      \fill [lightgray] plot [smooth] coordinates {(4.6,-0.6) (3.5,-0.8)
  (2.95,0.2) (4.45,0.5) (4.9,0.4) (4.7,-0.6) (3.45,-0.8)};      
    \draw (-6.5,0) node[]{\huge$\cdots$};
    \draw (6.5,0) node[]{\huge$\cdots$};
\draw [thick,dashed,<->](-2,-3.5) -- (2,-3.5)
 node[below,pos=0.5]{\small$2\pi$};
 \draw [thick,dashed,->,red](0,0) -- (1.2,0)
 node[right,pos=1]{\small$x_{1}$};
  \draw [thick,dashed,->,red](0,0) -- (0,1.2)
  node[above,pos=1]{\small$x_{2}$};
  \draw (0,-0.2) node[red]{\small$\bm{o}$};
  \draw (-0.55,-0.05) node {\small$\epsilon(\bm{x})$};
  \draw [thick] (-2,2) -- (2,2);
  \draw [thick] (-2,-2) -- (2,-2);
 \draw (0,3.25) node[]{\huge$\Omega$};  
 \draw (-1,2.5) node[]{$\Omega_{L}$};
 \draw (-1,-1) node[]{$\Omega_{0}$};
 \draw (-1,-2.8) node[]{$\Omega_{R}$};
 \draw [thick,->](1,2.4) -- (1,2.1)
 node[above,pos=0]{\small$\Gamma_{L}$};
 \draw [thick,->](1,-2.4) -- (1,-2.1)
 node[below,pos=0]{\small$\Gamma_{R}$};
 \draw [thick,->](-2.4,3.0) -- (-2.1,3.0)
 node[left,pos=0]{\small$\Gamma_{-}$};
\draw [thick,->](2.4,3.0) -- (2.1,3.0)
 node[right,pos=0]{\small$\Gamma_{+}$};
 \draw [thick,dashed,<->](-2.15,-2) -- (-2.15,2)
 node[left,pos=0.5]{\small$2d_{0}$};  
\end{tikzpicture}
\caption{One period of the structure $\Omega$ is
  partitioned into three subdomains: $\Omega_{L}$, $\Omega_{0}$, $\Omega_{R}$,
  separated by interfaces $\Gamma_{L}$ and $\Gamma_{R}$.  
The boundaries of $\Omega$ are denoted by $\Gamma_{-}$ and
$\Gamma_{+}$, respectively. A rectangular coordinate system is
introduced with its
origin $\bm{o}$ on the central line of $\Omega$. The dielectric function
of the structure is denoted by $\epsilon(\bm{x})$ and is equal to $1$ for large
$|x_{2}|$.}
\label{fig:sec2:periodicwg}
\end{figure}

Consider a lossless, two-dimensional dielectric structure that is
periodic with period $2\pi$ in one spatial direction.
A rectangular coordinate system is introduced, centered at a point
$\bm{o}$, with the $x_{1}$-axis parallel to the periodic
direction and the $x_{2}$-axis perpendicular to it. The dielectric function
$\epsilon(\bm{x})\in{L^{\infty}(\mathbb{R}^{2})}$ for $\bm{x}:=(x_{1},x_{2})$ satisfies
\begin{equation}
  \label{eq:sec2:refindex}
  \left\{
    \begin{aligned}
      &\epsilon(x_{1}+2m\pi,x_{2})=\epsilon(\bm{x}),\ &&\text{for}\ m\in\mathbb{Z};\\
      &\epsilon(\bm{x})=1,\ &&\text{if}\ |x_{2}|\ge{d_{0}};\\
      &0<\epsilon_{\mathrm{min}}\le{\epsilon(\bm{x})}\le{\epsilon_{\mathrm{max}}},       
    \end{aligned}\right.  
\end{equation}
where $d_{0},\epsilon_{\mathrm{min}},\epsilon_{\mathrm{max}}>0$ are constants
such that $\epsilon_{\mathrm{max}}>\epsilon_{\mathrm{min}}$. The domain
for one period of the structure is defined as
\begin{equation}
  \label{eq:sec2:Omega}
  \Omega=\{\bm{x}:-\pi<x_{1}<\pi,\ -\infty<x_{2}<\infty\}.
\end{equation}
As illustrated in Fig. \ref{fig:sec2:periodicwg}, $\Omega$ is
partitioned into the following subdomains:
\begin{enumerate}[label=(\arabic*).]
  \item $\Omega_{L}$: the semi-infinite domain
    $(-\pi,\pi)\times(d_{0},\infty)$;
  \item $\Omega_{R}$: the semi-infinite domain 
    $(-\pi,\pi)\times(-d_{0},-\infty)$;    
  \item $\Omega_{0}$: the bounded domain 
    $(-\pi,\pi)\times(-d_{0},d_{0})$;
  \item $\Gamma_{L}$: the interface between $\Omega_{L}$ and $\Omega_{0}$,
    given by $(-\pi,\pi)\times\{d_{0}\}$;
  \item $\Gamma_{R}$: the interface between $\Omega_{0}$ and $\Omega_{R}$,
    given by $(-\pi,\pi)\times\{-d_{0}\}$.
\end{enumerate}
The boundaries of $\Omega$ are denoted by $\Gamma_{-}$ and $\Gamma_{+}$, respectively.

We study an E-polarized, quasi-periodic, time-harmonic
field in $\Omega$ with a Bloch wavenumber
$\beta$ belonging to the Brillouin zone $[-1/2,1/2)$, which
has the topology of $\mathbb{S}^{1}$. 
Expressing the non-zero component of the electric field in the form
$u(\bm{x})e^{\bi{\beta}x_{1}}$ and defining the differential operator 
\begin{equation}
  \mathcal{L}:=-(\nabla+\bi\bm{\beta})\cdot(\nabla+\bi\bm{\beta}),\ \bm{\beta}:=(\beta,0),
\end{equation}
the function $u$ satisfies the following equations:
\begin{align}
  \mathcal{L}u-k^{2}\epsilon(\bm{x}){u}=&0\ \text{in}\ \Omega,\label{eq:sec2:Omega:gov}\\
  u|_{\Gamma_{+}}=&u|_{\Gamma_{-}},\label{eq:sec2:Omega:pbc1}\\
  \partial_{x_{1}}u|_{\Gamma_{+}}=&\partial_{x_{1}}u|_{\Gamma_{-}},\label{eq:sec2:Omega:pbc2}
\end{align}
where $k$ denotes the normalized frequency.
For simplicity, we also refer to $u$ as the field.
We seek scattering solutions that satisfy
\eqref{eq:sec2:Omega:gov}--\eqref{eq:sec2:Omega:pbc2}.
In the semi-infinite domains $\Omega_{L}$ and $\Omega_{R}$, the field admits the
expansions:
\begin{align}
  u(\bm{x})=&\sum_{m\in{Z_{0}}}a_{L,m}\alpha_{m}^{-1/2}e^{-\bi\alpha_{m}(x_{2}-d_{0})}\phi_{m}(x_{1})+\sum_{m\in{Z_{0}}}b_{L,m}\alpha_{m}^{-1/2}e^{\bi\alpha_{m}(x_{2}-d_{0})}\phi_{m}(x_{1})\notag\\
  &+\sum_{m\in\mathbb{Z}\backslash{Z_{0}}}c_{L,m}e^{\bi\alpha_{m}(x_{2}-d_{0})}\phi_{m}(x_{1})\
                   \text{in}\ \Omega_{L},\label{eq:sec2:Omegal:expan}\\
  u(\bm{x})=&\sum_{m\in{Z_{0}}}a_{R,m}\alpha_{m}^{-1/2}e^{\bi\alpha_{m}(x_{2}+d_{0})}\phi_{m}(x_{1})+\sum_{m\in{Z_{0}}}b_{R,m}\alpha_{m}^{-1/2}e^{-\bi\alpha_{m}(x_{2}+d_{0})}\phi_{m}(x_{1})\notag\\
  &+\sum_{m\in\mathbb{Z}\backslash{Z_{0}}}c_{R,m}e^{-\bi\alpha_{m}(x_{2}+d_{0})}\phi_{m}(x_{1})\
                   \text{in}\ \Omega_{R}.\label{eq:sec2:Omegar:expan}
\end{align}
Here, for each $m\in\mathbb{Z}$, we define 
$\alpha_{m}(\beta,k):=\sqrt{k^{2}-(m+\beta)^{2}}$ and $\phi_{m}(x_{1}):=e^{\bi{m}x_{1}}/\sqrt{2\pi}$.
The finite index set ${Z}_{0}\subset\mathbb{Z}$, of size $N_{0}$, satisfies
\begin{equation}
  \label{eq:sec2:m-m+}
  \alpha_{m}>0\ \text{if}\ m\in{Z_{0}},\ \text{while}\ \bi\alpha_{m}<0\
  \text{if}\ m\in\mathbb{Z}\backslash{Z_{0}}.
\end{equation}
The coefficients appearing in expansions
\eqref{eq:sec2:Omegal:expan}--\eqref{eq:sec2:Omegar:expan} represent
three types of field components:
\begin{itemize}[label=$\bullet$]
\item Incident field coefficients: $\{a_{\flat,m}\}$, with
  $\flat\in\{L,R\}$ and $m\in{Z_{0}}$;  
\item Outgoing field coefficients: $\{b_{\flat,m}\}$, with
  $\flat\in\{L,R\}$ and $m\in{Z_{0}}$;  
\item Evanescent field coefficients: $\{c_{\flat,m}\}$, with
  $\flat\in\{L,R\}$ and $m\in\mathbb{Z}\backslash{Z_{0}}$.  
\end{itemize}
In what follows, we collect the incident and scattered coefficients
into the column vectors
\begin{equation}
  \label{eq:sec2:incscavec}
  \bm{a}:=[a_{\flat,m}]_{\flat\in\{L,R\},m\in{Z_{0}}},\
  \bm{b}:=[b_{\flat,m}]_{\flat\in\{L,R\},m\in{Z_{0}}},
\end{equation}
which will be referred to as the incident and scattered coefficient
vectors, respectively.

Let the dielectric function $\epsilon(\cdot,\bm{\delta})$ depend
continuously on a parameter $\bm{\delta}$ in an open set
$W_{1}\subset\mathbb{R}^{N_{1}}$ for some $N_{1}\in\mathbb{N}$, and suppose $\epsilon(\cdot,\bm{\delta})$ satisfies
\eqref{eq:sec2:refindex} for every $\bm{\delta}\in{W_{1}}$. In the special case
$N_{1}=0$, we regard $\epsilon$ as fixed.
Define the parameter space $\Lambda\subset\mathbb{R}^{N_{1}+2}$ by
\begin{equation}
  \label{eq:sec2:Lambda}
  \Lambda:=\{(\beta,{\bm{\delta}},k):\beta\in[-1/2,1/2),\ {\bm{\delta}}\in{W}_{1},\ k>|\beta|\
  \text{and}\ \alpha_{m}\ne0,\ \forall{m\in\mathbb{Z}}\}.  
\end{equation}
The condition $k>|\beta|$ guarantees $Z_{0}\ne\emptyset$, thereby allowing 
propagating fields in $\Omega$, while the condition $\alpha_{m}\ne0$ excludes the
constant field in the $x_{2}$-direction. We consider the scattering
problem for parameters lying in each connected component of $\Lambda$, over
which the set $Z_{0}$ remains invariant.

We introduce precise definitions for a BIC and for a field governed by
a matrix.
\begin{definition}
  \label{def:sec2:BIC}
  Let $u$ be a nontrivial field satisfying system
  \eqref{eq:sec2:Omega:gov}--\eqref{eq:sec2:Omegar:expan} at a point
  $(\beta,\bm{\delta},k)\in\Lambda$.
  \begin{enumerate}[label=(\roman*).]
    \item The field $u$ is a BIC if its coefficient
      vectors vanish:      
\begin{equation}
  \label{eq:sec2:BIC}
  \bm{a}=\bm{0}\ \text{and}\ \bm{b}=\bm{0}.
\end{equation}
\item For a given matrix $\bm{M}$ in the unitary group
$U(2N_{0})$, the field \emph{$u$ is governed by $\bm{M}$} if
its coefficient vectors are related by
\begin{equation}
  \label{eq:sec2:FOM}
  \bm{b}=\bm{M}\bm{a}.
\end{equation}
  \end{enumerate}
\end{definition}
\begin{remark}
  It is clear that the BIC field decays exponentially as $x_{2}\to\pm\infty$.
  Moreover, while the definition of a BIC is independent of the length
  $d_{0}$ of the domain $\Omega_{0}$, the definition of a field
  governed by $\bm{M}$ depends on it. Note that a BIC is governed by
  every unitary matrix $\bm{M}\in{U}(2N_{0})$, since it trivially
  satisfies $\bm{b}=\bm{M}\bm{a}$ when $\bm{a}=\bm{b}=\bm{0}$.
\end{remark}

\begin{remark}
  Let $\bm{S}$ denote the scattering matrix of the wave
  system, which will be rigorously defined in Section
  \ref{sec:sca:scamat},  
  and let $\bm{I}_{2N_{0}}$ be the identity matrix in
  $U(2N_{0})$. Every field $u$ is governed by $\bm{S}$. If, in
  addition, $u$ is governed by $e^{\bi\theta}\bm{I}_{2N_{0}}$ for some  
  $\theta\in[0,2\pi)$ and satisfies $\bm{a}\ne\bm{0}$, then $\bm{b}$
  differs from $\bm{a}$ only by the phase factor $e^{\bi\theta}$.    
  From the identity
  \begin{equation}
    \label{eq:sec2:eigenvec:a}
    \bm{b}=\bm{S}\bm{a}=e^{\bi\theta}\bm{I}_{2N_{0}}\bm{a},
  \end{equation}  
  it also follows that $\bm{a}$ is an eigenvector of $\bm{S}$ with
  eigenvalue $e^{\bi\theta}$. 
\end{remark}

For a point
$\bm{z}\in\mathbb{R}^{N}$ with
$N\in\mathbb{N}$, let $B_{r}(\bm{z})$ denote the open ball in $\mathbb{R}^{N}$ and 
$\widehat{B}_{r}(\bm{z})$ the open ball in $\mathbb{C}^{N}$, each
centered at $\bm{z}$ with radius $r>0$.
\begin{definition}
  \label{sec2:def:simple}
  Let $u_{*}$ be a BIC at $(\beta_{*},\bm{\delta}_{*},k_{*})\in\Lambda$.
  We say $u_{*}$ is
\begin{enumerate}[label=(\roman*).]
\item \emph{simple} if it is the unique BIC supported at
  $(\beta_{*},\bm{\delta}_{*},k_{*})$; 
\item \emph{isolated} if there exists $r>0$ such that no other BIC
  lies in the punctured neighborhood
  $B_{r}((\beta_{*},\bm{\delta}_{*},k_{*}))\cap\Lambda\backslash\{(\beta_{*},\bm{\delta}_{*},k_{*})\}$.
\end{enumerate}  
\end{definition}

For any $\bm{M}\in{U(2N_{0})}$, we define the associated set $\lambda_{\bm{M}}\subset\Lambda$:
\begin{equation}
  \label{eq:sec2:PiM}
  \lambda_{\bm{M}}:=\left\{(\beta,{\bm{\delta}},k)\in{\Lambda}:\text{the\ system\ at}\ (\beta,{\bm{\delta}},k)\ \text{admits\ a\ field\ governed\ by}\ \bm{M}\right\}.
\end{equation}
All BIC points belong to $\lambda_{\bm{M}}$. Consider a simple BIC
$u_{*}$ located at $(\beta_{*},\bm{\delta}_{*},k_{*})\in\Lambda$ with scattering matrix
$\bm{S}_{0}$. We will prove that if $\bm{M}$ is chosen from
\begin{equation}
  \label{eq:sec2:U1}
    U_{1}:=\{\bm{M}\in{U}(2N_{0}):\mathrm{det}(\bm{S}_{0}-\bm{M})\ne0\},
  \end{equation}
then $\lambda_{\bm{M}}$ is locally the graph of a continuous function
$k(\beta,\bm{\delta})$ for $(\beta,\bm{\delta})$ in some ball
$B_{r}((\beta_{*},\bm{\delta}_{*}))$, $r>0$.
Furthermore, a continuous family of fields $u(\cdot,\beta,\bm{\delta})$ can be defined
on this ball such that $u(\cdot,\beta_{*},\bm{\delta}_{*})=u_{*}$.
Let $\bm{a}(\beta,\bm{\delta})$ denote the corresponding incident
coefficient vector. The continuous mapping
\begin{equation}
  \label{eq:sec2:PM}
 \mathcal{P}_{\bm{M},1}:{B}_{r}((\beta_{*},\bm{\delta}_{*}))\to\mathbb{C}^{2N_{0}},\
 \mathcal{P}_{\bm{M},1}(\beta,\bm{\delta}):=\bm{a}(\beta,\bm{\delta}),
\end{equation}
has zeros precisely at BIC points. 
In Sections \ref{sec:BICindex} and \ref{sec:BICsuffcond}, we show that if
the BIC is also
isolated and the dimensions of the domain and codomain of $\mathcal{P}_{\bm{M},1}$
match, its local robustness with respect to $(\beta,\bm{\delta})$ can be analyzed
by relating it to the
mapping degree (or winding number) of $\mathcal{P}_{\bm{M},1}$ over
$B_{r}((\beta_{*},\bm{\delta}_{*}))$ for sufficiently small $r$. (For degree
theory in Euclidean spaces,
see Chapter IV of \cite{Outerelo09}).

We consider four distinct cases classified by their fundamental
spatial symmetry:
\begin{enumerate}[label=\Roman*.]
\item No specific symmetry.
\item Reflection symmetry in $x_{1}$:
  \begin{equation}
    \label{eq:sec2:x1symmetry}
   \epsilon(x_{1},x_{2},\bm{\delta})=\epsilon(-x_{1},x_{2},\bm{\delta}),\ \text{for\ all}\
   \bm{\delta}\in{W}_{1},\ \bm{x}\in\Omega.
  \end{equation}
\item Reflection symmetry in $x_{2}$:
  \begin{equation}
    \label{eq:sec2:x2symmetry}
   \epsilon(x_{1},x_{2},\bm{\delta})=\epsilon(x_{1},-x_{2},\bm{\delta}),\ \text{for\ all}\
   \bm{\delta}\in{W}_{1},\ \bm{x}\in\Omega.
  \end{equation}  
\item Simultaneous reflection symmetry in $x_{1}$ and $x_{2}$:
  \begin{equation}
    \label{eq:sec2:x1x2symmetry}
   \epsilon(x_{1},x_{2},\bm{\delta})=\epsilon(-x_{1},x_{2},\bm{\delta})=\epsilon(x_{1},-x_{2},\bm{\delta}),\
   \text{for\ all}\   
   \bm{\delta}\in{W}_{1},\ \bm{x}\in\Omega.
  \end{equation}  
\end{enumerate}
Additionally, we define a special permutation of a matrix
$\bm{M}\in{U(2N_{0})}$ as
\begin{equation}
  \label{eq:sec2:matrix:permutation}
  \bm{M}^{P}:=\bm{R}_{2N_{0}}\bm{M}\bm{R}_{2N_{0}},
\end{equation}
where
\begin{equation}
  \label{eq:sec2:Rmatrix}
   \bm{R}_{2N_{0}}:=\begin{bmatrix}
    &\bm{I}_{N_{0}}\\
    \bm{I}_{N_{0}}&
  \end{bmatrix}.
\end{equation}

\section{The scattering problem in $\Omega_{0}$}
Following the approach in \cite{nedelec91,chen91,abboud93,bao95},
we truncate the scattering problem to a bounded domain $\Omega_{0}$ by
imposing Dirichlet-to-Neumann
(DtN) boundary conditions on $\Gamma_{L}$ and $\Gamma_{R}$. This yields a
bounded linear operator associated with the variational formulation,
which we analyze, and enables us to define the 
corresponding scattering matrix. Throughout the analysis,
we denote by $(\cdot,\cdot)_{D}$ the inner product over a domain $D$.

\subsection{Periodic function spaces}
We begin by introducing the Sobolev spaces used in this work.
For $d_{1},d_{2}\in\mathbb{R}$ with $d_{1}<d_{2}$, we define the domain
\begin{equation}
  \label{eq:sec3:Omegad1d2}
 \Omega_{d_{1},d_{2}}:=\{(x_{1},x_{2}):-\pi<x_{1}<\pi,\ d_{1}<x_{2}<d_{2}\}.
\end{equation}
and the following function spaces:
\begin{align*}
  C^{\infty}_{\mathrm{per},1}(\overline{\Omega_{d_{1},d_{2}}}):=&
    \left\{u\in{C^{\infty}(\overline{\Omega_{d_{1},d_{2}}})}:
      \partial_{x_{1}}^{n}u(-\pi,\cdot)=\partial_{x_{1}}^{n}u(\pi,\cdot)\ \text{for}\ n\in\mathbb{N}\right\},\\
  H^{1}_{\mathrm{per},1}(\Omega_{d_{1},d_{2}}):=&\left\{\text{the\
                                           completion\ of}\ 
    C^{\infty}_{\mathrm{per},1}(\overline{\Omega_{d_{1},d_{2}}})\ \text{in}\ 
    H^{1}(\Omega_{d_{1},d_{2}})\right\},\\
  H^{1}_{\mathrm{per},1,\mathrm{loc}}(\Omega_{L}):=&\left\{u\in{H^{1}_{\mathrm{loc}}(\Omega_{L})}:\ u|_{\Omega_{d_{1},d_{2}}}\in{H^{1}_{\mathrm{per},1}(\Omega_{d_{1},d_{2}})}\ \text{for}\
    {d_{1},d_{2}}\in(d_{0},\infty)\right\},\\
  H^{1}_{\mathrm{per},1,\mathrm{loc}}(\Omega_{R}):=&\left\{u\in{H^{1}_{\mathrm{loc}}(\Omega_{R})}:\
   u|_{\Omega_{d_{1},d_{2}}}\in{H^{1}_{\mathrm{per},1}(\Omega_{d_{1},d_{2}})}\ \text{for}\
    {d_{1},d_{2}}\in(-\infty,-d_{0})\right\},\\
C^{\infty}_{\mathrm{per}}([-\pi,\pi]):=&
  \left\{f\in{C^{\infty}([-\pi,\pi])}:d^{n}f(\pi)=d^{n}f(-\pi)\ \text{for}\ n\in\mathbb{N}\right\}.
\end{align*}
For any $s\in\mathbb{R}$, let $H^{s}_{\mathrm{per}}((-\pi,\pi))$ denote the
completion of $C^{\infty}_{\mathrm{per}}([-\pi,\pi])$ with respect to the norm
    \begin{equation}
      \label{eq:sec3:Hsnorm}
     \|f\|^{2}_{H^{s}_{\mathrm{per}}((-\pi,\pi))}:=\sum_{m=-\infty}^{\infty}(1+|m|^{2})^{s}|(f,\phi_{m})_{(-\pi,\pi)}|^{2},
    \end{equation}
    as defined in \cite[Section 3.6]{iorio01}.
The space $H^{-s}_{\mathrm{per}}((-\pi,\pi))$ is the dual of
$H^{s}_{\mathrm{per}}((-\pi,\pi))$. For $s=1/2$ the norm
\eqref{eq:sec3:Hsnorm} is
equivalent to the standard $H^{1/2}$-norm on $H^{1/2}_{\mathrm{per}}((-\pi,\pi))$
(cf. \cite{anne94}).
For convenience, we adopt the notation
\begin{equation*}
 H^{1}_{\mathrm{per},1}(\Omega_{0}):=H^{1}_{\mathrm{per},1}(\Omega_{-d_{0},d_{0}}),\  H^{s}_{\mathrm{per}}(\Gamma_{L})=H^{s}_{\mathrm{per}}(\Gamma_{R}):=H^{s}_{\mathrm{per}}((-\pi,\pi)).
\end{equation*}

\subsection{Variational formulation}
We first construct the DtN operator on $\Gamma_{L}$. In $\Omega_{L}$, we
solve the following problem via separation of variables:
\begin{align}
  (\nabla+\bi\bm{\beta})\cdot(\nabla+\bi\bm{\beta})u+k^{2}u=&0\ \text{in}\ \Omega_{L},\label{eq:sec2:Omegal:gov}\\
  u|_{\Gamma_{+}}=&u|_{\Gamma_{-}},\label{eq:sec2:Omegal:pbc1}\\
  \partial_{x_{1}}u|_{\Gamma_{+}}=&\partial_{x_{1}}u|_{\Gamma_{-}}.\label{eq:sec2:Omegal:pbc2}
\end{align}
This yields the following modes in $\Omega_{L}$:
\begin{equation}
\label{eq:sec3:Omegal:modes}  
w_{L,m}^{\pm}(x_{1},x_{2}):=\left\{
  \begin{aligned}
      &\alpha_{m}^{-1/2}e^{\pm\bi\alpha_{m}(x_{2}-d_{0})}\phi_{m}(x_{1}),\
        &&\text{for}\ m\in{Z_{0}};\\
      &e^{\pm\bi\alpha_{m}(x_{2}-d_{0})}\phi_{m}(x_{1}),\ &&\text{for}\ m\in{\mathbb{Z}\backslash{Z_{0}}}.
  \end{aligned}\right.
\end{equation}
For a given $(\beta,{\bm{\delta}},k)\in{\Lambda}$, the
modes $w_{L,m}^{+}$ and $w_{L,m}^{-}$ with $m\in{Z_{0}}$ propagate
forward and backward along $x_{2}$, respectively,
whereas modes with $m\in\mathbb{Z}\backslash{Z_{0}}$ are exponentially decaying and
growing in the $x_{2}$-direction.
Excluding the incoming field and all exponentially growing
modes, the scattered field in $\Omega_{L}$ can be written as:
\begin{equation}
  \label{eq:sec3:Omegal:expan}
  u^{\mathrm{sca}}=\sum_{m\in{Z_{0}}}b_{L,m}w^{+}_{L,m}+\sum_{m\in\mathbb{Z}\backslash{Z_{0}}}c_{L,m}w^{+}_{L,m}\
  \text{in}\ H^{1}_{\mathrm{per},1,\mathrm{loc}}(\Omega_{L}).
\end{equation}
Accordingly, the scattered field admits the following expansions on
$\Gamma_{L}$:
\begin{alignat}{2}
  u^{\mathrm{sca}}|_{\Gamma_{L}}=&\sum_{m\in\mathbb{Z}}(u^{\mathrm{sca}},\phi_{m})_{\Gamma_{L}}\phi_{m}
  &&\text{in}\ H^{1/2}_{\mathrm{per}}(\Gamma_{L}),\\  
  \partial_{x_{2}}u^{\mathrm{sca}}|_{\Gamma_{L}}=&\sum_{m\in\mathbb{Z}}\bi\alpha_{m}(u^{\mathrm{sca}},\phi_{m})_{\Gamma_{L}}\phi_{m}\quad &&\text{in}\ {H}^{-1/2}_{\mathrm{per}}(\Gamma_{L}).  
\end{alignat}
This allows us to define a DtN operator $\mathcal{D}_{L}:
H^{1/2}_{\mathrm{per}}(\Gamma_{L})\to{H}^{-1/2}_{\mathrm{per}}(\Gamma_{L})$ associated with
expansion
\eqref{eq:sec3:Omegal:expan} as
\begin{equation}
  \label{eq:sec3:Gammal:DtN}
  {\mathcal{D}}_{L}f:=\sum_{m\in\mathbb{Z}}\bi\alpha_{m}(f,\phi_{m})_{\Gamma_{L}}\phi_{m},\ \text{for}\ f\in{H^{1/2}_{\mathrm{per}}(\Gamma_{L})},
\end{equation}
which satisfies
$\mathcal{D}_{L}u^{\mathrm{sca}}|_{\Gamma_{L}}=\partial_{x_{2}}u^{\mathrm{sca}}|_{\Gamma_{L}}$.

The DtN operator on $\Gamma_{R}$ can be constructed similarly. In $\Omega_{R}$,
separation of variables yields the modes:
\begin{equation}
\label{eq:sec3:Omegar:modes}  
w_{R,m}^{\pm}(x_{1},x_{2}):=\left\{
  \begin{aligned}
      &\alpha_{m}^{-1/2}e^{\pm\bi\alpha_{m}(x_{2}+d_{0})}\phi_{m}(x_{1}),\
        &&\text{for}\ m\in{Z_{0}};\\
      &e^{\pm\bi\alpha_{m}(x_{2}+d_{0})}\phi_{m}(x_{1}),\ &&\text{for}\ m\in{\mathbb{Z}\backslash{Z_{0}}}.
  \end{aligned}\right.
\end{equation}
The scattered field in $\Omega_{R}$ is then expressed as
\begin{equation}
  \label{eq:sec3:Omegar:expan}
  u^{\mathrm{sca}}=\sum_{m\in{Z_{0}}}b_{R,m}w^{-}_{R,m}+\sum_{m\in\mathbb{Z}\backslash{Z_{0}}}c_{R,m}w^{-}_{R,m}\
  \text{in}\ H^{1}_{\mathrm{per},1,\mathrm{loc}}(\Omega_{R}).
\end{equation}
Accordingly, the DtN operator
$\mathcal{D}_{R}:H^{1/2}_{\mathrm{per}}(\Gamma_{R})\to{H}^{-1/2}_{\mathrm{per}}(\Gamma_{R})$
is given by
\begin{equation}
  \label{eq:sec3:Gammar:DtN}
  \mathcal{D}_{R}f:=\sum_{m\in\mathbb{Z}}\bi\alpha_{m}(f,\phi_{m})_{\Gamma_{R}}\phi_{m},\ \text{for}\ f\in{H^{1/2}_{\mathrm{per}}(\Gamma_{R})},
\end{equation}
and satisfies
$\mathcal{D}_{R}u^{\mathrm{sca}}|_{\Gamma_{R}}=-\partial_{x_{2}}u^{\mathrm{sca}}|_{\Gamma_{R}}$.

We now define a sesquilinear form using the DtN operators introduced above. For
any $u,v\in{H^{1}_{\mathrm{per},1}(\Omega_{0})}$, set
\begin{align}
  (\mathcal{A}u,v)_{\Omega_{0}}:=&((\nabla+\bi\bm{\beta}){u},(\nabla+\bi\bm{\beta}){v})_{\Omega_{0}}-k^{2}(\epsilon{u},v)_{\Omega_{0}}-(\mathcal{D}_{L}u,v)_{\Gamma_{L}}-(\mathcal{D}_{R}u,v)_{\Gamma_{R}}\notag\\
  =&(\nabla{u},\nabla{v})_{\Omega_{0}}-2\bi\beta(\partial_{x_{1}}{u},v)_{\Omega_{0}}+\beta^{2}(u,v)_{\Omega_{0}}-k^{2}(\epsilon{u},v)_{\Omega_{0}}\notag\\  
  &-(\mathcal{D}_{L}u,v)_{\Gamma_{L}}-(\mathcal{D}_{R}u,v)_{\Gamma_{R}}.  \label{eq:sec3:Omega:sesform}
\end{align}
Here, $\mathcal{A}$ represents the bounded linear operator
from $H^{1}_{\mathrm{per},1}(\Omega_{0})$ to its dual 
$({H}^{1}_{\mathrm{per},1}(\Omega_{0}))^{*}$ induced by this form, with
$L^{2}(\Omega_{0})$ as the pivot space.
We introduce two sets of linear functionals
$\{\widehat{\phi}_{L,m}\}_{m\in{Z_{0}}}$ and $\{\widehat{\phi}_{R,m}\}_{m\in{Z_{0}}}$
on $H^{1}_{\mathrm{per},1}(\Omega_{0})$, together with two sets of elements
$\{\widetilde{\phi}_{L,m}\}_{m\in{Z_{0}}}$ and $\{\widetilde{\phi}_{R,m}\}_{m\in{Z_{0}}}$
in $({H}^{1}_{\mathrm{per},1}(\Omega_{0}))^{*}$. 
For any $u\in{H^{1}_{\mathrm{per},1}(\Omega_{0})}$, their action is defined as
\begin{align}
  \widehat{\phi}_{L,m}u&:=(u,\phi_{m})_{\Gamma_{L}},\
  \widehat{\phi}_{R,m}u:=(u,\phi_{m})_{\Gamma_{R}}\ \text{for}\
    m\in{Z_{0}},  \label{eq:sec3:phitilde}\\
  (\widetilde{\phi}_{L,m},u)_{\Omega_{0}}&:=(\phi_{m},u)_{\Gamma_{L}},\
  (\widetilde{\phi}_{R,m},u)_{\Omega_{0}}:=(\phi_{m},u)_{\Gamma_{R}}\ \text{for}\ m\in{Z_{0}}.\label{eq:sec3:phitildestar}
\end{align}
Consider an incident field given by
\begin{equation}
  \label{eq:sec2:inc}
 \sum_{m\in{Z_{0}}}a_{L,m}w^{-}_{L,m}\ \text{in}\ \Omega_{L}\ \text{and}\ 
 \sum_{m\in{Z_{0}}}a_{R,m}w^{+}_{R,m}\ \text{in}\ \Omega_{R}.
\end{equation}
The corresponding scattering problem is to find a field
$u\in{H^{1}_{\mathrm{per},1}}(\Omega_{0})$
satisfying
\begin{equation}
\label{eq:sec3:Omegam:sca}  
\mathcal{A}u=\sum_{\flat\in\{L,R\}}\sum_{m\in{Z_{0}}}-2{\bi}a_{\flat,m}\alpha_{m}^{1/2}\widetilde{\phi}_{\flat,m}.
\end{equation}
For comparison, the solution to the associated adjoint problem,
\begin{equation}
\label{eq:sec3:Omegam:sca:adjoint}  
\mathcal{A}^{*}u=\sum_{\flat\in\{L,R\}}\sum_{m\in{Z_{0}}}2{\bi}b_{\flat,m}\alpha_{m}^{1/2}\widetilde{\phi}_{\flat,m}
\end{equation}
yields a field whose outgoing field components in $\Omega_{L}$ and $\Omega_{R}$ are
$\sum_{m\in{Z_{0}}}b_{L,m}w^{+}_{L,m}$ and
$\sum_{m\in{Z_{0}}}b_{R,m}w^{-}_{R,m}$, respectively.

\begin{remark}
  Each solution $u$ to the scattering problem
  \eqref{eq:sec3:Omegam:sca} extends to a solution of the full-domain
  problem  
  \eqref{eq:sec2:Omega:gov}--\eqref{eq:sec2:Omegar:expan} in $\Omega$.
  This extension is obtained by continuing $u$ into $\Omega_{L}$ and $\Omega_{R}$
  using the expansions \eqref{eq:sec3:Omegal:expan} and
  \eqref{eq:sec3:Omegar:expan}, respectively, together with the
  prescribed incident field \eqref{eq:sec2:inc} (see Proposition 3.1
  in \cite{anne94}). Likewise, the solution to the adjoint problem 
  \eqref{eq:sec3:Omegam:sca:adjoint} also defines a solution 
  of the full-domain problem
  through an analogous extension, with its incident field components 
  recovered via the inner products on $\Gamma_{L}$ and $\Gamma_{R}$. In what  
  follows, we do not distinguish between $u$ and its extension.  
\end{remark}

\subsection{Properties of $\mathcal{A}$}
To enable a local analytic continuation of $\mathcal{A}$ to
complex $k$, we define the square root in $\alpha_{m}$ with its
branch cut placed along the negative imaginary axis. The following
properties then hold for $\mathcal{A}$.

\begin{lemma}
  \label{lem:sec3:A}
  Consider the scattering problem \eqref{eq:sec3:Omegam:sca} at a
  point $(\beta_{0},\bm{\delta}_{0},k_{0})\in\Lambda$. The associated operator $\mathcal{A}$ defined
  via \eqref{eq:sec3:Omega:sesform} satisfies the 
  following properties:
  \begin{enumerate}[label=(\roman*).]
  \item $\mathcal{A}$ is Fredholm with zero index.
  \item $\mathrm{ker}(\mathcal{A})=\mathrm{ker}(\mathcal{A}^{*})$. Moreover, a
    function $u$ belongs to $\mathrm{ker}(\mathcal{A})$ if and only if $u$ is a
    BIC or $u=0$.    
  \item $\mathcal{A}$ is analytic in $k$. Furthermore, there exist radii
    $r_{1},r_{2}>0$    
    such that both $\mathcal{A}$ and $\partial_{k}\mathcal{A}$ are continuous on 
    $B_{r_{1}}((\beta_{0},\bm{\delta}_{0}))\times{\widehat{B}}_{r_{2}}(k_{0})$.
  \item $\mathcal{A}$ is $C^{1}$ in $\beta$. Furthermore, there exist radii
    $r_{1},r_{2}>0$ such that $\partial_{\beta}\mathcal{A}$ is continuous on 
    $B_{r_{1}}((\beta_{0},\bm{\delta}_{0}))\times{\widehat{B}}_{r_{2}}(k_{0})$.
  \item If the dielectric function $\epsilon$ is $C^{1}$ in
    $\bm{\delta}$, then $\mathcal{A}$ is also $C^{1}$ in $\bm{\delta}$. Furthermore, there
    exist radii    
    $r_{1},r_{2}>0$ such that $\partial_{\bm{\delta}}\mathcal{A}$ is continuous on 
    $B_{r_{1}}((\beta_{0},\bm{\delta}_{0}))\times{\widehat{B}}_{r_{2}}(k_{0})$.
  \end{enumerate}
  \begin{proof}
    We prove properties (i), (ii) and (iii). Properties
    (iv) and (v) follow by a similar argument.
    \begin{enumerate}[label=(\roman*).]
    \item We decompose $\mathcal{A}=\mathcal{A}_{1}+\mathcal{A}_{2}$ via their associated
      sesquilinear forms. For any      
      $u,v\in{H^{1}_{\mathrm{per},1}(\Omega_{0})}$, define
      \begin{align}
        (\mathcal{A}_{1}u,v)_{\Omega_{0}}:=&(\nabla{u},\nabla{v})_{\Omega_{0}}+(u,v)_{\Omega_{0}}-(\mathcal{D}_{L}u,v)_{\Gamma_{L}}-(\mathcal{D}_{R}u,v)_{\Gamma_{R}},\
                                       \label{eq:sec3:Omega:B1}\\
        (\mathcal{A}_{2}u,v)_{\Omega_{0}}:=&-2\bi\beta(\partial_{x_{1}}u,v)_{\Omega_{0}}+(\beta^{2}-1)(u,v)_{\Omega_{0}}-k^{2}(\epsilon{u},v)_{\Omega_{0}}.\label{eq:sec3:Omega:C}          
      \end{align}
      A direct computation shows that for any $u\in{H^{1}}(\Omega_{0})$
      \begin{equation}
        \label{eq:sec3:B:positive}
        \mathrm{Re}((\mathcal{A}_{1}u,u)_{\Omega_{0}})\ge\|u\|^{2}_{H^{1}_{\mathrm{per},1}(\Omega_{0})}.
      \end{equation}
      By the Lax-Milgram lemma (see Theorem 2.32 in \cite{mcl00}),
      this implies $\mathcal{A}_{1}$ is invertible. Furthermore, $\mathcal{A}_{2}$ is
      compact, because the inclusion
      $H^{1}_{\mathrm{per},1}(\Omega_{0}){\to}L^{2}(\Omega_{0})$ and its adjoint are
      compact. Consequently, $\mathcal{A}$ is the sum of an invertible operator
      and a compact operator, and is therefore Fredholm with zero index.
    \item Let $u$ be a BIC for the system
      \eqref{eq:sec2:Omega:gov}--\eqref{eq:sec2:Omegar:expan}. By
      definition, $u$ has vanishing incident and scattered     
      coefficients. This implies $u\in\mathrm{ker}(\mathcal{A})$ and
      $u\in\mathrm{ker}(\mathcal{A}^{*})$.      
      To prove the converse, suppose $u\in\mathrm{ker}(\mathcal{A})$ or
      $u\in\mathrm{ker}(\mathcal{A}^{*})$. A direct computation yields that    
      \begin{align}
        \mathrm{Im}((\mathcal{A}u,u)_{\Omega_{0}})=&\mathrm{Im}((u,\mathcal{A}^{*}u)_{\Omega_{0}})\notag\\
        =&\sum_{m\in{Z_{0}}}-\alpha_{m}|(u,\phi_{m})_{\Gamma_{R}}|^{2}+\sum_{m\in{Z_{0}}}-\alpha_{m}|(u,\phi_{m})_{\Gamma_{L}}|^{2}=0.\label{eq:sec3:Auu}           
      \end{align}
      This equation forces $(u,\phi_{m})_{\Gamma_{L}}=0$ and
      $(u,\phi_{m})_{\Gamma_{R}}=0$ for all $m\in{Z_{0}}$.      
      Therefore, $u$ is a BIC or $u=0$.      
      \item Referring to the expression for $\mathcal{A}$ in
        \eqref{eq:sec3:Omega:sesform}, it suffices to prove the
        holomorphic dependence on $k$ for the DtN operators
        $\mathcal{D}_{L}$ and $\mathcal{D}_{R}$ in a small neighborhood. We demonstrate
        this for $\mathcal{D}_{L}$. A direct        
        computation shows that for any $m_{1},m_{2}\in\mathbb{Z}$,
    \begin{equation}
      \label{eq:sec3:DtNexpress}
      (\mathcal{D}_{L}\phi_{m_{1}},\phi_{m_{2}})_{\Gamma_{L}}=\left\{
        \begin{aligned}
          &{\bi}\alpha_{m_{1}},\ &&\text{if}\ m_{1}=m_{2};\\
          &0,\ &&\text{if}\ m_{1}\ne{m_{2}}.
        \end{aligned}\right.
    \end{equation}
    Since $\alpha_{m}^{2}(\beta_{0},k_{0})=k^{2}_{0}-(m+\beta_{0})^{2}$
    decays as $|m|\to\infty$, 
    there exist radii $r_{1},r_{2}>0$ such that in the neighborhood
    ${B_{r_{1}}((\beta_{0},\bm{\delta}_{0})\times{\widehat{B}}_{r_{2}}(k_{0})}$:
    \begin{equation}
      \label{eq:sec3:etamto2}
      \mathrm{Re}(\alpha_{m}^{2})>\sigma_{0}^{2},\ \text{if}\ m\in{Z_{0}},\ \text{while}\
      \mathrm{Re}(\alpha_{m}^{2})<-\sigma_{0}^{2},\ \text{if}\ m\in\mathbb{Z}\backslash{Z_{0}},
    \end{equation}
    for some $\sigma_{0}>0$.
    This ensures that each $\alpha_{m}$ is holomorphic in $k$ in the
    neighborhood. Since the set $\{\phi_{m}\}$    
    is fundamental in    
    both $H^{1/2}_{\mathrm{per}}(\Gamma_{L})$ and    
    ${H}^{-1/2}_{\mathrm{per}}(\Gamma_{L})$, it follows that 
    $\mathcal{D}_{L}$ is holomorphic in $k$ by Theorem 3.12 in    
    \cite[Section III.3.1]{kato95}.
    We now establish the continuity. For any two points
    $(\beta_{1},\delta_{1},k_{1})$ and    
    $(\beta_{2},\delta_{2},k_{2})$ in this neighborhood, 
    \begin{align}
      \left|\alpha_{m}(k_{1},\beta_{1})-\alpha_{m}(k_{2},\beta_{2})\right|
      \le&\frac{\left|{k_{1}^{2}-k_{2}^{2}}-(\beta_{1}^{2}-\beta_{2}^{2})-2m(\beta_{1}-\beta_{2})\right|}{|\alpha_{m}(k_{1},\beta_{1})+\alpha_{m}(k_{2},\beta_{2})|}
         \notag\\      
      =&\mathcal{O}(|k_{1}-k_{2}|)+\mathcal{O}(|\beta_{1}-\beta_{2}|),
    \end{align}
    and
    \begin{align}
      \left|\partial_{k}\alpha_{m}(k_{1},\beta_{1})-\partial_{k}\alpha_{m}(k_{2},\beta_{2})\right|
      \le&\frac{|k_{1}-k_{2}|}{|\alpha_{m}(\beta_{1},k_{1})|}+\frac{|k_{2}(\alpha_{m}(\beta_{1},k_{1})-\alpha_{m}(\beta_{2},k_{2}))|}{|\alpha_{m}(k_{1},\beta_{1})\alpha_{m}(k_{2},\beta_{2})|}
         \notag\\      
      =&\mathcal{O}(|k_{1}-k_{2}|)+\mathcal{O}(|\beta_{1}-\beta_{2}|),
    \end{align}
    where the constants hidden in $\mathcal{O}$-terms are independent of $m$.
    Here, we have used \eqref{eq:sec3:etamto2} together with the
    boundedness of    
    $\beta$ and $k$ in the neighborhood. These estimates imply the 
    continuity of $\mathcal{D}_{L}$, $\mathcal{D}_{R}$, $\partial_{k}\mathcal{D}_{L}$ and
    $\partial_{k}\mathcal{D}_{R}$. Consequently, $\mathcal{A}$ and $\partial_{k}\mathcal{A}$ are continuous as
    well.    
    \end{enumerate}    
  \end{proof}
\end{lemma}

\subsection{Scattering matrix}
\label{sec:sca:scamat}
In this subsection, we define the scattering matrix and summarize its
properties, which have also been examined in
\cite{chesnel18,yuan19_4}.
Consider the scattering problem \eqref{eq:sec3:Omegam:sca}.
For $\flat\in\{L,R\}$ and $m\in{Z_{0}}$, define the function
\begin{equation}
  \label{eq:sec4:uLRm}
  v_{\flat,m}=-2\bi\alpha_{m}^{1/2}\mathcal{A}^{-1}\widetilde{\phi}_{\flat,m},
\end{equation}
which solves \eqref{eq:sec3:Omegam:sca} for an incident field
$w^{-}_{L,m}$ or $w_{R,m}^{+}$. The scattering matrix
$\bm{S}$ is then defined by
    \begin{equation}
    \label{eq:sec3:scaMat}
    \bm{S}:=
    \begin{bmatrix}
      [(v_{L,m_{2}},\alpha_{m_{1}}^{1/2}\phi_{m_{1}})_{\Gamma_{L}}]_{m_{1},m_{2}\in{Z_{0}}}&[(v_{L,m_{2}},\alpha_{m_{1}}^{1/2}\phi_{m_{1}})_{\Gamma_{R}}]_{m_{1},m_{2}\in{Z_{0}}}\\
[(v_{R,m_{2}},\alpha_{m_{1}}^{1/2}\phi_{m_{1}})_{\Gamma_{L}}]_{m_{1},m_{2}\in{Z_{0}}}&[(v_{R,m_{2}},\alpha_{m_{1}}^{1/2}\phi_{m_{1}})_{\Gamma_{R}}]_{m_{1},m_{2}\in{Z_{0}}}
    \end{bmatrix}-\bm{I}_{2N_{0}},
  \end{equation}
  with rows indexed by the pair $(\flat,m_{1})$ (for $\flat\in\{L,R\}$) and 
  columns indexed by $(\flat,m_{2})$.

\begin{lemma}
  \label{lem:sec3:scamat}
  For the scattering problem \eqref{eq:sec3:Omegam:sca} at a
  point $(\beta,\bm{\delta},k)\in\Lambda$, the following properties hold:
  \begin{enumerate}[label=(\roman*).]
  \item The scattering matrix $\bm{S}$ is well-defined and unitary;
  \item The incident and scattered coefficient vectors $\bm{a}$ and
    $\bm{b}$ of a solution $u$ are related by
    \begin{equation}
      \label{eq:sec3:Sab}
     \bm{b}=\bm{S}\bm{a}; 
   \end{equation}
  \item If the dielectric function $\epsilon$ has reflection symmetry in $x_{1}$,
    then $\bm{S}=\bm{S}^{T}$;
  \item If the dielectric function $\epsilon$ has reflection symmetry in $x_{2}$,
   then $\bm{S}=\bm{S}^{P}$.       
  \end{enumerate}
  \begin{proof}
    \begin{enumerate}[label=(\roman*).]
      \item If $\mathrm{ker}(\mathcal{A})=\emptyset$, the operator $\mathcal{A}$ is invertible and the
        scattering        
  matrix $\bm{S}$ is well-defined. Now, suppose
  $\mathrm{ker}(\mathcal{A})\ne\emptyset$. By Lemma \ref{lem:sec3:A} (ii),  
  $\mathrm{ker}(\mathcal{A})$ consists of only BICs and satisfies
  $\mathrm{ker}(\mathcal{A})=\mathrm{ker}(\mathcal{A}^{*})$. Consequently, the functions  
  $\widetilde{\phi}_{L,m}$ and $\widetilde{\phi}_{R,m}$ defined in
 \eqref{eq:sec3:phitildestar} are orthogonal to
 $\mathrm{ker}(\mathcal{A}^{*})$. The Fredholm property of $\mathcal{A}$ then guarantees
 that the scattering matrix \eqref{eq:sec3:scaMat} is well-defined. 
 The unitary property of $\bm{S}$ follows from an application of the
  second Green  
  identity (Theorem 4.4 in \cite{mcl00}):  
  \begin{equation}
    \label{eq:sec3:secGreen}
    (\mathcal{L}u_{1},u_{2})_{\Omega_{0}}-(u_{1},\mathcal{L}u_{2})_{\Omega_{0}}=(u_{1},\partial_{\bm{\nu}}u_{2})_{\partial{\Omega_{0}}}-(\partial_{\bm{\nu}}u_{1},u_{2})_{\partial{\Omega_{0}}},
  \end{equation}
  where $\bm{\nu}$ is the outward unit normal on $\partial\Omega_{0}$. The identity
  is applied by substituting $u_{1}$ and $u_{2}$ with the scattering
  solutions  
  $\mathcal{A}^{-1}\alpha_{m}^{1/2}\widetilde{\phi}_{\flat,m}$ for $\flat\in\{L,R\}$ and  
  $m\in{Z_{0}}$ (cf. \cite{chesnel18}). 
\item This follows directly from the definitions of the incident and
  scattered coefficients in the expansions 
  \eqref{eq:sec2:Omegal:expan}--\eqref{eq:sec2:Omegar:expan} and the
  scattering matrix in \eqref{eq:sec3:scaMat}.
\item Let $u(\bm{x})$ be a solution to
  \eqref{eq:sec3:Omegam:sca} with incident and scattered coefficient
  vectors $\bm{a}$ and $\bm{b}$. A direct
  computation shows that $\overline{u(-x_{1},x_{2})}$ is also a solution,
  with incident and scattered coefficient vectors
  $\overline{\bm{b}}$ and
  $\overline{\bm{a}}$. Therefore, by the definition of the
  scattering matrix,
  \begin{equation}
    \label{eq:sec3:Sconjba}
    \bm{S}\overline{\bm{b}}=\overline{\bm{a}}.
  \end{equation}
  Substituting $\bm{b}=\bm{S}\bm{a}$ into this relation gives 
  $\bm{S}\overline{\bm{S}}\overline{\bm{a}}=\overline{\bm{a}}$.
  Since the incident coefficient vector $\bm{a}$ can be chosen
  arbitrarily, it follows that
    \begin{equation}
    \label{eq:sec3:SSconj}
    \bm{S}\overline{\bm{S}}=\bm{I}_{2N_{0}}.
  \end{equation}
  This establishes the desired result.
  \item If $u(\bm{x})$ solves
  \eqref{eq:sec3:Omegam:sca} with incident and scattered coefficient
  vectors $\bm{a}$ and $\bm{b}$, then $u(x_{1},-x_{2})$ is
  also a solution whose corresponding coefficient vectors are
  $\bm{R}_{2N_{0}}\bm{a}$ and $\bm{R}_{2N_{0}}\bm{b}$. The remaining
  steps follow exactly as in the proof of (iii).
 \end{enumerate}
\end{proof}
\end{lemma}

\section{Local structure of $\lambda_{\emph{M}}$ near a simple BIC point}
We analyze the local structure of $\lambda_{\bm{M}}$ near a simple BIC point
by using the implicit function theorem. Consider a simple BIC, denoted
$u_{*}$, for the scattering problem
\eqref{eq:sec3:Omegam:sca} at the point $(\beta_{*},\bm{\delta}_{*},k_{*})\in\Lambda$. We
define the space $\mathbb{H}(\{u_{*}\})$ as
\begin{equation}
  \label{eq:sec4:Hspace}
  \mathbb{H}(\{u_{*}\}):=\left\{\psi\in{H^{1}_{\mathrm{per},1}(\Omega_{0})}:(u_{*},\psi)_{\Omega_{0}}=0\right\}.  
\end{equation}
Let $\bm{S}_{0}$ be the scattering matrix at 
$(\beta_{*},\bm{\delta}_{*},k_{*})$. For any matrix $\bm{M}\in{U(2N_{0})}$, 
we define the operator
$\widehat{\mathcal{A}}_{\bm{M}}:\Lambda\times\mathbb{H}(\{u_{*}\})\times\mathbb{C}^{2N_{0}}\to{(H^{1}_{\mathrm{per},1}(\Omega_{0}))^{*}}\times\mathbb{C}^{2N_{0}}$
by
\begin{equation}
     \label{eq:sec4:AM}
  \widehat{\mathcal{A}}_{\bm{M}}(\beta,{\bm{\delta}},k,\psi,\bm{a})
  :=\begin{bmatrix}
    \mathcal{A}(\beta,{\bm{\delta}},k)(\psi+u_{*})+2\bi\bm{a}\cdot[\alpha_{m}^{1/2}(\beta,k)\widetilde{\phi}_{\flat,m}]_{\flat\in\{L,R\},m\in{Z_{0}}}\\    
    [\alpha_{m}^{1/2}(\beta,k)\widehat{\phi}_{\flat,m}\psi]_{\flat\in\{L,R\},m\in{Z_{0}}}-(\bm{I}_{2N_{0}}+\bm{M})\bm{a}
  \end{bmatrix}. 
\end{equation}
Substituting $u=\psi+u_{*}$, the first row of \eqref{eq:sec4:AM}
corresponds to the standard scattering problem
\eqref{eq:sec3:Omegam:sca}, while the second row enforces that $u$ is
governed by $\bm{M}$.
The derivative of this operator with respect to the combined variable
$Y:=(k,\psi,\bm{a})$ is the bounded linear operator
\begin{equation}
    \label{eq:sec4:Aderivative}
    \partial_{Y}\widehat{\mathcal{A}}_{\bm{M}}:=
    \begin{bmatrix}
     \partial_{k}\widehat{\mathcal{A}}_{\bm{M}}&\partial_{\psi}\widehat{\mathcal{A}}_{\bm{M}}&\partial_{\bm{a}}\widehat{\mathcal{A}}_{\bm{M}}
    \end{bmatrix},
\end{equation}
which maps
$\mathbb{C}\times\mathbb{H}(\{u_{*}\})\times\mathbb{C}^{2N_{0}}$ to
${(H^{1}_{\mathrm{per},1}(\Omega_{0}))^{*}}\times\mathbb{C}^{2N_{0}}$ and is
defined at every point in the domain
${\Lambda}\times\mathbb{H}(\{u_{*}\})\times\mathbb{C}^{2N_{0}}$.

\begin{lemma}
  \label{lem:sec4:Ahatinvertible}
  Let $u_{*}$ be a simple BIC and $\bm{S}_{0}$ be the scattering matrix
  at the point
  $(\beta_{*},\bm{\delta}_{*},k_{*})\in\Lambda$. For every $\bm{M}\in{U}_{1}$,
  the derivative $\partial_{Y}\widehat{\mathcal{A}}_{\bm{M}}$, defined in
  \eqref{eq:sec4:Aderivative}, is invertible at the point
  $(\beta_{*},\bm{\delta}_{*},k_{*},0,\bm{0})$.  
  \begin{proof}
    We prove the invertibility of $\partial_{Y}\widehat{\mathcal{A}}_{\bm{M}}$ at
    $(\beta_{*},{\bm{\delta}}_{*},k_{*},0,\bm{0})$ by showing that it is both
    injective and surjective. The desired result then follows    
    from the open mapping theorem. For simplicity, we introduce the
    flowing notation:
    \begin{equation}
      \label{eq:sec4:AB:alphaB}
      \begin{aligned}
        \alpha_{m,*}:=&\alpha_{m}(\beta_{*},k_{*}),\
        \epsilon_{*}:=\epsilon(\cdot,\bm{\delta}_{*}),\
        \mathcal{A}_{0}:=\mathcal{A}(\beta_{*},\bm{\delta}_{*},k_{*}),\\
        \partial_{k}\mathcal{A}_{0}:=&\partial_{k}\mathcal{A}(\beta_{*},\bm{\delta}_{*},k_{*})\ \text{and}\
      \partial_{k}\mathcal{D}_{\flat,0}:=\partial_{k}\mathcal{D}_{\flat}(\beta_{*},k_{*}),\ \text{for}\ \flat\in\{L,R\}.
      \end{aligned}
    \end{equation}

    Suppose 
    \begin{equation}
      \label{eq:sec4:kerhatA:1}
     \partial_{k}\widehat{\mathcal{A}}_{\bm{M}}k+\partial_{\psi}\widehat{\mathcal{A}}_{\bm{M}}\psi+\partial_{\bm{a}}\widehat{\mathcal{A}}_{\bm{M}}\bm{a}=0,\
     \text{at}\ (\beta_{*},{\bm{\delta}}_{*},k_{*},0,\bm{0}),
   \end{equation}
   for some $(k,\psi,\bm{a})\in\mathbb{C}\times\mathbb{H}(\{u_{*}\})\times\mathbb{C}^{2N_{0}}$.
    This is equivalent to 
    \begin{equation}
      \label{eq:sec4:kerhatA:2}
      \left\{
        \begin{aligned}
          &\mathcal{A}_{0}\psi=-2\bi\bm{a}\cdot[\alpha_{m,*}^{1/2}\widetilde{\phi}_{\flat,m}]_{\flat\in\{L,R\},m\in{Z_{0}}}-k\partial_{k}\mathcal{A}_{0}u_{*};\\
          &[\alpha_{m,*}^{1/2}\widehat{\phi}_{\flat,m}\psi]_{\flat\in\{L,R\},m\in{Z_{0}}}=(\bm{I}_{2N_{0}}+\bm{M})\bm{a},            
        \end{aligned}\right.
    \end{equation}
    A direct computation yields
    \begin{align}
      (\partial_{k}\mathcal{A}_{0}u_{*},u_{*})_{\Omega_{0}}=&-2k_{*}({\epsilon}_{*}u_{*},u_{*})_{\Omega_{0}}-(\partial_{k}\mathcal{D}_{L,0}u_{*},u_{*})_{\Gamma_{L}}\notag\\
                                      &-(\partial_{k}\mathcal{D}_{R,0}u_{*},u_{*})_{\Gamma_{R}}\notag\\      
      <&0.      \label{eq:sec4:partialkAubic}
    \end{align}
    It is clear that the right-hand side of the first equation in
    \eqref{eq:sec4:kerhatA:2} lies in $\mathrm{ran}(\mathcal{A}_{0})$, which is
    orthogonal to $\mathrm{ker}(\mathcal{A}_{0}^{*})$.    
    The condition \eqref{eq:sec4:partialkAubic} and Lemma
    \ref{lem:sec3:A} (ii) together imply that $k=0$.
    Substituting $k=0$ into \eqref{eq:sec4:kerhatA:2} yields the
    reduced system:
    \begin{equation}
      \label{eq:sec4:kerhatA:3}
      \left\{
        \begin{aligned}
          &\mathcal{A}_{0}\psi=-2\bi\bm{a}\cdot[\alpha_{m,*}^{1/2}\widetilde{\phi}_{\flat,m}]_{\flat\in\{L,R\},m\in{Z_{0}}};\\
          &[\alpha_{m,*}^{1/2}\widehat{\phi}_{\flat,m}\psi]_{\flat\in\{L,R\},m\in{Z_{0}}}=(\bm{I}_{2N_{0}}+\bm{M})\bm{a}.
        \end{aligned}\right.
    \end{equation}
    The first equation in \eqref{eq:sec4:kerhatA:3} identifies $\psi$ as
    a scattering solution with    
    incident coefficient vector $\bm{a}$. By Lemma
    \ref{lem:sec3:scamat} (ii), the second
    equation can be rewritten as 
    \begin{equation}
      \label{eq:sec4:kerhatA:4}
      (\bm{S}_{0}-\bm{M})\bm{a}=0.
    \end{equation}
    Since $\bm{M}\in{U_{1}}$ which requires
    $\mathrm{det}(\bm{S}_{0}-\bm{M})\ne0$, it follows    
    that $\bm{a}=0$. The first equation in
    \eqref{eq:sec4:kerhatA:3} then reduces to $\mathcal{A}_{0}\psi=0$. As    
    $\psi\in\mathbb{H}(\{u_{*}\})$ is orthogonal to
    $\mathrm{ker}(\mathcal{A}_{0})$, we conclude that
    $\psi=0$. This establishes the injectivity.    

    We now prove that $\partial_{Y}\widehat{\mathcal{A}}_{\bm{M}}$ is surjective at the
    point $(\beta_{*},{\bm{\delta}}_{*},k_{*},0,\bm{0})$. Let    
    $f\in(H^{1}_{\mathrm{per},1}(\Omega_{0}))^{*}$ and
    $\bm{e}\in\mathbb{C}^{2N_{0}}$ be arbitrary.
     First, define the scalar $k$ as
    \begin{equation}
      \label{eq:sec4:invAextension:k}
      k=\frac{(f,u_{*})_{\Omega_{0}}}{(\partial_{k}\mathcal{A}_{0}u_{*},u_{*})_{\Omega_{0}}}.
    \end{equation}
    Next, let $\psi_{0}\in{\mathbb{H}(\{u_{*}\})}$ be the unique
    solution to the inhomogeneous problem
    \begin{equation}
      \label{eq:sec4:invAextension:u0}
      \psi_{0}:=\mathcal{A}_{0}^{-1}(f-k\partial_{k}\mathcal{A}_{0}u_{*}).
    \end{equation}
    With $k$ and $\psi_{0}$ defined, the problem of finding a preimage
    for $(f,\bm{e})$ reduces to solving the following system for $\psi$
    and $\bm{a}$:
    \begin{equation}
      \label{eq:sec4:ranhatA}
      \left\{
        \begin{aligned}
          &\mathcal{A}_{0}(\psi+\psi_{0})=-2\bi\bm{a}\cdot[\alpha_{m,*}^{1/2}\widetilde{\phi}_{\flat,m}]_{\flat\in\{L,R\},m\in{Z_{0}}};\\
          &[\alpha_{m,*}^{1/2}\widehat{\phi}_{\flat,m}\psi]_{\flat\in\{L,R\},m\in{Z_{0}}}=(\bm{I}_{2N_{0}}+\bm{M})\bm{a}+\bm{e}.
        \end{aligned}\right.
    \end{equation}
    Invoking the definition of the scattering matrix $\bm{S}_{0}$ from
    \eqref{eq:sec3:scaMat}, this system is equivalent to
    \begin{equation}
      \label{eq:sec4:imAextension:scamat}
      \left\{
        \begin{aligned}
          &\mathcal{A}_{0}(\psi+\psi_{0})=-2\bi\bm{a}\cdot[\alpha_{m,*}^{1/2}\widetilde{\phi}_{\flat,m}]_{\flat\in\{L,R\},m\in{Z_{0}}};\\
          &(\bm{S}_{0}-\bm{M})\bm{a}=[\alpha_{m,*}^{1/2}\widehat{\phi}_{\flat,m}\psi_{0}]_{\flat\in\{L,R\},m\in{Z_{0}}}+\bm{e}.
        \end{aligned}\right.
    \end{equation}
    Since $\bm{M}\in{U_{1}}$, the matrix $\bm{S}_{0}-\bm{M}$ is invertible. 
    We can therefore solve the system explicitly:
    \begin{align}
      \bm{a}=&(\bm{S}_{0}-\bm{M})^{-1}([\alpha_{m,*}^{1/2}\widehat{\phi}_{\flat,m}\psi_{0}]_{\flat\in\{L,R\},m\in{Z_{0}}}+\bm{e}),\label{eq:sec4:imAextension:a}\\
      \psi=&\mathcal{A}_{0}^{-1}(-2\bi\bm{a}\cdot[\alpha_{m,*}^{1/2}\widetilde{\phi}_{\flat,m}]_{\flat\in\{L,R\},m\in{Z_{0}}})-\psi_{0}.\label{eq:sec4:imAextension:u}      
    \end{align}
    This establishes the surjectivity.
  \end{proof}
\end{lemma}

\begin{lemma}
  \label{lem:sec4:local:st}
  Let $u_{*}$ be a simple BIC and $\bm{S}_{0}$ be the scattering matrix
  at the point
  $(\beta_{*},\bm{\delta}_{*},k_{*})\in\Lambda$. For every $\bm{M}\in{U}_{1}$, there
  exist radii $r_{3},r_{4}>0$ such that for all
  $(\beta,\bm{\delta})\in{B_{r_{3}}((\beta_{*},\bm{\delta}_{*}))}$ and
  $k\in{B}_{r_{4}}(k_{*})$, if $u$ is a scattering solution to
  \eqref{eq:sec3:Omegam:sca} at $(\beta,\bm{\delta},k)$
  governed by $\bm{M}$, then $(u,u_{*})_{\Omega_{0}}\ne0$.
  \begin{proof}
    Let $\{(\beta_{j},\bm{\delta}_{j},k_{j})\}_{j\in\mathbb{N}}$ be a sequence in $\Lambda$ such that
    \begin{equation}
      \label{eq:sec4:betan}
      (\beta_{j},\bm{\delta}_{j},k_{j})\to(\beta_{*},\bm{\delta}_{*},k_{*})\ \text{as}\ j\to\infty.
    \end{equation}
    Define $\epsilon_{j}:=\epsilon(\cdot,\bm{\delta}_{j})$ and let
    $u_{j}\in{H^{1}_{\mathrm{per},1}}(\Omega_{0})$ be a scattering solution to    
    \eqref{eq:sec3:Omegam:sca} at $(\beta_{j},\bm{\delta}_{j},k_{j})$,
    normalized by $\|u_{j}\|_{H^{1}_{\mathrm{per},1}(\Omega_{0})}=1$.
    Let $\bm{a}_{j}$ and $\bm{b}_{j}$ denote the incident and scattered
    coefficient vectors of $u_{j}$, respectively. Since 
    $\bm{a}_{j}+\bm{b}_{j}$ is related to the trace of $u_{j}$,
    the boundedness of $\{u_{j}\}$ in
    $H^{1}_{\mathrm{per},1}(\Omega_{0})$ implies that
    $\{\bm{a}_{j}+\bm{b}_{j}\}$ is bounded. Moreover, from the
    variational formulation
    \begin{equation}
      \label{eq:sec4:Auj}
    (\mathcal{A}(\beta_{j},\bm{\delta}_{j},k_{j})u_{j},v)_{\Omega_{0}}=-2\bi\bm{a}_{j}\cdot[\alpha_{m,j}^{1/2}(\phi_{m},v)_{\Gamma_{\flat}}]_{\flat\in\{L,R\},m\in{Z_{0}}}\
    \text{for}\ v\in{H^{1}_{\mathrm{per},1}(\Omega_{0})},
  \end{equation}
  where $\alpha_{m,j}:=\alpha_{m}(\beta_{j},k_{j})$ for $m\in{Z_{0}}$, we can also deduce
  the boundedness of  
    $\{\bm{a}_{j}-\bm{b}_{j}\}$. Consequently, both $\{\bm{a}_{j}\}$ and
    $\{\bm{b}_{j}\}$ are bounded sequences.

    We next prove that $\{u_{j}\}$ is relatively compact in
    $H^{1}_{\mathrm{per},1}(\Omega_{0})$.    
    For any $j_{1},j_{2}\in\mathbb{N}$, define the diagonal matrix
    \begin{equation}
      \label{eq:sec4:Dj1j2}
      \bm{D}_{j_{1},j_{2}}:=
      \begin{bmatrix}
        \mathrm{diag}\Big(\left\{\frac{\alpha_{m,j_{1}}}{\alpha_{m,j_{2}}}\right\}_{m\in{Z_{0}}}\Big)&\\                                                                  &\mathrm{diag}\Big(\left\{\frac{\alpha_{m,j_{1}}}{\alpha_{m,j_{2}}}\right\}_{m\in{Z_{0}}}\Big)                 
      \end{bmatrix}.
    \end{equation}
    A direct computation gives the identity
    \begin{align}
      (\nabla(u_{j_{1}}-u_{j_{2}}),\nabla(u_{j_{1}}-u_{j_{2}}))_{\Omega_{0}}=&(\nabla{u_{j_{1}}},\nabla(u_{j_{1}}-u_{j_{2}}))_{\Omega_{0}}-(\nabla{u_{j_{2}}},\nabla(u_{j_{1}}-u_{j_{2}}))_{\Omega_{0}}\notag\\
      =&\Pi_{1}+\Pi_{2}+\Pi_{3},      \label{eq:sec4:deltau:identity}
    \end{align}
    where
    \begin{align}
      \Pi_{1}:=&2\bi\beta_{j_{1}}(\partial_{x_{1}}u_{j_{1}},u_{j_{1}}-u_{j_{2}})_{\Omega_{0}}-\beta_{j_{1}}^{2}(u_{j_{1}},u_{j_{1}}-u_{j_{2}})_{\Omega_{0}}+k_{j_{1}}^{2}(\epsilon_{j_{1}}u_{j_{1}},u_{j_{1}}-u_{j_{2}})_{\Omega_{0}};\notag\\
             &-2\bi\beta_{j_{2}}(\partial_{x_{1}}u_{j_{2}},u_{j_{1}}-u_{j_{2}})_{\Omega_{0}}+\beta_{j_{2}}^{2}(u_{j_{2}},u_{j_{1}}-u_{j_{2}})_{\Omega_{0}}\notag\\
             &-k_{j_{2}}^{2}(\epsilon_{j_{2}}u_{j_{2}},u_{j_{1}}-u_{j_{2}})_{\Omega_{0}};\label{eq:sec4:Pi1}\\
      \Pi_{2}:=&\sum_{\flat\in\{L,R\}}\sum_{m\in{Z_{0}}}\bi\alpha_{m,j_{1}}(u_{j_{1}},\phi_{m})_{\Gamma_{\flat}}(\phi_{m},u_{j_{1}}-u_{j_{2}})_{\Gamma_{\flat}}\notag\\
             &-\sum_{\flat\in\{L,R\}}\sum_{m\in{Z_{0}}}\bi\alpha_{m,j_{2}}(u_{j_{2}},\phi_{m})_{\Gamma_{\flat}}(\phi_{m},u_{j_{1}}-u_{j_{2}})_{\Gamma_{\flat}}\notag\\
             &+\sum_{\flat\in\{L,R\}}\sum_{m\in{\mathbb{Z}{\backslash}Z_{0}}}\bi(\alpha_{m,j_{1}}-\alpha_{m,j_{2}})(u_{j_{2}},\phi_{m})_{\Gamma_{\flat}}(\phi_{m},u_{j_{1}}-u_{j_{2}})_{\Gamma_{\flat}}\notag\\
             &-2\bi\bm{a}_{j_{1}}\cdot[\alpha_{m,j_{1}}^{1/2}(\phi_{m},u_{j_{1}}-u_{j_{2}})_{\Gamma_{\flat}}]_{\flat\in\{L,R\},m\in{Z_{0}}}\notag\\
             &+2\bi\bm{a}_{j_{2}}\cdot[\alpha_{m,j_{2}}^{1/2}(\phi_{m},u_{j_{1}}-u_{j_{2}})_{\Gamma_{\flat}}]_{\flat\in\{L,R\},m\in{Z_{0}}};\label{eq:sec4:Pi2}\\
      \Pi_{3}:=&\sum_{\flat\in\{L,R\}}\sum_{m\in{\mathbb{Z}\backslash{Z_{0}}}}\bi\alpha_{m,j_{1}}(u_{j_{1}}-u_{j_{2}},\phi_{m})_{\Gamma_{\flat}}(\phi_{m},u_{j_{1}}-u_{j_{2}})_{\Gamma_{\flat}}.
    \end{align}
    Using the boundedness of the sequences $\{(\beta_{j},\bm{\delta}_{j},k_{j})\}$,
    $\{u_{j}\}$, $\{\bm{a}_{j}\}$ and $\{\bm{b}_{j}\}$, we obtain the
    estimates:
    \begin{align}
      \Pi_{1}=&\mathcal{O}(\|u_{j_{1}}-u_{j_{2}}\|_{L^{2}(\Omega_{0})});\label{eq:sec4:Pi1:est}\\
      \Pi_{2}=&\bi\|\bm{a}_{j_{1}}+\bm{b}_{j_{1}}-\bm{a}_{j_{2}}-\bm{b}_{j_{1}}\|_{\mathbb{C}^{2N_{0}}}^{2}+\bi((\bm{D}_{j_{1},j_{2}}^{1/2}-\bm{I}_{2N_{0}})(\bm{a}_{j_{1}}+\bm{b}_{j_{1}}))\cdot\overline{(\bm{a}_{j_{2}}+\bm{b}_{j_{2}})}\notag\\
            &+\bi((\bm{D}_{j_{1},j_{2}}^{1/2}-\bm{I}_{2N_{0}})(\bm{a}_{j_{2}}+\bm{b}_{j_{2}}))\cdot\overline{(\bm{a}_{j_{1}}+\bm{b}_{j_{1}})}\notag\\      
             &+\sum_{\flat\in\{L,R\}}\sum_{m\in{\mathbb{Z}{\backslash}Z_{0}}}\bi(\alpha_{m,j_{1}}-\alpha_{m,j_{2}})(u_{j_{2}},\phi_{m})_{\Gamma_{\flat}}(\phi_{m},u_{j_{1}}-u_{j_{2}})_{\Gamma_{\flat}}\notag\\
            &-2\bi(\bm{a}_{j_{1}}-\bm{D}_{j_{1},j_{2}}^{-1/2}\bm{a}_{j_{2}})\cdot\overline{(\bm{a}_{j_{1}}+\bm{b}_{j_{1}}-\bm{D}_{j_{1},j_{2}}(\bm{a}_{j_{2}}+\bm{b}_{j_{2}}))}\notag\\
      =&\mathcal{O}(\|\bm{a}_{j_{1}}+\bm{b}_{j_{1}}-\bm{a}_{j_{2}}-\bm{b}_{j_{1}}\|_{\mathbb{C}^{2N_{0}}})+\mathcal{O}(|k_{j_{1}}-k_{j_{2}}|)+\mathcal{O}(|\beta_{j_{1}}-\beta_{j_{2}}|);\label{eq:sec4:Pi2:est}\\
      \Pi_{3}<&0\label{eq:sec4:Pi3:est},
    \end{align}
    where the constants hidden in $\mathcal{O}$-terms are independent of
    $j_{1}$ and $j_{2}$.    
    Since the inclusion $H^{1}_{\mathrm{per},1}(\Omega_{0})\to{L^{2}(\Omega_{0})}$
    is compact, the sequence $\{u_{j}\}$ is relatively    
    compact in $L^{2}(\Omega_{0})$. Furthermore, the boundedness of
    $\{\bm{a}_{j}\}$ and $\{\bm{b}_{j}\}$ implies they
    are also relatively compact in $\mathbb{C}^{2N_{0}}$.
    From the estimates above together with 
    the identity \eqref{eq:sec4:deltau:identity}, we conclude that
    $\{u_{j}\}$ is relatively compact in $H^{1}_{\mathrm{per},1}(\Omega_{0})$.

    Finally, we prove the lemma by contradiction. Assume that $u_{j}$ is
    governed by $\bm{M}$ and that $(u_{j},u_{*})_{\Omega_{0}}=0$ for all
    $j\in\mathbb{N}$. From the preceding analysis, $\{u_{j}\}$ admits a
    subsequence $\{u_{j_{n}}\}$ that     
    converges in $H^{1}_{\mathrm{per},1}(\Omega_{0})$. Let $u_{\dagger}$ denote its
    limit, and let $\bm{a}_{\dagger}$
    and $\bm{b}_{\dagger}$ be the corresponding incident and scattered
    coefficient vectors. Because the convergence holds in
    $H^{1}_{\mathrm{per},1}(\Omega_{0})$, we obtain
    \begin{equation}
      \label{eq:sec4:udagger}
      (u_{\dagger},u_{*})_{\Omega_{0}}=0,\
      \|u_{\dagger}\|_{H^{1}_{\mathrm{per},1}(\Omega_{0})}=1\ \text{and}\
      \bm{b}_{\dagger}=\bm{M}\bm{a}_{\dagger}.
    \end{equation}
    If $u_{\dagger}$ is a BIC, then \eqref{eq:sec4:udagger} contradicts the
    assumption that $u_{*}$ is simple. If $u_{\dagger}$ is a propagating
    field, then \eqref{eq:sec4:udagger} together with the scattering
    relation $\bm{b}_{\dagger}=\bm{S}_{0}\bm{a}_{\dagger}$ implies
    \begin{equation}
      \label{eq:sec4:bdagger}
      (\bm{S}_{0}-\bm{M})\bm{a}_{\dagger}=\bm{0},
    \end{equation}
    which contradicts the hypothesis $\bm{M}\in{U_{1}}$.
  \end{proof}
\end{lemma}

The following theorem is the central result of this work.
\begin{theorem}
  \label{thm:sec4:PiM:bic}
  Let $u_{*}$ be a simple BIC and $\bm{S}_{0}$ be the scattering matrix
  at the point
  $(\beta_{*},{\bm{\delta}}_{*},k_{*})\in\Lambda$. Given any $\bm{M}\in{U}_{1}$,
  there exist radii $r_{5},r_{6}>0$ such that for every
  $(\beta,{\bm{\delta}})\in{B_{r_{5}}((\beta_{*},{\bm{\delta}}_{*}))}$, we can find unique  
  $k(\beta,{\bm{\delta}})\in{B_{r_{6}}(k_{*})}$,
  $u(\cdot,{\beta,{\bm{\delta}}})\in{H^{1}_{\mathrm{per},1}(\Omega_{0})}$ and
  $\bm{a}({\beta,\bm{\delta}})\in\mathbb{C}^{2N_{0}}$  
  satisfying  
  \begin{equation}
    \label{eq:sec4:ku:condition}
    \left\{
      \begin{aligned}       
        &(\beta,{\bm{\delta}},k(\beta,\bm{\delta}))\in\lambda_{\bm{M}};\\
        &u(\cdot,\beta,\bm{\delta})\ \text{satisfies}\
          \eqref{eq:sec3:Omegam:sca}\ \text{with\ incident\
          coefficient\ vector}\ \bm{a}(\beta,\bm{\delta});\\          
        &u(\cdot,\beta,\bm{\delta})\ \text{is\ governed\ by}\ \bm{M}  \text{and}\
          (u(\cdot,\beta,\bm{\delta}),u_{*})_{\Omega_{0}}=\|u_{*}\|_{L^{2}(\Omega_{0})}^{2}.
      \end{aligned}\right.
  \end{equation}
  Furthermore, 
  $k(\beta,\bm{\delta})$, $u(\cdot,\beta,\bm{\delta})$ and $\bm{a}(\beta,\bm{\delta})$ depend
  continuously on $(\beta,{\bm{\delta}})$ and the
  following conditions hold:
  \begin{equation}
    \label{eq:sec4:ku:condition2}
    k(\beta_{*},\bm{\delta}_{*})=k_{*},\
    u(\cdot,\beta_{*},\bm{\delta}_{*})=u_{*}\ \text{and}\ \bm{a}(\beta_{*},\bm{\delta}_{*})=\bm{0}.
  \end{equation}  
  \begin{proof}
    By the definition of $\lambda_{\bm{M}}$ in \eqref{eq:sec2:PiM}, we have
    $(\beta_{*},\bm{\delta}_{*},k_{*})\in\lambda_{\bm{M}}$. Lemma \ref{lem:sec3:A} (iii)
    guarantees the existence of radii 
    $r_{1},r_{2}>0$ such that both $\widehat{\mathcal{A}}_{\bm{M}}$ and
    $\partial_{k}\widehat{\mathcal{A}}_{\bm{M}}$ are    
    continuous on
    $B_{r_{1}}((\beta_{*},{\bm{\delta}}_{*}))\times\widehat{B}_{r_{2}}(k_{*})\times\mathbb{H}(\{u_{*}\})\times\mathbb{C}^{2N_{0}}$. A
    direct computation confirms that
    \begin{equation}
      \label{eq:sec4:Ahat:0}
      \widehat{\mathcal{A}}_{\bm{M}}(\beta,\bm{\delta},k,\psi,\bm{a})=
      \begin{bmatrix}
        0\\
        \bm{0}
      \end{bmatrix},
    \end{equation}
    at the point $(\beta_{*},\bm{\delta}_{*},k_{*},0,\bm{0})$.
    Since Lemma \ref{lem:sec4:Ahatinvertible} establishes the
    invertibility of the derivative $\partial_{Y}\widehat{\mathcal{A}}_{\bm{M}}$ at this
    point, the hypotheses of the implicit function theorem
    (cf. Theorem 1.2.1 in    
    \cite{chang03}) are satisfied. Hence, there exist $r_{5},r_{6}>0$
    such that for every $(\beta,{\bm{\delta}})\in{B_{r_{5}}((\beta_{*},{\bm{\delta}}_{*}))}$,
    \eqref{eq:sec4:Ahat:0} admits a unique solution
    \begin{equation}
      \label{eq:sec4:Ahat:solution}
      (\beta,\bm{\delta},k(\beta,\bm{\delta}),\psi(\cdot,\beta,\bm{\delta}),\bm{a}(\beta,\bm{\delta}))
    \end{equation}
    with $k(\beta,\bm{\delta})\in{\widehat{B}_{r_{6}}(k_{*})}$,
    $\|\psi(\cdot,\beta,\bm{\delta})\|_{\mathbb{H}(\{u_{*}\})}\le{r_{6}}$ and
    $\bm{a}(\beta,\bm{\delta})\in{\widehat{B}_{r_{6}}(\bm{0})}$. These functions
    depend continuously on $(\beta,\bm{\delta})$. Defining
    \begin{equation}
      \label{eq:sec4:ubetadelta}
     u(\cdot,\beta,\bm{\delta}):= \psi(\cdot,\beta,\bm{\delta})+u_{*},
   \end{equation}
   the triple $(k,u,\bm{a})$ satisfies
    \eqref{eq:sec4:ku:condition} and
    \eqref{eq:sec4:ku:condition2}. 

  We now prove that the frequency $k(\beta,\bm{\delta})$ is real for a
  sufficiently small radius $r_{5}$.  
  Let $\bm{b}(\beta,\bm{\delta})$ be the scattered coefficient vector
  corresponding to $u(\cdot,\beta,\bm{\delta})$. Applying the second Green identity
  \eqref{eq:sec3:secGreen} with $u_{1}=u_{2}=u(\cdot,\beta,\bm{\delta})$ yields: 
    \begin{align}
      &4\mathrm{Re}(k(\beta,\bm{\delta}))({\epsilon}(\cdot,\bm{\delta})u(\cdot,\beta,\bm{\delta}),u(\cdot,\beta,\bm{\delta}))_{\Omega_{0}}\mathrm{Im}(k(\beta,\bm{\delta}))\bi\notag\\      
      =&-(\bm{a}(\beta,\bm{\delta})+\bm{b}(\beta,\bm{\delta}))\cdot\overline{(\bm{b}(\beta,\bm{\delta})-\bm{a}(\beta,\bm{\delta}))}\bi-(\bm{b}(\beta,\bm{\delta})-\bm{a}(\beta,\bm{\delta}))\cdot\overline{(\bm{a}(\beta,\bm{\delta})+\bm{b}(\beta,\bm{\delta}))}\bi\notag\\      
      &+(\|\bm{a}(\beta,\bm{\delta})\|^{2}_{\mathbb{C}^{2N_{0}}}+\|\bm{b}(\beta,\bm{\delta})\|^{2}_{\mathbb{C}^{2N_{0}}})\mathcal{O}(\mathrm{Im}(k(\beta,\bm{\delta})))+\mathcal{O}(\mathrm{Im}(k(\beta,\bm{\delta}))^{2})\notag\\
        =&(\|\bm{a}(\beta,\bm{\delta})\|^{2}_{\mathbb{C}^{2N_{0}}}+\|\bm{b}(\beta,\bm{\delta})\|^{2}_{\mathbb{C}^{2N_{0}}})\mathcal{O}(\mathrm{Im}(k(\beta,\bm{\delta})))+\mathcal{O}(\mathrm{Im}(k(\beta,\bm{\delta}))^{2}),\label{eq:sec4:Imk}
    \end{align}
    where we have used the relation
    $\bm{b}(\beta,\bm{\delta})=\bm{M}\bm{a}(\beta,\bm{\delta})$.    
    As $(\beta,{\bm{\delta}})\to(\beta_{*},{\bm{\delta}}_{*})$, we have 
    $u(\cdot,\beta,\bm{\delta})\to{u_{*}}$,
    $\mathrm{Re}(k(\beta,\bm{\delta}))\to{k_{*}}$,
    $\bm{a}(\beta,\bm{\delta})\to\bm{0}$ and $\bm{b}(\beta,\bm{\delta})\to\bm{0}$ by continuity.
    Consequently, comparing the leading-order terms in
    $\mathrm{Im}(k(\beta,\bm{\delta}))$ on both sides of \eqref{eq:sec4:Imk} shows
    that $\mathrm{Im}(k(\beta,\bm{\delta}))=0$ for sufficiently small $r_{5}$,
    confirming that $k(\beta,\bm{\delta})$ is real.

    Finally, by Lemma \ref{lem:sec4:local:st}, for sufficiently small
    $r_{5}$ and $r_{6}$, the functions $u$ and $\bm{a}$ are also
    unique in ${H^{1}_{\mathrm{per},1}(\Omega_{0})}$ and
    $\mathbb{C}^{2N_{0}}$, respectively.    
  \end{proof}
\end{theorem}

\begin{corollary}
  \label{coro:sec4:local:st}
  Let $u_{*}$ be a simple BIC and $\bm{S}_{0}$ be the scattering matrix
  at the point $(\beta_{*},\bm{\delta}_{*},k_{*})\in\Lambda$. For every
  $\bm{M}\in{U}_{1}$, the point
  $(\beta_{*},\bm{\delta}_{*},k_{*})$ belongs to $\lambda_{\bm{M}}$. Furthermore, in a  
  neighborhood of this point, 
  $\lambda_{\bm{M}}$ is the graph of a continuous function $k(\beta,\bm{\delta})$ on
  $B_{r}((\beta_{*},\bm{\delta}_{*}))$ for some $r>0$.
  \begin{proof}
    This follows directly from Lemma \ref{lem:sec4:local:st} and Theorem
    \ref{thm:sec4:PiM:bic}.
  \end{proof}
\end{corollary}

\begin{remark}
  The above corollary implies that, for every $\bm{M}\in{U_{1}}$,
  $\lambda_{\bm{M}}$ is locally a
  hypersurface in $\Lambda$. Hence, given any phase factor $e^{\bi\theta}$ with
  $\theta\in[0,2\pi)$  
  such that $e^{\bi\theta}$ is not an eigenvalue of $\bm{S}_{0}$, the set
  $\lambda_{\bm{M}}$ for $\bm{M}=e^{\bi\theta}\bm{I}_{2N_{0}}$ is also locally a  
  hypersurface. If the BIC is isolated, then in every neighborhood of
  the BIC point we can find a field of which the 
  incident and scattered coefficient vectors satisfy
  $\bm{b}=e^{\bi\theta}\bm{a}$ and $\bm{a}\ne\bm{0}$ for almost every $\theta$. This
  observation explains the phase singularity associated with the BIC.
\end{remark}

\section{Index for BIC robustness}\label{sec:BICindex}
In this section, we investigate the local robustness of a simple and
isolated BIC $u_{*}$ located at $(\beta_{*},\bm{\delta}_{*},k_{*})\in\Lambda$. For a
given $\bm{M}\in{U_{1}}$,
Theorem \ref{thm:sec4:PiM:bic} yields a continuous mapping $\mathcal{P}_{\bm{M},1}$
defined in \eqref{eq:sec2:PM}.
We demonstrate that, under the symmetry assumptions introduced in Section
\ref{sec2}, this mapping admits lower-dimensional reductions.
The local robustness of the BIC with respect to $(\beta,\bm{\delta})$ is
therefore related to the
mapping degree of
$\mathcal{P}_{\bm{M},1}$ and its reductions in a neighborhood of
$(\beta_{*},\bm{\delta}_{*})$, provided their domain and codomain dimensions
match. We thus introduce the BIC index as the local mapping degree and
prove that it remains invariant under different choices of the matrix
$\bm{M}$ and the length $d_{0}$ of the domain $\Omega_{0}$.
  
\subsection{Symmetry reduction of  $\mathcal{P}_{\mathbf{\emph{M}},1}$}
We begin by summarizing the properties of a unitary matrix $\bm{M}$
satisfying either $\bm{M}=\bm{M}^{P}$ or $\bm{M}=\bm{M}^{T}$.
\begin{lemma}
  \label{lem:sec5:MP}
  Let $\bm{M}\in{U}({2N_{0}})$. Then $\bm{M}=\bm{M}^{P}$
  if and only if $\bm{M}$ admits the decomposition
  \begin{equation}
    \label{eq:sec5:MP:decomp}
    \bm{M}=
    \frac{1}{\sqrt{2}}\begin{bmatrix}
                        \bm{I}_{N_{0}}&\bm{I}_{N_{0}}\\
                        \bm{I}_{N_{0}}&-\bm{I}_{N_{0}}
                      \end{bmatrix}
                      \begin{bmatrix}
                        \bm{B}_{1}&\\
                        &\bm{B}_{2}
                    \end{bmatrix}
\frac{1}{\sqrt{2}}\begin{bmatrix}
                        \bm{I}_{N_{0}}&\bm{I}_{N_{0}}\\
                        \bm{I}_{N_{0}}&-\bm{I}_{N_{0}}
                  \end{bmatrix},
  \end{equation}
  where $\bm{B}_{1},\bm{B}_{2}\in{U}(N_{0})$.
  If $\bm{M}=\bm{M}^{P}$, the following hold:
\begin{enumerate}[label=(\roman*).]
  \item For any $\bm{B}\in{U}(2N_{0})$ also satisfying
    $\bm{B}=\bm{B}^{P}$,
    \begin{equation}
      \label{eq:sec5:BMP}
      \bm{B}\bm{M}=(\bm{B}\bm{M})^{P}\ \text{and}\
    \bm{B}+\bm{M}=(\bm{B}+\bm{M})^{P}.
    \end{equation}
  \item $\bm{M}^{-1}=(\bm{M}^{-1})^{P}$ and 
    $\bm{M}^{1/2}$ can be chosen to satisfy
    $\bm{M}^{1/2}=(\bm{M}^{1/2})^{P}$.    
  \end{enumerate}
  If $\bm{M}=\bm{M}^{T}$, then:
  \begin{enumerate}[label=(\roman*).]
    \setcounter{enumi}{2}
  \item $\bm{M}^{-1}=(\bm{M}^{-1})^{T}$ and $\bm{M}^{1/2}$ can be
    chosen to satisfy    
    $\bm{M}^{1/2}=(\bm{M}^{1/2})^{T}$.    
  \end{enumerate}
  If $\bm{M}=\bm{M}^{P}=\bm{M}^{T}$, then:
  \begin{enumerate}[label=(\roman*).]
    \setcounter{enumi}{3}
    \item $\bm{M}^{1/2}$ can be chosen to satisfy
    $\bm{M}^{1/2}=(\bm{M}^{1/2})^{P}=(\bm{M}^{1/2})^{T}$.    
  \end{enumerate}  
  \begin{proof}
    Suppose $\bm{M}$ admits the decomposition in
    \eqref{eq:sec5:MP:decomp}. A    
    direct computation shows that $\bm{M}\in{U}(2N_{0})$ and satisfies
    $\bm{M}=\bm{M}^{P}$.    
    Conversely, assume $\bm{M}=\bm{M}^{P}$. Then $\bm{M}$ can be written
    in the block form
    \begin{equation}
      \label{eq:sec5:MP}
      \bm{M}=
      \begin{bmatrix}
        \bm{M}_{1}&\bm{M}_{2}\\
        \bm{M}_{2}&\bm{M}_{1}
      \end{bmatrix},
    \end{equation}
    where $\bm{M}_{1}$ and $\bm{M}_{2}$ are $N_{0}\times{N_{0}}$
    matrices. A straightforward computation yields
    \begin{equation}
      \label{eq:sec5:MP:2}      
      \frac{1}{\sqrt{2}}\begin{bmatrix}
                        \bm{I}_{N_{0}}&\bm{I}_{N_{0}}\\
                        \bm{I}_{N_{0}}&-\bm{I}_{N_{0}}
                      \end{bmatrix}
      \begin{bmatrix}
        \bm{M}_{1}&\bm{M}_{2}\\
        \bm{M}_{2}&\bm{M}_{1}
      \end{bmatrix}
      \frac{1}{\sqrt{2}}\begin{bmatrix}
                        \bm{I}_{N_{0}}&\bm{I}_{N_{0}}\\
                        \bm{I}_{N_{0}}&-\bm{I}_{N_{0}}
                      \end{bmatrix}=
      \begin{bmatrix}
        \bm{M}_{1}+\bm{M}_{2}&\\
        &\bm{M}_{1}-\bm{M}_{2}
      \end{bmatrix}.      
    \end{equation}
   Defining
    \begin{equation}
      \label{eq:sec5:N1N2}
      \bm{B}_{1}:=
      \bm{M}_{1}+\bm{M}_{2},\ \bm{B}_{2}:=
      \bm{M}_{1}-\bm{M}_{2},
    \end{equation}
    we recover the decomposition \eqref{eq:sec5:MP:decomp}.
    Statement (i) follows from direct verification. For (ii),
    writing $\bm{M}^{-1}$ and $\bm{M}^{1/2}$ as polynomials in
    $\bm{M}$ and applying (i) gives the desired result. 
    Statement (iii) and (iv) can be proved analogously.
  \end{proof}
\end{lemma}

For any function $u\in{L^{1}_{\mathrm{loc}}(\Omega)}$, we introduce the
following operators:
\begin{equation}
  \label{eq:sec5:reflections}
    \mathcal{T}_{1}u(x_{1},x_{2}):=\overline{u(-x_{1},x_{2})}\ \text{and}\ 
  \mathcal{T}_{2}u(x_{1},x_{2}):=u(x_{1},-x_{2}).
\end{equation}

\begin{lemma}
  \label{lem:sec5:reflections}
  Let $u$ be a solution to the scattering problem
  \eqref{eq:sec3:Omegam:sca} at a point $(\beta,\bm{\delta},k)\in\Lambda$, with
  incident and scattered coefficient vectors $\bm{a}$
  and $\bm{b}$, respectively.
  \begin{itemize}[label=$\bullet$]
  \item Suppose the dielectric function $\epsilon$ has reflection symmetry in
    $x_{1}$. Then the following properties hold:      
  \begin{enumerate}[label=(\roman*).]
    \item $\mathcal{T}_{1}u$ is also a solution to
      \eqref{eq:sec3:Omegam:sca}. Its incident and scattered
      coefficient vectors are $\overline{\bm{b}}$ and
      $\overline{\bm{a}}$, respectively.
    \item If $u$ is a simple BIC, then $\mathcal{T}_{1}u=e^{\bi\eta}u$ for some
      $\eta\in[0,2\pi)$.      
    \item If $u$ is governed by a matrix
      $\bm{M}\in{U}(2N_{0})$, then $\mathcal{T}_{1}u$ is governed by
      $\bm{M}^{T}$.      
      Furthermore, if $\bm{M}=\bm{M}^{T}$ and we let $\bm{a}_{\eta}$
      denote the incident coefficient vector of
      $(u+e^{-\bi\eta}\mathcal{T}_{1}u)/2$ for some $\eta\in[0,2\pi)$, then 
      \begin{equation}
        \label{eq:sec5:reala}
        e^{\bi\eta/2}\bm{M}^{1/2}\bm{a}_{\eta}\in\mathbb{R}^{2N_{0}},
      \end{equation}
      where $\bm{M}^{1/2}$ is chosen to satisfy
      $\bm{M}^{1/2}=(\bm{M}^{1/2})^{T}$.      
    \end{enumerate}
  \item Suppose the dielectric function $\epsilon$ has reflection symmetry in
    $x_{2}$. Then the following
      properties hold:      
      \begin{enumerate}[label=(\roman*).]
        \setcounter{enumi}{3}
    \item $\mathcal{T}_{2}u$ is also a solution to
      \eqref{eq:sec3:Omegam:sca}. Its incident and scattered
      coefficient vectors are $\bm{R}_{2N_{0}}\bm{a}$ and
      $\bm{R}_{2N_{0}}\bm{b}$, respectively.      
    \item If $u$ is a simple BIC, then $\mathcal{T}_{2}u=u$ 
      or $\mathcal{T}_{2}u=-u$.
    \item If $u$ is governed by a matrix
      $\bm{M}\in{U}(2N_{0})$, then $\mathcal{T}_{2}u$ is governed by
      $\bm{M}^{P}$. Furthermore, if $\bm{M}=\bm{M}^{P}$ and we let
      $\bm{a}_{e}$ and $\bm{a}_{o}$ denote the      
      incident coefficient vectors of $(u+\mathcal{T}_{2}u)/2$ and
      $(u-\mathcal{T}_{2}u)/2$, respectively, then
      \begin{equation}
        \label{eq:sec5:even:odd}
        \bm{a}_{e}=\bm{R}_{2N_{0}}\bm{a}_{e}\ \text{and}\
        \bm{a}_{o}=-\bm{R}_{2N_{0}}\bm{a}_{o}.        
      \end{equation}
    \end{enumerate}    
  \end{itemize}    
  \begin{proof}
    We now prove statement (iii); the other results follow from direct
    computations.
    By Definition \ref{def:sec2:BIC}, the incident and scattered
    coefficient vectors $\bm{a}$ and $\bm{b}$ of the field $u$ satisfy    
    \begin{equation}
      \label{eq:sec5:Hsymmetry:bMa}
      \bm{b}=\bm{M}\bm{a},
    \end{equation}
    which implies
    \begin{equation}
      \label{eq:sec5:Hsymmetry:aMb}
      \overline{\bm{a}}=\bm{M}^{T}\overline{\bm{b}}.
    \end{equation}
    From statement (i), the field $\mathcal{T}_{1}u$ has incident coefficient
    vector $\overline{\bm{b}}$ and scattered coefficient vector
    $\overline{\bm{a}}$. Equation
    \eqref{eq:sec5:Hsymmetry:aMb} therefore shows that $\mathcal{T}_{1}u$ is
    governed by    
    $\bm{M}^{T}$. If we now assume $\bm{M}=\bm{M}^{T}$, the linear
    combination    
    $(u+e^{-\bi\eta}\mathcal{T}_{1}u)/2$ is also a field governed by
    $\bm{M}$. Denote its incident coefficient vector by
    $\bm{a}_{\eta}$, which can be expressed as:
    \begin{equation}
     \label{eq:sec5:Hsymmetry:abtheta}       
      \bm{a}_{\eta}=(\bm{a}+e^{-\bi\eta}\overline{\bm{b}})/2.
    \end{equation}
    A direct computation then yields
    \begin{equation}
      \label{eq:sec5:Hsymmetry:athetaMbtheta}
      e^{\bi\eta}\bm{M}\bm{a}_{\eta}=\overline{\bm{a}_{\eta}}.
    \end{equation}
    Multiplying both sides of \eqref{eq:sec5:Hsymmetry:athetaMbtheta}
    by $e^{-\bi\eta/2}\overline{\bm{M}}^{1/2}$ and using the fact that
    $\bm{M}$ is    
    unitary and symmetric, we obtain
    \begin{equation}
      \label{eq:sec5:Hsymmetry:athetaMbtheta:real}
      e^{\bi\eta/2}\bm{M}^{1/2}\bm{a}_{\eta}=\overline{e^{\bi\eta/2}\bm{M}^{1/2}\bm{a}_{\eta}}.  \end{equation}
    The identity \eqref{eq:sec5:Hsymmetry:athetaMbtheta:real}
    shows that the vector
    $e^{\bi\eta/2}\bm{M}^{1/2}\bm{a}_{\eta}$ is real.
    \end{proof}
\end{lemma}

Applying Lemma \ref{lem:sec5:reflections}, we exploit the spatial
symmetry to reduce the mapping $\mathcal{P}_{\bm{M},1}$. Let $u_{*}$ denote the
simple BIC at the
point $(\beta_{*},\bm{\delta}_{*},k_{*})\in\Lambda$, and let $u(\cdot,\beta,\bm{\delta})$ be the implicit
function from Theorem \ref{thm:sec4:PiM:bic}, defined for
$(\beta,\bm{\delta})\in{B_{r_{5}}((\beta_{*},\bm{\delta}_{*}))}$. We consider the symmetry
cases defined in Section \ref{sec2}:
\begin{itemize}
\item Case II. Assume $\mathcal{T}_{1}u_{*}=e^{\bi\eta}u_{*}$ for some  
  $\eta\in[0,2\pi)$ and let $\bm{M}=\bm{M}^{T}$. By the uniqueness of
  $u(\cdot,\beta,\bm{\delta})$, it follows that  
  $\mathcal{T}_{1}u(\cdot,\beta,\bm{\delta})=e^{\bi\eta}u(\cdot,\beta,\bm{\delta})$. We then define
  \begin{equation}
    \label{eq:sec5:PHM}
    \mathcal{P}_{\bm{M},2}:{B}_{r_{5}}((\beta_{*},\bm{\delta}_{*}))\to\mathbb{R}^{2N_{0}},\
    \mathcal{P}_{\bm{M},2}:=e^{\bi\eta/2}\bm{M}^{1/2}\mathcal{P}_{\bm{M},1},
  \end{equation}
  where $\bm{M}^{1/2}$ is chosen to satisfy
  $\bm{M}^{1/2}=(\bm{M}^{1/2})^{T}$.  
\item Case III. Assume $\mathcal{T}_{2}u_{*}=Cu_{*}$ for some $C\in\{-1,1\}$ and
  let $\bm{M}=\bm{M}^{P}$. By the uniqueness of $u(\cdot,\beta,\bm{\delta})$, it
  follows that $\mathcal{T}_{2}u(\cdot,\beta,\bm{\delta})=Cu(\cdot,\beta,\bm{\delta})$. We then define
  \begin{equation}
    \label{eq:sec5:PVM}
    \mathcal{P}_{\bm{M},3}:{B}_{r_{5}}((\beta_{*},\bm{\delta}_{*}))\to\mathbb{C}^{N_{0}},\
    \mathcal{P}_{\bm{M},3}:=\frac{1}{2}
  \begin{bmatrix}
   \bm{I}_{N_{0}}&C\bm{I}_{N_{0}} 
  \end{bmatrix}\mathcal{P}_{\bm{M},1}.
\end{equation}
  
\item Case IV. Assume $\mathcal{T}_{1}u_{*}=e^{\bi\eta}u_{*}$ and
  $\mathcal{T}_{2}u_{*}=Cu_{*}$ for some $\eta\in[0,2\pi)$ and $C\in\{-1,1\}$ and let  
   $\bm{M}=\bm{M}^{T}=\bm{M}^{P}$. Combining the above   
   constructions, we define   
  \begin{equation}
    \label{eq:sec5:PVHM}
    \mathcal{P}_{\bm{M},4}:{B}_{r_{5}}((\beta_{*},\bm{\delta}_{*}))\to\mathbb{R}^{N_{0}},\
    \mathcal{P}_{\bm{M},4}:=\frac{1}{2}
  \begin{bmatrix}
   \bm{I}_{N_{0}}&C\bm{I}_{N_{0}} 
  \end{bmatrix}e^{\bi\eta/2}\bm{M}^{1/2}\mathcal{P}_{\bm{M},1},
\end{equation}
where $\bm{M}^{1/2}$ is chosen to satisfy
$\bm{M}^{1/2}=(\bm{M}^{1/2})^{T}=(\bm{M}^{1/2})^{P}$.
\end{itemize}
The mappings $\mathcal{P}_{\bm{M},2}$, $\mathcal{P}_{\bm{M},3}$ and $\mathcal{P}_{\bm{M},4}$ are 
continuous and their zeros correspond precisely to BICs.

We introduce the following subsets of the unitary group $U(2N_{0})$:
\begin{align}
  U_{2}:=&\{\bm{M}\in{U}(2N_{0}):\mathrm{det}(\bm{S}_{0}-\bm{M})\ne0\ \text{and}\
  \bm{M}=\bm{M}^{T}\},  \label{eq:sec5:U2}\\  
  U_{3}:=&\{\bm{M}\in{U}(2N_{0}):\mathrm{det}(\bm{S}_{0}-\bm{M})\ne0\ \text{and}\
           \bm{M}=\bm{M}^{P}\},    \label{eq:sec5:U3}\\
  U_{4}:=&\{\bm{M}\in{U}(2N_{0}):\mathrm{det}(\bm{S}_{0}-\bm{M})\ne0,\
           \bm{M}=\bm{M}^{T}\ \text{and}\ \bm{M}=\bm{M}^{P}\}.    \label{eq:sec5:U4}           
\end{align}
For the symmetry cases under consideration, the following dimensional
constraints are imposed:
  \begin{equation}
    \label{eq:sec5:dimension}
    \left\{
      \begin{aligned}
        &1+N_{1}=4N_{0}\ &&\text{in case I};\\
        &1+N_{1}=2N_{0}\ &&\text{in case II};\\
        &1+N_{1}=2N_{0}\ &&\text{in case III};\\
        &1+N_{1}=N_{0}\ &&\text{in case IV}.                 
      \end{aligned}
      \right.
    \end{equation}
    These relations coincide with those found in \cite{abdrabou23} .
    We now define the BIC index, which characterizes the local
    robustness of a BIC with respect to $(\beta,\bm{\delta})$. This definition
    relies on the    
    identification $\mathbb{C}^{N}\cong\mathbb{R}^{2N}$ ($N\in\mathbb{Z}_{+})$
    and employs the mappings
    $\mathcal{P}_{\bm{M},1},\mathcal{P}_{\bm{M},2},\mathcal{P}_{\bm{M},3},\mathcal{P}_{\bm{M},4}$,
    corresponding to symmetry cases I, II, III and IV, respectively.
\begin{definition}
    \label{def:sec5:index}
    Let $u_{*}$ be a simple and isolated BIC and $\bm{S}_{0}$ be the
    scattering matrix at the point
    $(\beta_{*},\bm{\delta}_{*},k_{*})\in\Lambda$. For a given symmetry case satisfying
    the dimensional constraint \eqref{eq:sec5:dimension} and for a
    matrix $\bm{M}\in{U}_{j}$ with $j\in\{1,2,3,4\}$, we define the BIC
    index as the mapping degree (see Definition 2.1 in Chapter IV of
    \cite{Outerelo09} and Definition 3.1.6 in \cite{chang03})    
    \begin{equation}
  \label{eq:sec5:degree}
 \mathrm{ind}_{j}((\beta_{*},\bm{\delta}_{*},k_{*})):=\mathrm{deg}(\mathcal{P}_{\bm{M},j},B_{r}((\beta_{*},\bm{\delta}_{*})),\bm{0}),\
 j\in\{1,2,3,4\}, 
\end{equation}
where $r>0$ is chosen sufficiently small such that $(\beta_{*},\bm{\delta}_{*})$ is
the unique zero of $\mathcal{P}_{\bm{M},j}$ in $\overline{B_{r}((\beta_{*},\bm{\delta}_{*}))}$.
\end{definition}

\begin{remark}
The BIC index introduced above coincides with the Brouwer index (see
Definition 3.2.6 in \cite{chang03}). By restricting $\mathcal{P}_{\bm{M},j}$ to
the the boundary
$\partial{B}_{r}((\beta_{*},\bm{\delta}_{*}))$, the index can also be expressed as a
winding number (see Definition 4.1 in Chapter IV of \cite{Outerelo09})
\begin{equation}
  \label{eq:sec5:windingnum}
  \mathrm{ind}_{j}((\beta_{*},\bm{\delta}_{*},k_{*}))=w(\mathcal{P}_{\bm{M},j}|_{\partial{B}_{r}((\beta_{*},\bm{\delta}_{*}))},\bm{0}).
\end{equation}
\end{remark}

\subsection{Invariance of the BIC index}
In this subsection, we prove that the BIC index introduced in
Definition \ref{def:sec5:index} remains invariant under
different choices of the matrix $\bm{M}$ and the length $d_{0}$ of
$\Omega_{0}$.

Given $t_{1}>t_{0}$, we define a family of unitary matrices
\begin{equation}
  \label{eq:sec5:familyM1}
 \bm{M}(t,\beta,\bm{\delta},k)\in{U}(2N_{0}),\ (t,\beta,\bm{\delta},k)\in[t_{0},t_{1}]\times\Lambda.
\end{equation}
Consider a BIC at $(\beta_{*},\bm{\delta}_{*},k_{*})\in\Lambda$ with scattering
matrix $\bm{S}_{0}$. For some radii $r_{1},r_{2}>0$, we impose the
following conditions on $\bm{M}$:
\begin{equation}
  \label{eq:sec5:familyM2}
  \left\{
    \begin{aligned}
      &\bm{M}(t,\beta_{*},\bm{\delta}_{*},k_{*})\in{U}_{1}\ \text{for}\
        t\in[t_{0},t_{1}];\\      
      &\bm{M}\ \text{is\ analytic\ in}\ k;\\      
      &\bm{M}\ \text{and}\ \partial_{k}\bm{M}\ \text{are\ continuous\ on}\
        [t_{0},t_{1}]\times{B}_{r_{1}}((\beta_{*},\bm{\delta}_{*}))\times{\widehat{B}_{r_{2}}(k_{*})}.        
    \end{aligned}\right.
\end{equation}
By assumption, $\bm{M}(t,\beta,\bm{\delta},k)$ is a homotopy between the
endpoint matrices $\bm{M}(t_{0},\beta,\bm{\delta},k)$ and
$\bm{M}(t_{1},\beta,\bm{\delta},k)$. The following lemma demonstrates that this
homotopy
induces a corresponding homotopy between the incident coefficient
vectors $\bm{a}(t_{0},\beta,\bm{\delta})$ and $\bm{a}(t_{1},\beta,\bm{\delta})$.

\begin{lemma}
  \label{lem:sec5:homotopy}
  Let $u_{*}$ be a simple BIC at 
  $(\beta_{*},\bm{\delta}_{*},k_{*})\in\Lambda$ with scattering matrix $\bm{S}_{0}$.
  Given a family of unitary matrices $\bm{M}(t,\beta,\bm{\delta},k)$ satisfying
  \eqref{eq:sec5:familyM1}--\eqref{eq:sec5:familyM2}, there
  exist radii $r_{7},r_{8}>0$ such that for every $t\in[t_{0},t_{1}]$ and 
  $(\beta,\bm{\delta})\in{B_{r_{7}}}((\beta_{*},\bm{\delta}_{*}))$, we can find unique
  $k(t,\beta,\bm{\delta})\in{B_{r_{8}}(k_{*})}$, 
  $u(\cdot,t,\beta,\bm{\delta})\in{H^{1}_{\mathrm{per},1}}(\Omega_{0})$ and 
  $\bm{a}(t,\beta,\bm{\delta})\in\mathbb{C}^{2N_{0}}$  
  satisfying
  \begin{equation}
    \label{eq:sec5:ku:condition}
    \left\{
      \begin{aligned}       
        &(\beta,{\bm{\delta}},k(t,\beta,{\bm{\delta}}))\in\lambda_{\bm{M}(t,\beta,\bm{\delta},k(t,\beta,\bm{\delta}))};
        \\
        &u(\cdot,t,\beta,\bm{\delta})\ \text{satisfies}\
          \eqref{eq:sec3:Omegam:sca}\ \text{with\ incident\
          coefficient\ vector}\ \bm{a}(t,\beta,\bm{\delta});\\         
        &u(\cdot,t,\beta,\bm{\delta})\ \text{is\ governed\ by}\
          \bm{M}(t,\beta,\bm{\delta},k(t,\beta,\bm{\delta}));\\
        &(u(\cdot,t,\beta,{\bm{\delta}}),u_{*})_{\Omega_{0}}=\|u_{*}\|_{L^{2}(\Omega_{0})}^{2}.
      \end{aligned}\right.
  \end{equation}
  Furthermore, 
  $k(t,\beta,{\bm{\delta}})$, $u(\cdot,t,\beta,{\bm{\delta}})$ and $\bm{a}(t,\beta,\bm{\delta})$ depend
  continuously on $t\in[t_{0},t_{1}]$ and
  $(\beta,{\bm{\delta}})\in{B_{r_{7}}((\beta_{*},{\bm{\delta}}_{*}))}$  
  and the following conditions hold:
  \begin{equation}
    \label{eq:sec5:ku:condition2}
    k(t,\beta_{*},{\bm{\delta}}_{*})=k_{*},\
    u(\cdot,t,\beta_{*},{\bm{\delta}}_{*})=u_{*}\ \text{and}\
    \bm{a}(t,\beta_{*},{\bm{\delta}}_{*})=\bm{0},\
    \text{for}\ t\in[t_{0},t_{1}].
  \end{equation}
  \begin{proof}
    Define the operator
    \begin{equation}
      \label{eq:sec5:F}
      \mathcal{F}(t,\beta,\bm{\delta},k,\psi,\bm{a}):=\widehat{\mathcal{A}}_{\bm{M}(t,\beta,\bm{\delta},k)}(\beta,\bm{\delta},k,\psi,\bm{a}),        
    \end{equation}
    where $\widehat{\mathcal{A}}_{\bm{M}}$ is given in
    \eqref{eq:sec4:AM}. A direct computation shows that
    \begin{equation}
      \label{eq:sec5:F:0}
      \mathcal{F}(t,\beta,\bm{\delta},k,\psi,\bm{a})=
      \begin{bmatrix}
        0\\
        \bm{0}
      \end{bmatrix}
    \end{equation}
    at the point $(s,\beta_{*},\bm{\delta}_{*},k_{*},0,\bm{0})$ for every
    $s\in[t_{0},t_{1}]$. For a fixed $s$, we apply the implicit function
    theorem to $\mathcal{F}$    
    following the same procedure as in Theorem
    \ref{thm:sec4:PiM:bic}. This yields     
    radii $r_{s},r_{5,s},r_{6,s}>0$ such that for every
    $t\in{B_{r_{s}}(s)\cap[t_{0},t_{1}]}$ and
    $(\beta,\bm{\delta})\in{B_{r_{5,s}}((\beta_{*},\bm{\delta}_{*}))}$,
    \eqref{eq:sec5:F:0} admits a unique solution     
    \begin{equation}
      \label{eq:sec5:F:solution}
      (t,\beta,\bm{\delta},k(t,\beta,\bm{\delta}),\psi(\cdot,t,\beta,\bm{\delta}),\bm{a}(t,\beta,\bm{\delta}))
    \end{equation}
    with $k(t,\beta,\bm{\delta})\in{\widehat{B}_{r_{6,s}}(k_{*})}$,
    $\|\psi(\cdot,t,\beta,\bm{\delta})\|_{\mathbb{H}(\{u_{*}\})}\le{r_{6,s}}$ and
    $\bm{a}(t,\beta,\bm{\delta})\in{\widehat{B}_{r_{6,s}}(\bm{0})}$.
    Defining 
    \begin{equation}
      \label{eq:sec5:ubetadelta}
      u(\cdot,t,\beta,\bm{\delta}):= \psi(\cdot,t,\beta,\bm{\delta})+u_{*},
    \end{equation}
    the triple $(k,u,\bm{a})$ satisfies \eqref{eq:sec5:ku:condition} and
    \eqref{eq:sec5:ku:condition2}. Using \eqref{eq:sec5:familyM2} and
    adapting the proofs of Lemma
    \ref{lem:sec4:local:st} and Theorem \ref{thm:sec4:PiM:bic}, we
    also confirm that for sufficiently small $r_{s}$ and $r_{5,s}$,    
    $k$ is real, and $u$ and $\bm{a}$ are unique in
    $H^{1}_{\mathrm{per},1}(\Omega_{0})$ and $\mathbb{C}^{2N_{0}}$,
    respectively.
    
    Since $\{B_{r_{s}}(s):s\in[t_{0},t_{1}]\}$ is an open cover of
    $[t_{0},t_{1}]$, compactness guarantees a
    finite subcover centered at points $\{s_{j}\}_{j=1,\ldots,N}$ with
    radii $\{r_{s_{j}}\}_{j=1,\ldots,N}$ 
    for some $N\in\mathbb{Z}_{+}$. The desired result follows by setting
    \begin{equation}
      \label{eq:sec5:r7r8}
     r_{7}:=\min(\{r_{5,s_{j}}\}_{j=1,\ldots,N}),\
     r_{8}:=\min(\{r_{6,s_{j}}\}_{j=1,\ldots,N}).     
    \end{equation}
  \end{proof}
\end{lemma}

\begin{lemma}
      \label{lem:sec5:pathconectivity}
    Assume $\bm{S}_{0}\in{U}(2N_{0})$. Then the set $U_{1}$ is
    path-connected. If $\bm{S}_{0}=\bm{S}_{0}^{T}$, then $U_{2}$ is also
    path-connected.    
    Likewise, if $\bm{S}_{0}=\bm{S}_{0}^{P}$, then $U_{3}$ is
    path-connected. Finally, if $\bm{S}_{0}$ satisfies both
    $\bm{S}_{0}=\bm{S}_{0}^{T}$ and    
    $\bm{S}_{0}=\bm{S}_{0}^{P}$, then ${U}_{4}$ is path-connected.
    \begin{proof}
      We first introduce the following four subsets in $U(2N_{0})$:
\begin{align}
  V_{1}:=&\{\bm{M}\in{U}(2N_{0}):\mathrm{det}(\bm{I}_{2N_{0}}-\bm{M})\ne0\},  \label{eq:sec5:V1}\\
  V_{2}:=&\{\bm{M}\in{U}(2N_{0}):\mathrm{det}(\bm{I}_{2N_{0}}-\bm{M})\ne0\ \text{and}\
  \bm{M}=\bm{M}^{T}\},  \label{eq:sec5:V2}\\  
  V_{3}:=&\{\bm{M}\in{U}(2N_{0}):\mathrm{det}(\bm{I}_{2N_{0}}-\bm{M})\ne0\ \text{and}\
           \bm{M}=\bm{M}^{P}\},    \label{eq:sec5:V3}\\
  V_{4}:=&\{\bm{M}\in{U}(2N_{0}):\mathrm{det}(\bm{I}_{2N_{0}}-\bm{M})\ne0,\
           \bm{M}=\bm{M}^{T}\ \text{and}\ \bm{M}=\bm{M}^{P}\}.    \label{eq:sec5:V4}           
\end{align}      
      For any $\bm{M}\in{U}(2N_{0})$, we have the identity
      \begin{equation}
        \label{eq:sec5:VSVI}
        \mathrm{det}(\bm{S}_{0}-\bm{M})=\mathrm{det}(\bm{S}_{0}^{1/2})\mathrm{det}(\bm{I}_{2N_{0}}-\bm{S}_{0}^{-1/2}\bm{M}\bm{S}_{0}^{-1/2})\mathrm{det}(\bm{S}_{0}^{1/2}).        
      \end{equation}
      By Lemma \ref{lem:sec5:MP}, this implies that the mapping     
      $\bm{M}\to\bm{S}_{0}^{-1/2}\bm{M}\bm{S}_{0}^{-1/2}$ is a    
      homeomorphism from      
      $U_{j}$ to $V_{j}$ for $j\in\{1,2,3,4\}$ if $\bm{S}_{0}$ satisfies
      corresponding properties and $\bm{S}_{0}^{1/2}$ is chosen
      appropriately.      
      
      We now prove that $V_{1}$ is
      path-connected. From the definition of $V_{1}$ in
      \eqref{eq:sec5:V1}, a matrix $\bm{M}$ belongs to $V_{1}$ if and
      only if $1$ is not an eigenvalue of $\bm{M}$. We prove that
      every      
      matrix in $V_{1}$ is path-connected to $-\bm{I}_{2N_{0}}$.      
      Since $\bm{M}$ is unitary, it can be diagonalized as
      \begin{equation}
        \label{eq:sec5:Mdiag}
        \bm{M}=\bm{Q}\mathrm{diag}(\{e^{\bi\theta_{n}}\}_{n=1,\ldots,2N_{0}})\bm{Q}^{*},\        \bm{Q}\in{U}(2N_{0}),        
      \end{equation}
      with $\theta_{n}\in(0,2\pi)$ for $n=1,\ldots,2N_{0}$ (the exclusion of 0
      follows from the condition that 1 is not an eigenvalue).
      The set $(0,2\pi)^{2N_{0}}$      
      is path-connected. Therefore, we can construct a continuous path
      in $V_{1}$
      \begin{equation}
        \label{eq:sec5:Mhat}
       \widehat{\bm{M}}(t)=\bm{Q}\mathrm{diag}(\{e^{\bi\widehat{\theta}_{n}(t)}\}_{n=1,\ldots,2N_{0}})\bm{Q}^{*},\
       t\in[0,1],       
     \end{equation}
     where $\{\widehat{\theta}_{n}(t)\}$ are continuous functions such that
      \begin{equation}
        \label{eq:sec5:thetahat}
        \widehat{\theta}_{n}(0)=\theta_{n},\ \widehat{\theta}_{n}(1)=\pi,\ \text{for}\ n=1,\ldots,2N_{0}.
      \end{equation}
      Then $\widehat{\bm{M}}(0)=\bm{M}$,
      $\widehat{\bm{M}}(1)=-\bm{I}_{2N_{0}}$ and $\widehat{\bm{M}}(t)$      
      remains unitary with no eigenvalue equal to 1 for all $t$. 
      Thus $V_{1}$ is path-connected.

      The same method shows $V_{2}$ is path-connected when 
      $\bm{Q}$ is chosen real. Path-connectedness of $V_{3}$ is
      established by setting
      \begin{equation}
        \label{eq:sec5:Q:V3}
        \bm{Q}=\frac{1}{\sqrt{2}}
        \begin{bmatrix}
          \bm{I}_{N_{0}}&\bm{I}_{N_{0}}\\
          \bm{I}_{N_{0}}&-\bm{I}_{N_{0}}
        \end{bmatrix}
        \begin{bmatrix}
          \bm{B}_{1}&\\
          &\bm{B}_{2}
        \end{bmatrix},
      \end{equation}
      with $\bm{B}_{1},\bm{B}_{2}\in{U(N_{0})}$. Taking
      $\bm{B}_{1}$ and $\bm{B}_{2}$ in \eqref{eq:sec5:Q:V3} to be
      real then proves that $V_{4}$ is path-connected as well.
    \end{proof}
\end{lemma}  

To study the invariance of the BIC index with respect to changes in
the domain length $d_{0}$ of $\Omega_{0}$, we introduce the diagonal matrix
    \begin{equation}
      \label{eq:sec5:Tmatrix}
      \bm{T}(t,\beta,k):=
      \begin{bmatrix}
        \mathrm{diag}(\{e^{-\bi\alpha_{m}(\beta,k)t}\}_{m\in{Z_{0}}})&\\
        &\mathrm{diag}(\{e^{-\bi\alpha_{m}(\beta,k)t}\}_{m\in{Z_{0}}})
      \end{bmatrix}.
    \end{equation}
Let $u$ solve the scattering problem
\eqref{eq:sec3:Omegam:sca} in $\Omega_{-d_{0},d_{0}}$ at the point
$(\beta,\bm{\delta},k)\in\Lambda$, with incident and
scattered coefficient vectors
$\bm{a}$ and $\bm{b}$, and scattering matrix $\bm{S}$. Using the
expansions in
\eqref{eq:sec2:Omegal:expan}--\eqref{eq:sec2:Omegar:expan}, the same
function $u$ also satisfies \eqref{eq:sec3:Omegam:sca} in any wider
domain $\Omega_{-d_{1},d_{1}}$ with $d_{1}>d_{0}$. In this extended domain,
the coefficient vectors become
\begin{equation}
  \label{eq:sec5:trans:coefvec}
 \bm{T}(d_{1}-d_{0},\beta,k)\bm{a}\ \text{and}\ \bm{T}(d_{0}-d_{1},\beta,k)\bm{b},
\end{equation}
which leads to the transformed scattering matrix
\begin{equation}
  \label{eq:sec5:trans:scamat}
  \bm{T}(d_{0}-d_{1},\beta,k)\bm{S}\bm{T}(d_{0}-d_{1},\beta,k).
\end{equation}

\begin{theorem}
    \label{thm:sec5:localrobustness}
    Let $u_{*}$ be a simple and isolated BIC at
    the point $(\beta_{*},\bm{\delta}_{*},k_{*})\in\Lambda$ with scattering matrix
    $\bm{S}_{0}$. For each symmetry case satisfying the dimensional
    constraint \eqref{eq:sec5:dimension}, the corresponding BIC
    index $\mathrm{ind}_{j}((\beta_{*},\bm{\delta}_{*},k_{*}))$ with
    $j\in\{1,2,3,4\}$, defined in Definition
    \ref{def:sec5:index}, is    
    well-defined and exhibits the following    
    invariance properties:    
    \begin{enumerate}[label=(\roman*).]
      \item It is independent of the choice of matrix
        $\bm{M}\in{U_{j}}$.        
      \item It is unchanged when the scattering problem is
        considered in $\Omega_{-d_{1},d_{1}}$ for any
        $d_{1}>d_{0}$.       
    \end{enumerate}
    \begin{proof}
      We demonstrate the result for symmetry case I; the other
      cases follow similarly.
      
      First, we show the BIC index is well-defined. By Theorem     
      \ref{thm:sec4:PiM:bic}, for any $\bm{M}\in{U}_{1}$, there exist
      radii $r_{5},r_{6}>0$ such that we can find unique 
      $k(\beta,\bm{\delta})\in{B_{r_{6}}(k_{*})}$,
      $u(\cdot,\beta,\bm{\delta})\in{H^{1}_{\mathrm{per},1}(\Omega_{0})}$ and
      $\bm{a}(\beta,\bm{\delta})\in\mathbb{C}^{2N_{0}}$ depending      
      continuously on $(\beta,\bm{\delta})\in{B}_{r_{5}}((\beta_{*},\bm{\delta}_{*}))$
      which satisfy \eqref{eq:sec4:ku:condition} and      
      \eqref{eq:sec4:ku:condition2}. Since      
      $u_{*}$ is simple and isolated, we can choose
      $r_{5}$ and $r_{6}$ sufficiently small such that
      $(\beta_{*},\bm{\delta}_{*},k_{*})$ is the      
      only BIC point in
      ${\overline{B_{r_{5}}((\beta_{*},\bm{\delta}_{*}))}\times{\overline{B_{r_{6}}(k_{*})}}}$. Consequently,
      $(\beta_{*},\bm{\delta}_{*})$ is also the unique zero of $\mathcal{P}_{\bm{M},1}$
      on
      $\overline{B_{r_{5}}((\beta_{*},\bm{\delta}_{*}))}$. Provided
      \eqref{eq:sec5:dimension} holds, the index
      $\mathrm{ind_{1}((\beta_{*},\bm{\delta}_{*},k_{*}))}$ is therefore
      well-defined.      
      
      Next, we prove the index is independent of the choice of matrix
      in ${U}_{1}$. For any $\bm{M}_{0},\bm{M}_{1}\in{U_{1}}$, Lemma
      \ref{lem:sec5:pathconectivity} guarantees a
      continuous path $\bm{G}(t)\subset{U}_{1}$ with $t\in[0,1]$ such that
      \begin{equation}
        \label{eq:sec5:M0M1}
        \bm{G}(0)=\bm{M}_{0}\ \text{and}\
        \bm{G}(1)=\bm{M}_{1}.        
      \end{equation}
      By Lemma \ref{lem:sec5:homotopy}, there exists radius $r_{7}>0$
      for which we can
      define a homotopy
      \begin{equation}
        \label{eq:sec5:H1}
        \mathcal{H}_{1}:[0,1]\times{\overline{{B}_{r_{7}}((\beta_{*},\bm{\delta}_{*}))}}\to{\mathbb{C}^{2N_{0}}},\
        \mathcal{H}_{1}(t,\beta,\bm{\delta}):=\mathcal{P}_{\bm{G}(t),1}(\beta,\bm{\delta}).        
      \end{equation}
      where $r_{7}$ is chosen sufficiently small such that
      $(\beta_{*},\bm{\delta}_{*})$      
      remains the only zero of $\mathcal{P}_{\bm{G}(t),1}$ for every
      $t\in[0,1]$. This implies
      $\bm{0}\notin\mathcal{H}_{1}([0,1]\times\partial{B_{r_{7}}((\beta_{*},\bm{\delta}_{*}))})$. 
      Hence, by Proposition 2.4 in Chapter IV of
      \cite{Outerelo09}, we have
      \begin{equation}
        \label{eq:sec5:homotopy:degreeM}
        \mathrm{deg}(\mathcal{P}_{\bm{M}_{0},1},B_{r_{7}}((\beta_{*},\bm{\delta}_{*})),\bm{0})=\mathrm{deg}(\mathcal{P}_{\bm{M}_{1},1},B_{r_{7}}((\beta_{*},\bm{\delta}_{*})),\bm{0}).
      \end{equation}
      Thus the index
      $\mathrm{ind}_{1}((\beta_{*},\bm{\delta}_{*},k_{*}))$ is uniquely defined for
      any $\bm{M}\in{U}_{1}$.      

      Finally, we show that the index is preserved when the
      scattering problem is considered in a larger domain
      $\Omega_{-d_{1},d_{1}}$ for any $d_{1}>d_{0}$.
      Given $\bm{M}_{2}\in{U_{1}}$ and $d_{1}>d_{0}$, we
      define a family of unitary matrices for $t\in[0,d_{1}-d_{0}]$ and
      $(\beta,\bm{\delta},k)\in\Lambda$ by      
      \begin{equation}
        \label{eq:sec5:familyMd}
        \bm{W}(t,\beta,\bm{\delta},k):=\bm{T}(t,\beta,k)\bm{T}(-t,\beta_{*},k_{*})\bm{M}_{2}\bm{T}(-t,\beta_{*},k_{*})\bm{T}(t,\beta,k).        
      \end{equation}
      One can verify that $\bm{W}(t,\beta,\bm{\delta},k)$ satisfies 
      \eqref{eq:sec5:familyM1} and \eqref{eq:sec5:familyM2}. By Lemma
      \ref{lem:sec5:homotopy}, there      
      exists radii $r_{7},r_{8}>0$ such that we obtain unique continuous
      $k(t,\beta,\bm{\delta})\in{B_{r_{8}}(k_{*})}$,
      $u(\cdot,t,\beta,\bm{\delta})\in{H^{1}_{\mathrm{per},1}(\Omega_{0})}$ and
      $\bm{a}(t,\beta,\bm{\delta})\in\mathbb{C}^{2N_{0}}$      
      satisfying      
      \eqref{eq:sec5:ku:condition} and \eqref{eq:sec5:ku:condition2}
      for $t\in[0,d_{1}-d_{0}]$ and
      $(\beta,\bm{\delta})\in{B_{r_{7}}((\beta_{*},\bm{\delta}_{*}))}$.
      Set $d(t):=d_{0}+t$, $\bm{T}_{0}(t):=\bm{T}(t,\beta_{*},k_{*})$ and
     \begin{equation}
       \label{eq:sec5:Teta}
      \zeta(t,\beta,\bm{\delta}):=\frac{\|u_{*}\|^{2}_{L^{2}(\Omega_{-d(t),d(t)})}}{(u(\cdot,t,\beta,\bm{\delta}),u_{*})_{\Omega_{-d(t),d(t)}}}.      
     \end{equation}
     From \eqref{eq:sec5:ku:condition2}, 
      \begin{equation}
        \label{eq:sec5:innerprod:fix}
       (u(\cdot,t,\beta_{*},\bm{\delta}_{*}),u_{*})_{\Omega_{-d(t),d(t)}}=\|u_{*}\|^{2}_{L^{2}(\Omega_{-d(t),d(t)})},\
       \text{for}\ t\in[0,d_{1}-d_{0}],       
     \end{equation}
     so $\zeta$ is well-defined for sufficiently small $r_{7}$.
     Now consider the scattering problem in $\Omega_{-d(t),d(t)}$.    
     For each fixed $t$, the triple $(k,\zeta{u},\zeta\bm{T}\bm{a})$
     satisfies     
     \eqref{eq:sec4:ku:condition} and \eqref{eq:sec4:ku:condition2},
     where the field $u$ is governed by
     $\bm{T}_{0}(-t)\bm{M}_{2}\bm{T}_{0}(-t)$. This yields a
     continuous mapping
      \begin{equation}
        \label{eq:sec5:H2}
        \begin{aligned}
          \mathcal{H}_{2}&:[0,d_{1}-d_{0}]\times{\overline{B_{r_{7}}((\beta_{*},\bm{\delta}_{*}))}}\to{\mathbb{C}^{2N_{0}}},\\          
          \mathcal{H}_{2}(t,\beta,\bm{\delta})&:=\zeta(t,\beta,\bm{\delta})\bm{T}(t,\beta,k(t,\beta,\bm{\delta}))\mathcal{P}_{\bm{W}(t,\beta,k(t,\beta,\bm{\delta})),1}(\beta,\bm{\delta})\\
          &=\mathcal{P}_{\bm{T}_{0}(-t)\bm{M}_{2}\bm{T}_{0}(-t),1}(\beta,\bm{\delta}).            
        \end{aligned}
      \end{equation}
      Consequently, $\mathcal{H}_{2}$ constitutes a homotopy between $\mathcal{P}_{\bm{M}_{2},1}$
      for the scattering problem in $\Omega_{-d_{0},d_{0}}$ and
     $\mathcal{P}_{\bm{T}_{0}(d_{0}-d_{1})\bm{M}_{2}\bm{T}_{0}(d_{0}-d_{1}),1}$ in
     $\Omega_{-d_{1},d_{1}}$, where
     $\bm{0}\notin\mathcal{H}_{2}([0,1]\times\partial{B_{r_{7}}((\beta_{*},\bm{\delta}_{*}))})$. The
     homotopy invariance of the degree therefore implies the BIC index
     is preserved. 
  \end{proof}
\end{theorem}

To investigate the local robustness of BICs under perturbations,
we assume
the dielectric function $\epsilon(\cdot,\bm{\delta},\bm{\gamma})$ also depends on an
additional parameter $\bm{\gamma}\in{W_{2}}\subset\mathbb{R}^{N_{2}}$ for some
$N_{2}\in\mathbb{Z}_{+}$. Without loss of generality, let $W_{2}$ be an open set,
$\bm{0}\in{W_{2}}$ and $\epsilon(\cdot,\bm{\delta},\bm{0})$ represent 
the unperturbed structure.
For a given symmetry case, we assume the following conditions hold
for all $(\bm{\delta},\bm{\gamma})\in{{W}_{1}\times{W_{2}}}$:
\begin{equation}
  \label{eq:sec5:dielectric:cond}
  \left\{
    \begin{aligned}
      &\epsilon\ \text{is\ continuous\ in}\ (\bm{\delta},\bm{\gamma})\ \text{and\
      satisfies}\ \eqref{eq:sec2:refindex};\\
      &\epsilon\ \text{still\ satisfies\ the\ corresponding\ symmetry condition}.
    \end{aligned}\right.
\end{equation}
The following corollary establishes a robustness result.

\begin{corollary}
  \label{coro:sec5:BIC:sta}  
  Let $u_{*}$ be a simple and isolated BIC at the point
  $(\beta_{*},\bm{\delta}_{*},k_{*})\in\Lambda$ with scattering matrix $\bm{S}_{0}$.
  Assume that $\mathrm{ind}_{j}((\beta_{*},\bm{\delta}_{*},k_{*}))\ne0$ for some  
  $j\in\{1,2,3,4\}$ in the corresponding symmetry case.
  If the dielectric function $\epsilon$ is perturbed by an additional
  parameter $\bm{\gamma}$ satisfying \eqref{eq:sec5:dielectric:cond},
  then there exists $r>0$ such that for every $\bm{\gamma}\in{B}_{r}(\bm{0})$,
  a BIC $u_{\dag}$ exists at some $(\beta_{\dag},\bm{\delta}_{\dag},k_{\dag})\in\Lambda$. Here
  $u_{\dag}$, $\beta_{\dag}$, $\bm{\delta}_{\dag}$ and $k_{\dag}$ depend continuously
  on $\bm{\gamma}$.
  \begin{proof}
    We demonstrate the result for symmetry case I; the other cases follow
    similarly.

    By treating $(\bm{\delta},\bm{\gamma})$ as a combined parameter,
    we can apply Theorem \ref{thm:sec4:PiM:bic} directly.
  Given $\bm{M}\in{U}_{1}$, then there exist radii
  $r_{5,1},r_{5,2},r_{6}>0$ such that for every   
  $(\beta,\bm{\delta})\in{B_{r_{5,1}}}((\beta_{*},\bm{\delta}_{*}))$ and
  $\bm{\gamma}\in{B_{r_{5,2}}}(\bm{0})$ we obtain unique 
  $k(\beta,\bm{\delta},\bm{\gamma})\in{B_{r_{6}}}(k_{*})$,
  $u(\cdot,\beta,\bm{\delta},\bm{\gamma})\in{H^{1}_{\mathrm{per},1}(\Omega_{0})}$ and
  $\bm{a}(\beta,\bm{\delta},\bm{\gamma})\in\mathbb{C}^{2N_{0}}$ satisfying 
  \begin{equation}
    \label{eq:sec5:ku:gammacondition}
    \left\{
      \begin{aligned}       
        &u(\cdot,\bm{\gamma},\beta,\bm{\delta})\ \text{satisfies}\
          \eqref{eq:sec3:Omegam:sca}\ \text{with\ incident\
          coefficient\ vector}\ \bm{a}(\bm{\gamma},\beta,\bm{\delta});\\          
        &u(\cdot,\bm{\gamma},\beta,\bm{\delta})\ \text{is\ governed\ by}\ \bm{M}\ \text{and}\
          (u(\cdot,\bm{\gamma},\beta,\bm{\delta}),u_{*})_{\Omega_{0}}=\|u_{*}\|_{L^{2}(\Omega_{0})}^{2}.  
      \end{aligned}\right.
  \end{equation}
  Furthermore, 
  $k(\beta,\bm{\delta},\bm{\gamma})$, $u(\cdot,\beta,\bm{\delta},\bm{\gamma})$ and
  $\bm{a}(\beta,\bm{\delta},\bm{\gamma})$ depend continuously on
  $(\beta,{\bm{\delta}},\bm{\gamma})$, and 
  \begin{equation}
    \label{eq:sec5:ku:gammacondition2}
    k(\beta_{*},\bm{\delta}_{*},\bm{0})=k_{*},\
    u(\cdot,\beta_{*},\bm{\delta}_{*},\bm{0})=u_{*}\ \text{and}\
    \bm{a}(\beta_{*},\bm{\delta}_{*},\bm{0})=\bm{0}.    
  \end{equation}  
  Define $\mathcal{P}_{\bm{M},\bm{\gamma},1}(\beta,\bm{\delta}):=\bm{a}(\beta,\bm{\delta},\bm{\gamma})$.
  Because $u_{*}$ is isolated, we can choose $r_{5,1}$ sufficiently
  small such that $(\beta_{*},\bm{\delta}_{*})$ is the unique zero of  
     $\mathcal{P}_{\bm{M},\bm{0},1}$ in
     $\overline{B_{r_{5,1}}((\beta_{*},\bm{\delta}_{*}))}$. 
     By continuity of $\bm{a}$, we may also choose sufficiently small
     $r_{5,2}$ such that
     \begin{equation}
       \label{eq:sec5:PMgamma:boundary}
       \mathcal{P}_{\bm{M},\bm{\gamma},1}(\beta,\bm{\delta})\ne0\ \text{for}\
       (\beta,\bm{\delta})\in\partial{B_{r_{5,1}}((\beta_{*},\bm{\delta}_{*}))}\ \text{and}\
       \bm{\gamma}\in{B_{r_{5,2}}(\bm{0})}.
     \end{equation}
Now fix $\bm{\gamma}$ and consider the homotopy
      \begin{equation}
        \label{eq:sec5:H3}
        \mathcal{H}_{3}:[0,1]\times{\overline{{B}_{r_{5,1}}((\beta_{*},\bm{\delta}_{*}))}}\to{\mathbb{C}^{2N_{0}}},\
        \mathcal{H}_{3}(t,\beta,\bm{\delta}):=\mathcal{P}_{\bm{M},t\bm{\gamma},1}(\beta,\bm{\delta}).        
      \end{equation}
      The above choice of radii ensures that
      $\bm{0}\notin\mathcal{H}_{3}([0,1]\times\partial{B}_{r_{5,1}}((\beta_{*},\bm{\delta}_{*})))$. 
      The homotopy invariance of the degree therefore gives
\begin{equation}
  \label{eq:sec5:degree:invariance}
 \mathrm{deg}(\mathcal{P}_{\bm{M},\bm{0},1},B_{r_{5,1}}((\beta_{*},\bm{\delta}_{*})),\bm{0})=\mathrm{deg}(\mathcal{P}_{\bm{M},\bm{\gamma},1},B_{r_{5,1}}((\beta_{*},\bm{\delta}_{*})),\bm{0})\
 \text{for}\ \bm{\gamma}\in{B_{r_{5,2}}(\bm{0})}.
\end{equation}      
From Definition \ref{def:sec5:index} and the hypothesis,
    \begin{equation}
  \label{eq:sec5:degree:2}
 \mathrm{ind}_{1}((\beta_{*},\bm{\delta}_{*},k_{*}))=\mathrm{deg}(\mathcal{P}_{\bm{M},\bm{0},1},B_{r_{5,1}}((\beta_{*},\bm{\delta}_{*})),\bm{0})\ne0.
\end{equation}
Hence $\mathcal{P}_{\bm{M},\bm{\gamma},1}$ has a zero 
$(\beta_{\dag},\bm{\delta}_{\dag})\in{B_{r_{5,1}}((\beta_{*},\bm{\delta}_{*}))}$. Setting
\begin{equation}
  \label{eq:kudag}
 k_{\dag}:=k(\bm{\gamma},\beta_{\dag},\bm{\delta}_{\dag}),\ u_{\dag}:=u(\cdot,\bm{\gamma},\beta_{\dag},\bm{\delta}_{\dag})
\end{equation}
gives the desired perturbed BIC.
  \end{proof}
\end{corollary}

\section{Sufficient conditions for a nonzero BIC
  index}\label{sec:BICsuffcond}
We now derive sufficient conditions for a BIC to have
a nonzero index, assuming the dielectric function $\epsilon$ is $C^{1}$ in
$\bm{\delta}$. For $\flat\in\{L,R\}$ and $m\in{Z_{0}}$, define 
\begin{equation}
  \label{eq:sec4:vLRm}
  \widehat{v}_{\flat,m}=2\bi\alpha_{m}^{1/2}(\mathcal{A}^{*})^{-1}\widetilde{\phi}_{\flat,m},
\end{equation}
which satisfies \eqref{eq:sec3:Omegam:sca:adjoint} with the outgoing
field $w^{+}_{L,m}$ or $w^{-}_{R,m}$. Throughout the rest of the section, we 
adopt the combined notation $X=(\beta,\bm{\delta})$ to simplify expressions.

\begin{theorem}
  \label{thm:sec6:PiM:bic:smooth}
  Under the same hypotheses as Theorem \ref{thm:sec4:PiM:bic}, and
  assuming additionally that the dielectric function
  $\epsilon$ is $C^{1}$ in $\bm{\delta}$, there exists $r_{5}>0$ such that
  $k(\beta,\bm{\delta})$, $u(\cdot,\beta,\bm{\delta})$ and $\bm{a}(\beta,\bm{\delta})$ are 
  $C^{1}$ in $(\beta,\bm{\delta})$ on ${B_{r_{5}}((\beta_{*},\bm{\delta}_{*}))}$. Their
  derivatives are given by:
  \begin{equation}
    \label{eq:sec6:kua:derivative}
    \begin{aligned}
    &\begin{bmatrix}
      \partial_{X}k(\beta,\bm{\delta})&\partial_{X}u(\cdot,\beta,\bm{\delta})&\partial_{X}\bm{a}(\beta,\bm{\delta})
     \end{bmatrix}\\
      =&-(\partial_{Y}\widehat{\mathcal{A}}_{\bm{M}}(\beta_{*},\bm{\delta}_{*},k_{*},0,\bm{0}))^{-1}\partial_{X}\widehat{\mathcal{A}}_{\bm{M}}(\beta,\bm{\delta},k(\beta,\bm{\delta}),u(\cdot,\beta,\bm{\delta})-u_{*},\bm{a}(\beta,\bm{\delta})).         
    \end{aligned}
  \end{equation}
  Let $\epsilon_{*}:=\epsilon(\cdot,\bm{\delta}_{*})$, $\partial_{\bm{\delta}}\epsilon_{*}:=\partial_{\bm{\delta}}\epsilon(\cdot,\bm{\delta}_{*})$,
  $\partial_{X}k_{*}:=\partial_{X}k(\beta_{*},\bm{\delta}_{*})$ and $\partial_{X}\bm{a}_{*}:=\partial_{X}\bm{a}(\beta_{*},\bm{\delta}_{*})$.
  Then the following explicit formulas hold:
  \begin{itemize}
  \item Frequency derivatives:
      \begin{align}
    \partial_{\beta}k_{*}=&-\frac{\bi(\partial_{x_{1}}u_{*},u_{*})_{\Omega}-\beta_{*}(u_{*},u_{*})_{\Omega}}{k_{*}(\epsilon_{*}u_{*},u_{*})_{\Omega}},\label{eq:sec6:partialkappa:k}\\
        \partial_{\bm{\delta}}k_{*}=&-\frac{k_{*}(\partial_{\bm{\delta}}\epsilon_{*}u_{*},u_{*})_{\Omega}}{2(\epsilon_{*}u_{*},u_{*})_{\Omega}};\label{eq:sec6:partialdelta:k}                  
      \end{align}
    \item Incident coefficient vector derivatives:
        \begin{align}
    \partial_{\beta}\bm{a}_{*}=&(\bm{S}_{0}-\bm{M})^{-1}\big[-(\partial_{x_{1}}u_{*},\widehat{v}_{\flat,m,*})_{\Omega}-\bi\beta_{*}(u_{*},\widehat{v}_{\flat,m,*})_{\Omega}\notag\\
                                                      &\qquad\qquad\qquad+\bi{k_{*}}\partial_{\beta}k_{*}(\epsilon_{*}u_{*},\widehat{v}_{\flat,m,*})_{\Omega}\big]_{\flat\in\{L,R\},m\in{Z_{0}}},\label{eq:sec6:partialkappa:a}\\
    \partial_{\bm{\delta}}\bm{a}_{*}=&\frac{1}{2}(\bm{S}_{0}-\bm{M})^{-1}\big[\bi{k_{*}^{2}}(\partial_{\bm{\delta}}\epsilon_{*}u_{*},\widehat{v}_{\flat,m,*})_{\Omega}\notag\\    
                            &\qquad\qquad\qquad\quad+2\bi{k}_{*}\partial_{\bm{\delta}}k_{*}(\epsilon_{*}u_{*},\widehat{v}_{\flat,m,*})_{\Omega}\big]_{\flat\in\{L,R\},m\in{Z_{0}}},\label{eq:sec6:partialdelta:a}
  \end{align}
  \end{itemize}
where $\widehat{v}_{\flat,m,*}$ is defined in \eqref{eq:sec4:vLRm} at  
  $(\beta_{*},\bm{\delta}_{*},k_{*})$.
  \begin{proof}
    The formula \eqref{eq:sec6:kua:derivative} follows immediately from
    Theorem 1.2.1 in \cite{chang03}. Using the computations in Lemma    
    \ref{lem:sec4:Ahatinvertible}, the derivatives of $k(\beta,\bm{\delta})$
    and $\bm{a}(\beta,\bm{\delta})$ at $(\beta_{*},\bm{\delta}_{*})$ can be expressed as    
    \begin{align}
    \partial_{X}k_{*}=&-\frac{(\partial_{X}\mathcal{A}_{0}u_{*},u_{*})_{\Omega_{0}}}{(\partial_{k}\mathcal{A}_{0}u_{*},u_{*})_{\Omega_{0}}},\label{eq:sec6:partialX:k}\\
      \partial_{X}\bm{a}_{*}=&(\bm{S}_{0}-\bm{M})^{-1}\big[-\alpha_{m,*}^{1/2}\widehat{\phi}_{\flat,m}\mathcal{A}_{0}^{-1}(\partial_{X}\mathcal{A}_{0}u_{*}+\partial_{X}k_{*}\partial_{k}\mathcal{A}_{0}u_{*})\big]_{\flat\in\{L,R\},m\in{Z_{0}}}, \label{eq:sec6:partialX:a}    
  \end{align}
where $\alpha_{m,*}$, $\mathcal{A}_{0}$ and $\partial_{k}\mathcal{A}_{0}$ are as in
\eqref{eq:sec4:AB:alphaB} and
$\partial_{X}\mathcal{A}_{0}:=\partial_{X}\mathcal{A}(\beta_{*},\bm{\delta}_{*},k_{*})$.
Let $v$ denote a solution to the scattering problem
\eqref{eq:sec3:Omegam:sca} at $(\beta_{*},\bm{\delta}_{*},k_{*})$ for some
incident field. Using the definition of $\mathcal{A}$
from \eqref{eq:sec3:Omega:sesform}, we obtain:
\begin{align}
  (\partial_{\beta}\mathcal{A}_{0}u_{*},v)_{\Omega_{0}}=&-2\bi(\partial_{x_{1}}u_{*},v)_{\Omega_{0}}+2\beta_{*}(u_{*},v)_{\Omega_{0}}-(\partial_{\beta}\mathcal{D}_{L,0}u_{*},v)_{\Gamma_{L}}\notag\\
                              &-(\partial_{\beta}\mathcal{D}_{R,0}u_{*},v)_{\Gamma_{R}},\label{eq:sec6:partialkappa:A}\\
  (\partial_{\bm{\delta}}\mathcal{A}_{0}u_{*},v)_{\Omega_{0}}=&-k^{2}_{*}(\partial_{\bm{\delta}}\epsilon_{*}u_{*},v)_{\Omega_{0}},\label{eq:sec6:partialdelta:A}\\
  (\partial_{k}\mathcal{A}_{0}u_{*},v)_{\Omega_{0}}=&-2k_{*}(\epsilon_{*}u_{*},v)_{\Omega_{0}}-(\partial_{k}\mathcal{D}_{L,0}u_{*},v)_{\Gamma_{L}}-(\partial_{k}\mathcal{D}_{R,0}u_{*},v)_{\Gamma_{R}},\label{eq:sec6:partialk:A}  
\end{align}
where $\partial_{\beta}\mathcal{D}_{L,0}$, $\partial_{\beta}\mathcal{D}_{R,0}$, $\partial_{k}\mathcal{D}_{L,0}$ and
$\partial_{k}\mathcal{D}_{R,0}$ denote the derivatives of $\mathcal{D}_{L}$ and $\mathcal{D}_{R}$ at
$(\beta_{*},k_{*})$. Since $\partial_{\bm{\delta}}\epsilon=\bm{0}$ for $|x_{2}|>d_{0}$,
the inner product in \eqref{eq:sec6:partialdelta:A}
can be extended to the full domain $\Omega$. Expanding $u_{*}$ and $v$ in
$\Omega_{L}$ and $\Omega_{R}$ via 
\eqref{eq:sec2:Omegal:expan} and \eqref{eq:sec2:Omegar:expan}
and computing the inner products yield
\begin{align}
(\partial_{\beta}\mathcal{D}_{L,0}u_{*},v)_{\Gamma_{L}}  
=&2\bi(\partial_{x_{1}}u_{*},v)_{\Omega_{L}}-2\beta_{*}(u_{*},v)_{\Omega_{L}},\label{eq:sec6:partialkappa:DL}\\
(\partial_{\beta}\mathcal{D}_{R,0}u_{*},v)_{\Gamma_{R}}=&2\bi(\partial_{x_{1}}u_{*},v)_{\Omega_{R}}-2\beta_{*}(u_{*},v)_{\Omega_{R}},\label{eq:sec6:partialkappa:DR}\\
  (\partial_{k}\mathcal{D}_{L,0}u_{*},v)_{\Gamma_{L}}=
  &2k_{*}(\epsilon_{*}u_{*},v)_{\Omega_{L}},\label{eq:sec6:partialk:DL}\\
  (\partial_{k}\mathcal{D}_{R,0}u_{*},v)_{\Gamma_{R}}=&2k_{*}(\epsilon_{*}u_{*},v)_{\Omega_{R}}.\label{eq:sec6:partialk:DR}  
\end{align}
Substituting these into \eqref{eq:sec6:partialkappa:A} and
\eqref{eq:sec6:partialk:A} yields
\begin{align}
  (\partial_{\beta}\mathcal{A}_{0}u_{*},v)_{\Omega_{0}}=&-2\bi(\partial_{x_{1}}u_{*},v)_{\Omega}+2\beta_{*}(u_{*},v)_{\Omega},\label{eq:sec6:partialkappa:A:Omega}\\
  (\partial_{k}\mathcal{A}_{0}u_{*},v)_{\Omega_{0}}=&-2k_{*}(\epsilon_{*}u_{*},v)_{\Omega}.\label{eq:sec6:partialk:A:Omega}  
\end{align}
Setting $v=u_{*}$ and inserting these identities together with
\eqref{eq:sec6:partialdelta:A} into \eqref{eq:sec6:partialX:k} gives
the frequency derivatives
\eqref{eq:sec6:partialkappa:k}--\eqref{eq:sec6:partialdelta:k}.

Now take $f\in{(H^{1}_{\mathrm{per},1}(\Omega_{0}))^{*}}$ with
$(f,u_{*})_{\Omega_{0}}=0$. Then
\begin{align}
  \alpha_{m,*}^{1/2}\widehat{\phi}_{\flat,m}\mathcal{A}_{0}^{-1}f=&(\mathcal{A}_{0}^{-1}f,\alpha_{m,*}^{1/2}\phi_{m})_{\Gamma_{\flat}}=\overline{(\alpha_{m,*}^{1/2}\phi_{m},\mathcal{A}_{0}^{-1}f)_{\Gamma_{\flat}}}\notag\\
  =&\overline{(\alpha_{m,*}^{1/2}\widetilde{\phi}_{\flat,m},\mathcal{A}_{0}^{-1}f)_{\Omega_{0}}}=\overline{(\alpha_{m,*}^{1/2}(\mathcal{A}_{0}^{-1})^{*}\widetilde{\phi}_{\flat,m},f)_{\Omega_{0}}}\notag\\
  =&\frac{\bi}{2}\overline{(\widehat{v}_{\flat,m,*},f)_{\Omega_{0}}}=\frac{\bi}{2}(f,\widehat{v}_{\flat,m,*})_{\Omega_{0}}.\label{eq:sec6:fphi}
\end{align}
Finally, taking $v=\widehat{v}_{\flat,m,*}$ in the identities for
$(\partial_{\beta}\mathcal{A}_{0}u_{*},v)_{\Omega_{0}}$, $(\partial_{\bm{\delta}}\mathcal{A}_{0}u_{*},v)_{\Omega_{0}}$ and 
$(\partial_{k}\mathcal{A}_{0}u_{*},v)_{\Omega_{0}}$ and substituting them together with the
relation above into \eqref{eq:sec6:partialX:a}, we obtain the
coefficient derivatives
\eqref{eq:sec6:partialkappa:a}--\eqref{eq:sec6:partialdelta:a}.
\end{proof}
\end{theorem}

According to the theorem, the derivative $\partial_{X}k_{*}$ is independent of
the choice of $\bm{M}\in{U}_{1}$. Consequently, all sets
$\lambda_{\bm{M}}$ for $\bm{M}\in{U_{1}}$ are mutually tangent at
the simple BIC point $(\beta_{*},\bm{\delta}_{*},k_{*})\in\Lambda$. This further
clarifies the characteristic property of the phase singularity
associated with BICs.

Suppose the dimensional constraint \eqref{eq:sec5:dimension} for
symmetry case I holds, and let $u_{*}$ be a simple BIC at
$(\beta_{*},\bm{\delta}_{*},k_{*})\in\Lambda$. From degree theory, the condition
\begin{equation}
  \label{eq:sec6:suffcond:pre}
  \mathrm{det}\left(
    \begin{bmatrix}
      \mathrm{Re}(\nabla\mathcal{P}_{\bm{M},1}(\beta_{*},\bm{\delta}_{*}))\\      
      \mathrm{Im}(\nabla\mathcal{P}_{\bm{M},1}(\beta_{*},\bm{\delta}_{*}))
    \end{bmatrix}
    \right)=
  \mathrm{det}\left(
    \begin{bmatrix}
      \mathrm{Re}(\partial_{\beta}\bm{a}_{*})&\mathrm{Re}(\partial_{\bm{\delta}}\bm{a}_{*})\\
      \mathrm{Im}(\partial_{\beta}\bm{a}_{*})&\mathrm{Im}(\partial_{\bm{\delta}}\bm{a}_{*})
    \end{bmatrix}
  \right)\ne0
\end{equation}
ensures $u_{*}$ has a well-defined nonzero index
$\mathrm{ind}_{1}((\beta_{*},\bm{\delta}_{*},k_{*}))$. However, the theorem above
indicates that this determinant may depend on the choice of
$\bm{M}\in{U}_{1}$. Our
objective is to obtain a condition that remains invariant under 
$\bm{M}\in{U}_{j}$ for $j\in\{1,2,3,4\}$ in each symmetry case. To this end,
we define 
\begin{equation}
  \label{eq:sec6:eta}
  \bm{\xi}:=(\bm{S}_{0}-\bm{M})
              \begin{bmatrix}
                \partial_{\beta}\bm{a}_{*}&\partial_{\bm{\delta}}\bm{a}_{*}
              \end{bmatrix},
\end{equation}
which is independent of $\bm{M}$. We then introduce the
matrices:
\begin{equation}
  \label{eq:sec6:muj}
    \bm{\mu}_{1}:=\begin{bmatrix}
                    \mathrm{Re}(\bm{\xi})\\
                    \mathrm{Im}(\bm{\xi})
                  \end{bmatrix},\ \bm{\mu}_{2}:=\bm{\xi},\
  \bm{\mu}_{3}:=\begin{bmatrix}
                    \mathrm{Re}(\bm{L}\bm{\xi})\\
                    \mathrm{Im}(\bm{L}\bm{\xi})
                  \end{bmatrix},\
  \bm{\mu}_{4}:=\bm{L}\bm{\xi},
\end{equation}
where
\begin{equation}
  \label{eq:sec6:Leo}
  \bm{L}:=\frac{1}{2}
  \begin{bmatrix}
    \bm{I}_{N_{0}}&C\bm{I}_{N_{0}}
  \end{bmatrix},
\end{equation}
and $C\in\{-1,1\}$ satisfies $\mathcal{T}_{2}u_{*}=Cu_{*}$.

\begin{corollary}
  \label{coro:sec6:sufficond}
    Let $u_{*}$ be a simple BIC at
    the point $(\beta_{*},\bm{\delta}_{*},k_{*})\in\Lambda$ with scattering matrix
    $\bm{S}_{0}$.    
    Assume the dielectric function $\epsilon$ is $C^{1}$ in $\bm{\delta}$, and let
    $\bm{M}\in{U_{j}}$ for $j\in\{1,2,3,4\}$ in each symmetry case. Then
    the corresponding mapping $\mathcal{P}_{\bm{M},j}$ is $C^{1}$ in $(\beta,\bm{\delta})$.     
    Under the dimensional constraint in \eqref{eq:sec5:dimension}, the
    index $\mathrm{ind}_{j}((\beta_{*},\bm{\delta}_{*},k_{*}))$ is well-defined
    and nonzero if $\mathrm{det}(\bm{\mu}_{j})\ne0$.
    \begin{proof}
      The differentiability of $\mathcal{P}_{\bm{M},j}$ for $j\in\{1,2,3,4\}$
      follows directly from Theorem \ref{thm:sec6:PiM:bic:smooth}.

      We now establish the sufficient conditions for each symmetry
      case.
      \begin{itemize}
        \item Case I. Let $\bm{B}:=(\bm{S}_{0}-\bm{M})^{-1}$. From the
          definition of $\mathcal{P}_{\bm{M},1}$ we have
       \begin{equation}
        \label{eq:sec6:Pm1u}
        \nabla\mathcal{P}_{\bm{M},1}(\beta_{*},\bm{\delta}_{*})=\bm{B}\bm{\xi}.
       \end{equation}
      A direct computation gives
      \begin{equation}
        \label{eq:sec6:ReIm}
        \begin{bmatrix}
          \mathrm{Re}(\bm{B}\bm{\xi})\\
          \mathrm{Im}(\bm{B}\bm{\xi})
        \end{bmatrix}=
        \begin{bmatrix}
          \mathrm{Re}(\bm{B})&-\mathrm{Im}(\bm{B})\\
          \mathrm{Im}(\bm{B})&\mathrm{Re}(\bm{B})
        \end{bmatrix}
        \begin{bmatrix}
          \mathrm{Re}(\bm{\xi})\\
          \mathrm{Im}(\bm{\xi})
        \end{bmatrix}.        
      \end{equation}
      Using the identity
      \begin{equation}
        \label{eq:sec6:Btrans}
        \begin{bmatrix}
          \mathrm{Re}(\bm{B})&-\mathrm{Im}(\bm{B})\\
          \mathrm{Im}(\bm{B})&\mathrm{Re}(\bm{B})
        \end{bmatrix}=
        \frac{1}{\sqrt{2}}\begin{bmatrix}
          \bm{I}_{N_{0}}&-\bi\bm{I}_{N_{0}}\\
          -\bi\bm{I}_{N_{0}}&\bm{I}_{N_{0}}
                          \end{bmatrix}
                          \begin{bmatrix}
                            \bm{B}&\\
                            &\overline{\bm{B}}\\
                          \end{bmatrix}
                          \frac{1}{\sqrt{2}}\begin{bmatrix}
          \bm{I}_{N_{0}}&\bi\bm{I}_{N_{0}}\\
          \bi\bm{I}_{N_{0}}&\bm{I}_{N_{0}}
                          \end{bmatrix},
      \end{equation}
      we obtain
      \begin{equation}
        \label{eq:sec6:det1}
        \mathrm{det}\left(    \begin{bmatrix}
      \mathrm{Re}(\nabla\mathcal{P}_{\bm{M},1}(\beta_{*},\bm{\delta}_{*}))\\      
      \mathrm{Im}(\nabla\mathcal{P}_{\bm{M},1}(\beta_{*},\bm{\delta}_{*}))
    \end{bmatrix}\right)=|\mathrm{det}(\bm{B})|^{2}
                        \mathrm{det}(\bm{\mu}_{1}).
      \end{equation}
      Consequently, if $\mathrm{det}(\bm{\mu}_{1})\ne0$, then
      $(\beta_{*},\bm{\delta}_{*})$ is      
      a regular point of $\mathcal{P}_{\bm{M},1}$. By Proposition 3.2 in Chapter
      IV of \cite{Outerelo09}, we conclude
      \begin{equation}
        \label{eq:sec6:index:det1}
        \mathrm{ind}_{1}((\beta_{*},\bm{\delta}_{*},k_{*}))=\mathrm{sign}(\mathrm{det}(\bm{\mu}_{1}))\ne0.
      \end{equation}
      \item Case II. Assume $\mathcal{T}_{1}u_{*}=e^{\bi\eta}u_{*}$ for some
        $\eta\in[0,2\pi)$. From \eqref{eq:sec5:PHM}, we obtain 
       \begin{equation}
        \label{eq:sec6:Pm2u}
        \nabla\mathcal{P}_{\bm{M},2}(\beta_{*},\bm{\delta}_{*})=e^{\bi\eta/2}\bm{M}^{1/2}\bm{B}\bm{\xi}.
       \end{equation}
       Taking determinants yields
       \begin{equation}
        \label{eq:sec6:det2}
        \mathrm{det}(\nabla\mathcal{P}_{\bm{M},2}(\beta_{*},\bm{\delta}_{*}))=\mathrm{det}(e^{\bi\eta/2}\bm{M}^{1/2}\bm{B})\mathrm{det}(\bm{\mu}_{2}).
      \end{equation}
      Hence, $\mathrm{det}(\bm{\mu}_{2})\ne0$ implies
      $\mathrm{det}(\nabla\mathcal{P}_{\bm{M},2}(\beta_{*},\bm{\delta}_{*}))\ne0$.
      \item Case III. Assume $\mathcal{T}_{2}u_{*}=Cu_{*}$ with $C\in\{-1,1\}$. By
        Lemma \ref{lem:sec5:MP}, we have $\bm{B}=\bm{B}^{P}$. A direct
        computation gives
        \begin{equation}
          \label{eq:sec6:Pm3u}                  
        \nabla\mathcal{P}_{\bm{M},3}(\beta_{*},\bm{\delta}_{*})=\bm{L}\bm{B}\bm{\xi}=2\bm{L}\bm{B}\bm{L}^{T}\bm{L}\bm{\xi}. 
       \end{equation}
       Using the same transformation as in case I, we obtain
       \begin{equation}
         \label{eq:sec6:det3}
         \mathrm{det}\left(
           \begin{bmatrix}
             \mathrm{Re}(\nabla\mathcal{P}_{\bm{M},3}(\beta_{*},\bm{\delta}_{*}))\\      
             \mathrm{Im}(\nabla\mathcal{P}_{\bm{M},3}(\beta_{*},\bm{\delta}_{*}))
           \end{bmatrix}\right)
         =|\mathrm{det}(2\bm{L}\bm{B}\bm{L}^{T})|^{2}\mathrm{det}(\bm{\mu}_{3}).
       \end{equation}

       \item Case IV. Combining the results of cases II and III, we
         find
         \begin{equation}
       \nabla\mathcal{P}_{\bm{M},4}(\beta_{*},\bm{\delta}_{*})=e^{\bi\eta/2}\bm{L}\bm{M}^{1/2}\bm{B}\bm{\xi}=2e^{\bi\eta/2}\bm{L}\bm{M}^{1/2}\bm{B}\bm{L}^{T}\bm{L}\bm{\xi}. \label{eq:sec6:Pm4u}       
     \end{equation}
     Thus,
      \begin{equation}
        \label{eq:sec6:det4}
        \mathrm{det}(\nabla\mathcal{P}_{\bm{M},4}(\beta_{*},\bm{\delta}_{*}))=\mathrm{det}(2e^{\bi\eta/2}\bm{L}\bm{M}^{1/2}\bm{B}\bm{L}^{T})\mathrm{det}(\bm{\mu}_{4}). 
      \end{equation}
      \end{itemize}
    \end{proof}
\end{corollary}
\begin{remark}
  The choice of $\widehat{v}_{\flat,m,*}$ in the derivatives
  \eqref{eq:sec6:partialkappa:a}--\eqref{eq:sec6:partialdelta:a} is not
  unique: adding any term $C_{0}u_{*}$ (with $C_{0}\in\mathbb{C}$)
  yields an equally valid choice. However, the derivatives themselves
  remain unchanged. By selecting $\widehat{v}_{\flat,m,*}$ 
  such that $(\epsilon_{*}{u_{*}},\widehat{v}_{\flat,m,*})_{\Omega}=0$ for $\flat\in\{L,R\}$ and
  $m\in{Z_{0}}$, the sufficient conditions we obtain in symmetry cases
  III and IV  
  coincide precisely with those derived via perturbation theory in
  \cite{yuan17_2,yuan20_3} for $N_{0}=1$ and $N_{1}=0\
  \text{or}\ 1$.
\end{remark}

\begin{remark}
  Consider a simple and symmetry-protected BIC $u_{*}$ at
  $(0,\bm{\delta}_{*},k_{*})\in\Lambda$  
  in symmetry case II, with $N_{0}=1$ and $N_{1}=1$. A direct
  computation gives
  \begin{equation}
    \label{eq:sec6:sypbic}
    \mathrm{det}(\bm{\mu}_{2})=\mathrm{det}\left(
    \begin{bmatrix}
      -(\partial_{x_{1}}u_{*},\widehat{v}_{L,0,*})_{\Omega}&0\\
      -(\partial_{x_{1}}u_{*},\widehat{v}_{R,0,*})_{\Omega}&0
    \end{bmatrix}\right)=0.
\end{equation}
The determinant vanishes because additional conjugate symmetry is
present in the underlying PDE when $\beta=0$. Moreover, the variation
of $\epsilon$ under reflection symmetry in $x_{1}$ is not sufficiently
generic to isolate $u_{*}$ under variations in $(\beta,\bm{\delta},k)$. 
Nonetheless, the proposed framework remains applicable by
incorporating the additional symmetry. And the BIC index introduced in
this work can still be employed to characterize the robustness of
symmetry-protected BICs in symmetry case IV with $N_{1}=0$.
\end{remark}

\section{Numerical experiments}
When a numerical method has a precision on the order of $10^{-6}$, it
cannot reliably distinguish a resonance with an imaginary part of
order $10^{-7}$ from a BIC. To resolve this ambiguity, we
introduce a numerical criterion that confirms a BIC by verifying the
existence of a nonzero index, specifically for symmetry
cases II and III with $N_{0}=1$ and $N_{1}=1$ and symmetry case IV with
$N_{0}=1$ and $N_{1}=0$. 

\subsection{Methods}
Our method relies on the identity \eqref{eq:sec5:windingnum} between
the BIC index and the winding number.
Given a candidate BIC point at $(\beta_{\ddag},\bm{\delta}_{\ddag},k_{\ddag})\in\Lambda$, a BIC
located at $(\beta_{*},\bm{\delta}_{*},k_{*})\approx(\beta_{\ddag},\bm{\delta}_{\ddag},k_{\ddag})$ can be
detected as follows:
\begin{itemize}
  \item In symmetry cases II and III, compute the winding number of
    $\mathcal{P}_{\bm{M},j}$ (with $j=2$ and $j=3$, respectively) along the boundary
    $\partial{B}_{r}((\beta_{\ddag},\bm{\delta}_{\ddag}))$ for some $r>0$.
  \item In symmetry case IV, detect a sign change of $\mathcal{P}_{\bm{M},4}$ on
    $\partial{B}_{r}(\beta_{\ddag})$.    
\end{itemize}

\subsubsection{Symmetry case II}
For symmetry case II, we adopt the following numerical procedure:
\begin{enumerate}
\item \textbf{Parameter selection}. Fix a radius $r>0$ sufficiently
  small, an  
  integer $N$ sufficiently large, and an angle $\theta\in[0,2\pi)$.
\item \textbf{Sampling around the point of interest}. For each
  $n=0,\ldots,N-1$, define the sample points
    \begin{equation}
      \label{eq:sec7:sc2:nodes}
     (\beta_{n},\bm{\delta}_{n}):=(\beta_{\ddag},\bm{\delta}_{\ddag})+(r\cos(2n\pi/N),r\sin(2n\pi/N)).
   \end{equation}
 \item \textbf{Eigenvalue tracking}. Let $\bm{S}(\beta,\bm{\delta},k)$ denote
   the scattering matrix. For each $n$, compute $k_{n}\approx{k_{\ddag}}$ that
   satisfies   
  \begin{equation}
    \label{eq:sec7:sc2:et}
    \mathrm{det}(\bm{S}(\beta_{n},\bm{\delta}_{n},k_{n})-e^{\bi\theta}\bm{I}_{2N_{0}})=0,  
  \end{equation}
  and record the associated eigenvector $\bm{a}_{n}\in\mathbb{R}^{2}$. The
  scattering solution $u_{n}$ is then constructed using $\bm{a}_{n}$
  as the incident coefficient vector.  
\item \textbf{Normalization and phase alignment}.
  Normalize $u_{n}$ (and correspondingly $\bm{a}_{n}$) such that
  $\|u_{n}\|_{L^{2}(\Omega_{0})}=1$, and adjust the phase of $u_{n}$ (and
  correspondingly $\bm{a}_{n}$) to satisfy  
    \begin{equation}
      \label{eq:sec7:sc2:scale}
      (u_{0},u_{n})_{\Omega_{0}}>0,\ \text{for}\ n=1,\ldots,N-1.
    \end{equation}
  \item \textbf{Winding number calculation}. For $n=0,\ldots,N-1$, let
    $\omega_{n}\in(-\pi,\pi]$ denote the signed angle between $\bm{a}_{n+1}$
    and $\bm{a}_{n}$, with the identification $\bm{a}_{N}=\bm{a}_{0}$.
    The quantity
    \begin{equation}
      \label{eq:sec7:sc2:ind}
      {D}_{2}:=\frac{1}{2\pi}\sum_{n=0}^{N-1}\omega_{n}.
    \end{equation}
    is an integer. If $D_{2}\ne0$, this provides numerical
    evidence of a BIC inside the ball $B_{r}((\beta_{\ddag},\bm{\delta}_{\ddag}))$ at a
    frequency near $k_{\ddag}$.
\end{enumerate}

Assume a simple BIC $u_{*}$ exists at
$(\beta_{*},\bm{\delta}_{*},k_{*})\approx(\beta_{\ddag},\bm{\delta}_{\ddag},k_{\ddag})$, with
$(\beta_{*},\bm{\delta}_{*})\in{B_{r}((\beta_{\ddag},\bm{\delta}_{\ddag}))}$ and nonzero index
$\mathrm{ind}_{2}((\beta_{*},\bm{\delta}_{*},k_{*}))$. Let $\bm{S}_{0}$
be the scattering matrix at $(\beta_{*},\bm{\delta}_{*},k_{*})$. Provided $e^{\bi\theta}$
is not an eigenvalue of $\bm{S}_{0}$, Theorem
\ref{thm:sec4:PiM:bic} ensures that Step 3 yields
a unique frequency $k_{n}\approx{k_{\ddag}}$ for each $n=0,\ldots,N-1$. Moreover,
when $r$ is sufficiently small,
\begin{equation}
  \label{eq:sec7:sc2:unu*}
 \|u_{n}-C_{n}u_{*}\|_{L^{2}(\Omega_{0})}\ll\|u_{*}\|_{L^{2}(\Omega_{0})},\
 \text{where}\ C_{n}:=\frac{(u_{n},u_{*})_{\Omega_{0}}}{(u_{*},u_{*})_{\Omega_{0}}}.
\end{equation}
Since $u_{*}$ is not known in advance, the scaling condition
\begin{equation}
  \label{eq:sec7:scaling}
  (u_{n},u_{*})_{\Omega_{0}}=\|u_{*}\|^{2}_{L^{2}(\Omega_{0})}
\end{equation}
as in \eqref{eq:sec4:ku:condition} is not applicable. We therefore
introduce the normalization and phase alignment in Step 4
instead.
Because the structure has reflection symmetry in $x_{1}$, the
eigenvectors $\bm{a}_{n}$ can be chosen in $\mathbb{R}^{2}$ and the inner
product in \eqref{eq:sec7:sc2:scale} is real.
Hence, the normalization and phase alignment in Step 4
yields $C_{n}\approx{C_{0}}$, which leads to
\begin{equation}
  \label{eq:sec7:sc2:Pm2}
  \bm{a}_{n}\approx{C_{0}}\mathcal{P}_{\bm{M},2}((\beta_{n},\bm{\delta}_{n})),\ \text{where}\ \bm{M}=e^{\bi\theta}\bm{I}_{2N_{0}}.
\end{equation}
Thus, Step 5 computes the winding number
\begin{equation}
  \label{eq:sec7:sc2:wn}
 w(\mathcal{P}_{\bm{M},2}|_{\partial{B_{r}((\beta_{\ddag},\bm{\delta}_{\ddag}))}},\bm{0})=D_{2}.
\end{equation}
Conversely, if no BIC exists within $B_{r}((\beta_{\ddag},\bm{\delta}_{\ddag}))$ and
the above procedure completes successfully, then $D_{2}$
must be zero. 

\subsubsection{Symmetry cases III and IV}
For symmetry case III, only the following adjustments are required:
\begin{itemize}
\item In Step 1, select a constant $C\in\{-1,1\}$.
\item In Step 3, record the eigenvector $\bm{a}_{n}\in\mathbb{C}^{2}$
  that satisfies $\bm{R}_{2}\bm{a}_{n}=C{\bm{a}_{n}}$.  
\item In Step 5, define $\widehat{a}_{n}:=[1\ C]\bm{a}_{n}/2$ and 
  compute $\widehat{\omega}_{n}\in(-\pi,\pi]$ as the signed angle
  \begin{equation}
    \label{eq:sec7:sm3:omega}
    \widehat{\omega}_{n}:=\mathrm{Im}(\ln(\widehat{a}_{n+1})-\ln(\widehat{a}_{n})).    
  \end{equation}
  Then introduce the quantity
    \begin{equation}
      \label{eq:sec7:sc3:ind}
      D_{3}:=\frac{1}{2\pi}\sum_{n=0}^{N-1}\widehat{\omega}_{n}.
    \end{equation}
    A nonzero $D_{3}$ indicates that our method predicts a
    BIC $u_{*}$ in the neighborhood satisfying $\mathcal{T}_{2}u_{*}=Cu_{*}$. 
\end{itemize}

For symmetry case IV, modify the steps as follows:
\begin{itemize}
\item In Step 1, set $C\in\{-1,1\}$ and the integer $N=2$.
\item In Step 2, define the two sampling points as
  \begin{equation}
    \label{eq:sec7:sm4:nodes}
    \beta_{0}:=\beta_{\ddag}+r,\ \beta_{1}:=\beta_{\ddag}-r.
  \end{equation}
\item In Step 3, record the eigenvector $\bm{a}_{n}\in\mathbb{R}^{2}$
  satisfying $\bm{R}_{2}\bm{a}_{n}=C{\bm{a}_{n}}$.  
\item In Step 5, define $\widehat{a}_{n}:=[1\ C]\bm{a}_{n}/2$ and
  the quantity
    \begin{equation}
      \label{eq:sec7:sc4:ind}
      D_{4}:=\left\{
        \begin{aligned}
          0,\ &\text{if}\ \widehat{a}_{1}\widehat{a}_{0}>0;\\
          1,\ &\text{if}\ \widehat{a}_{1}\widehat{a}_{0}<0\
                  \text{and}\ \widehat{a}_{1}<0;\\
          -1,\ &\text{if}\ \widehat{a}_{1}\widehat{a}_{0}<0\
                  \text{and}\ \widehat{a}_{1}>0.\\
        \end{aligned}\right.
    \end{equation}
    A nonzero $D_{4}\ne0$ suggests the existence of a BIC
    $u_{*}$ in the neighborhood with $\mathcal{T}_{2}u_{*}=Cu_{*}$. 
\end{itemize}

\subsection{Numerical examples}
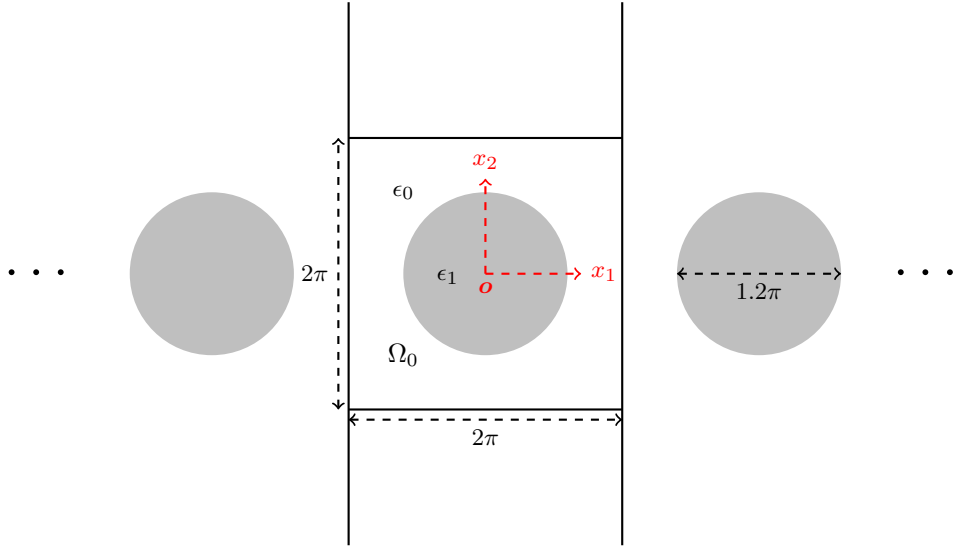
\begin{figure}[htb]
  \centering
  \begin{tikzpicture}[scale=0.9]
    \draw [thick](2,4) -- (2,-4);
    \draw [thick](-2,4) -- (-2,-4);
    \fill [lightgray] (0,0) circle(1.2);
    \fill [lightgray] (4,0) circle(1.2);
    \fill [lightgray] (-4,0) circle(1.2);
      \fill [lightgray] plot [smooth] coordinates {(0.6,-0.6) (-0.5,-0.8)
        (-1.05,0.2) (0.45,0.5) (0.9,0.4) (0.7,-0.6) (-0.55,-0.8)};
      \fill [lightgray] plot [smooth] coordinates {(-3.4,-0.6) (-4.5,-0.8)
        (-5.05,0.2) (-3.55,0.5) (-3.1,0.4) (-3.3,-0.6) (-4.55,-0.8)};
      \fill [lightgray] plot [smooth] coordinates {(4.6,-0.6) (3.5,-0.8)
  (2.95,0.2) (4.45,0.5) (4.9,0.4) (4.7,-0.6) (3.45,-0.8)};      
    \draw (-6.5,0) node[]{\huge$\cdots$};
    \draw (6.5,0) node[]{\huge$\cdots$};
\draw [thick,dashed,<->](-2,-2.15) -- (2,-2.15)
node[below,pos=0.5]{\small$2\pi$};
\draw [thick,dashed,<->](2.8,0) -- (5.2,0)
node[below,pos=0.5]{\small$1.2\pi$};
  \draw (-0.55,-0.05) node {\small$\epsilon_{1}$};
  \draw (-1.2,1.2) node {\small$\epsilon_{0}$}; 
  \draw [thick] (-2,2) -- (2,2);
  \draw [thick] (-2,-2) -- (2,-2);
 \draw (-1.2,-1.2) node[]{$\Omega_{0}$};
 \draw [thick,dashed,<->](-2.15,-2) -- (-2.15,2)
 node[left,pos=0.5]{\small$2\pi$};
 \draw [thick,dashed,->,red](0,0) -- (1.4,0)
 node[right,pos=1]{\small$x_{1}$};
  \draw [thick,dashed,->,red](0,0) -- (0,1.4)
  node[above,pos=1]{\small$x_{2}$};
  \draw (0,-0.2) node[red]{\small$\bm{o}$}; 
\end{tikzpicture}
\caption{A periodic array of circles of diameter $1.2\pi$. A rectangular
  coordinate system in defined at the center of a circle. The
  dielectric function $\epsilon(\bm{x})$ is piecewise constant, taking the
  value $\epsilon(\bm{x})=\epsilon_{1}$ inside the circles and $\epsilon(\bm{x})=\epsilon_{0}$ in the
  surrounding medium. The length of the domain $\Omega_{0}$ is set to $2\pi$.} 
\label{fig:sec7:periodiccircle}
\end{figure}

We consider a periodic array of circles of diameter $1.2\pi$, as
illustrated in Fig. \ref{fig:sec7:periodiccircle}. The
dielectric function $\epsilon(\bm{x})$ is periodic and piecewise constant.
Within one period, $\epsilon(\bm{x})$ is given by
\begin{equation}
  \label{eq:sec7:dielecfunc}
  \epsilon(\bm{x})=\left\{
    \begin{aligned}      
      \epsilon_{1}\ &\text{if}\ \rho<0.6\pi,\\
      \epsilon_{0}\ &\text{if}\ \rho>0.6\pi,
    \end{aligned}\right.
\end{equation}
where $\epsilon_{0}=1$, $\epsilon_{1}=10$, and $\rho=(x_{1}^{2}+x_{2}^{2})^{1/2}$.
The scattering problem is formulated in $\Omega_{0}$ with
$d_{0}=\pi$. It is solved using the boundary integral equations method,
and the secant method is employed in Step 3 to determine the
frequency $k_{n}$ satisfying \eqref{eq:sec7:sc2:et}. 

\subsubsection{Example 1}
\begin{figure}[htp]
  \includegraphics[width=1\textwidth]{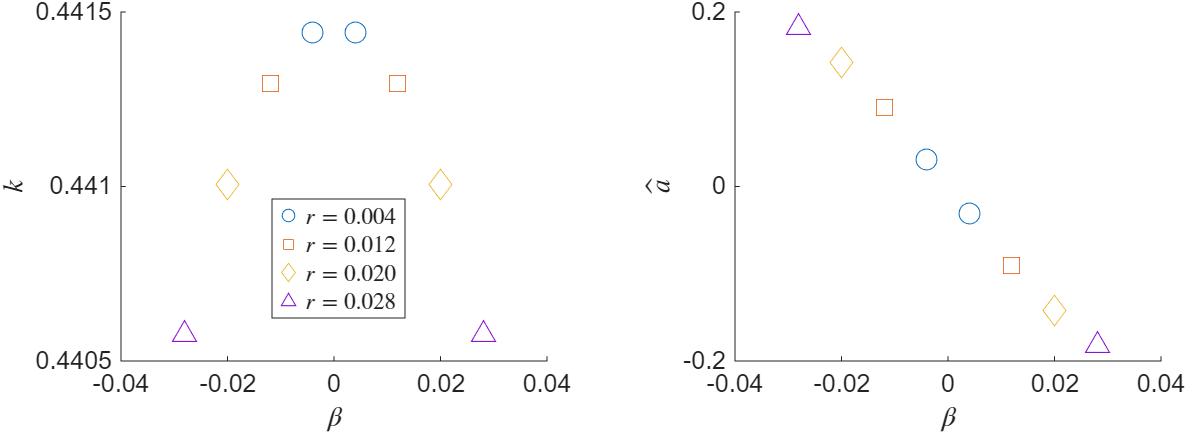}
  \caption{Simulation results for $C=1$, $\theta=\pi$ and varying $r$.
    For each $r$, the left
    graph depicts the frequencies $k_{0}$ and $k_{1}$, and
    the right graph shows $\widehat{a}_{0}$ and $\widehat{a}_{1}$.
    A consistent sign change between $\widehat{a}_{0}$ and
    $\widehat{a}_{1}$ is apparent. 
  } 
\label{fig:sec7:exm1}
\end{figure}

At the point $(\beta_{*},\bm{\delta}_{*},k_{*})\approx(0,\bm{0},0.4414)\in\Lambda$,
a symmetry-protected BIC is known to exist in this structure
\cite{yuan20_3}. Let $(\beta_{\ddag},\bm{\delta}_{\ddag},k_{\ddag})=(0,\bm{0},0.4414)$.
With $C=1$ and $\theta=\pi$, we apply our method to symmetry case IV
for each $r\in\{0.004,0.012,0.020,0.028\}$. The resulting frequencies
$k$ and computed values $\widehat{a}$ are presented in
Fig. \ref{fig:sec7:exm1}. The sign change in $\widehat{a}$ at each $r$
is unambiguous and implies $D_{4}=-1$. Thus, a BIC is
confirmed, as $D_{4}\ne0$.

\subsubsection{Example 2}
\begin{figure}[htp]
  \includegraphics[width=1\textwidth]{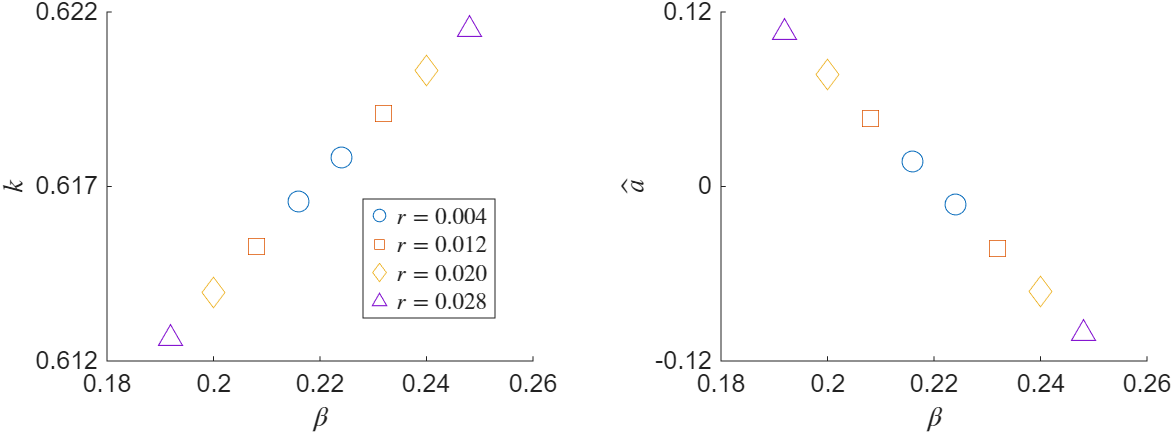}
  \caption{Simulation results for $C=1$, $\theta=\pi$ and varying $r$.
    For each $r$, the left
    graph depicts the frequencies $k_{0}$ and $k_{1}$, and
    the right graph shows $\widehat{a}_{0}$ and $\widehat{a}_{1}$.
    A consistent sign change between $\widehat{a}_{0}$ and
    $\widehat{a}_{1}$ is apparent.} 
\label{fig:sec7:exm2}
\end{figure}

At the point $(\beta_{*},\bm{\delta}_{*},k_{*})\approx(0.2206,\bm{0},0.6173)\in\Lambda$,
a propagating BIC is known to exist in this structure
\cite{yuan20_3}. Let $(\beta_{\ddag},\bm{\delta}_{\ddag},k_{\ddag})=(0.2206,\bm{0},0.6173)$.
With $C=1$ and $\theta=\pi$, we apply our method to symmetry case IV
for each $r\in\{0.004,0.012,0.020,0.028\}$. The resulting frequencies
$k$ and computed values $\widehat{a}$ are presented in
Fig. \ref{fig:sec7:exm2}. The sign change in $\widehat{a}$ at each $r$
is unambiguous and implies $D_{4}=-1$. Thus, a BIC is
confirmed, as $D_{4}\ne0$. 

\subsubsection{Example 3}
\begin{figure}[htp]
  \includegraphics[width=1\textwidth]{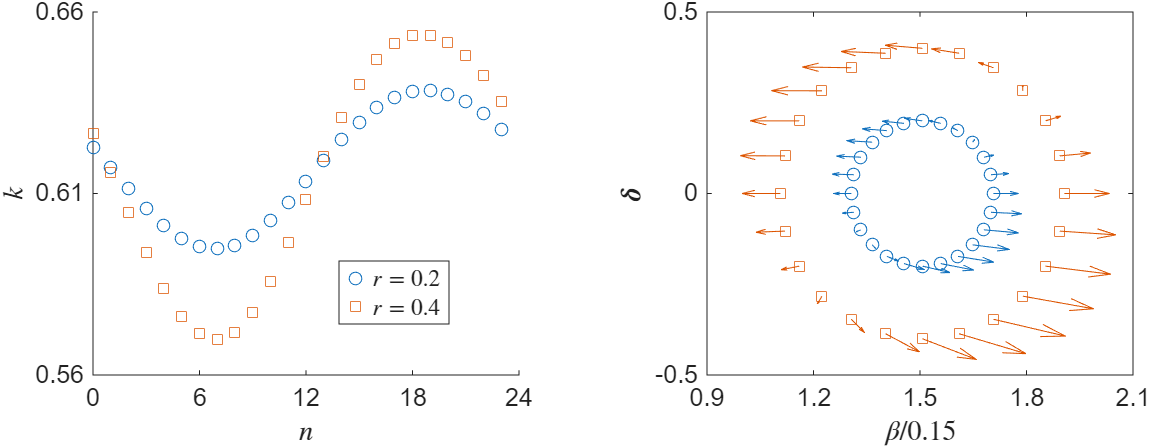}
  \caption{Simulation results for $C=1$, $N=24$, $\theta=\pi$ and varying
    $r$. For each $r$, the left
    graph shows the frequencies $k_{n}$ ($n=0,\ldots,23$), and    
    the right graph displays $\widehat{a}_{n}$ at the corresponding
    points $(\beta_{n}/0.15,\bm{\delta}_{n})$. A consistent nontrivial winding
    of $\widehat{a}$ is clearly visible.} 
\label{fig:sec7:exm3}
\end{figure}

Let the dielectric function $\epsilon$ in \eqref{eq:sec7:dielecfunc} depend
on $\bm{\delta}$ as
\begin{equation}
  \label{eq:sec7:dielecfunc:sm3}
  \epsilon(\bm{x},\bm{\delta})=\left\{
    \begin{aligned}      
      \epsilon_{1}\ &\text{if}\ \rho<0.6\pi(1+\bm{\delta}e^{-10(\tau-\pi)^{2}}),\\
      \epsilon_{0}\ &\text{if}\ \rho>0.6\pi(1+\bm{\delta}e^{-10(\tau-\pi)^{2}}),
    \end{aligned}\right.
\end{equation}
where  $(\rho,\tau)$ are the polar coordinates of $\bm{x}$ with $\tau\in[0,2\pi)$.
We again consider the BIC near the point
$(\beta_{\ddag},\bm{\delta}_{\ddag},k_{\ddag})=(0.2206,\bm{0},0.6173)$. With $C=1$, $N=24$
and $\theta=\pi$, we apply our method to symmetry case III for
$r\in\{0.2,0.4\}$. In Step 2, the sample points are defined by
    \begin{equation}
      \label{eq:sec7:sc2:nodes}
      (\beta_{n},\bm{\delta}_{n}):=(\beta_{\ddag},\bm{\delta}_{\ddag})+(0.15r\cos(n\pi/12)+0.0054,r\sin(n\pi/12)),
    \end{equation}
    for $n=0,\ldots,23$. Fig. \ref{fig:sec7:exm3} displays the resulting
    frequencies $k$ and the computed complex values $\widehat{a}$
    (represented by arrows). The winding of $\widehat{a}$ is evident
    for each $r$ and yields $D_{3}=1$, confirming the BIC and
    its local robustness with respect to $(\beta,\bm{\delta})$ in
    symmetry case III.

\section{Conclusion and future work}
In this work, we have established a rigorous theory describing how a
simple BIC continuously deforms into a
propagating field governed by a fixed unitary matrix $\bm{M}$ under
continuous parameter variation. The admissible choices of $\bm{M}$
include all diagonal matrices of the form
$e^{\bi\theta}\bm{I}_{2N_{0}}$, provided $e^{\bi\theta}$ is not an
eigenvalue of the scattering matrix at the BIC point. This
arbitrariness clarifies the phase singularity associated with
BICs. The set of parameters admitting such fields, denoted
$\lambda_{\bm{M}}$, locally forms a hypersurface in the parameter space $\Lambda$
near the BIC point. Moreover, the incident coefficients of the
propagating field define a continuous local vector field $\mathcal{P}$ on this
hypersurface.

We have examined four distinct symmetry cases, in three of which $\mathcal{P}$
can be reduced to a lower-dimensional mapping. When a BIC is
isolated and the
domain and codomain dimensions of $\mathcal{P}$ (or its reductions) coincide, its
local robustness with respect to given parameters can be characterized
by the mapping degree of $\mathcal{P}$ in a small neighborhood. This
characterization also yields a practical numerical criterion for
detecting and confirming
BICs. Furthermore, if the scattering problem is $C^{1}$ in the
parameters, the implicit function theorem ensures that $\mathcal{P}$ is also
$C^{1}$. This regularity allows us to derive sufficient robustness
conditions for a BIC via the determinant of the corresponding Jacobian
matrix. Numerical examples validating our theoretical findings are
also provided.

Several questions remain open for future investigation:
\begin{itemize}
\item In this work, we have examined the local structure of
  $\lambda_{\bm{M}}$ near simple BIC points. Its global structure is not yet fully
  understood---in particular, whether $\lambda_{\bm{M}}$ can be extended to
  the boundary of $\Lambda$, whether bifurcations occur, or whether it forms
  a manifold.
\item The singular case $\bm{M}\notin{U}_{1}$, in which the implicit
  function theorem fails, has not been addressed in this study and
  warrants further analysis.  
\end{itemize}

\section{Data availability statement}
The data that support the findings of this study are available from
the corresponding author upon reasonable request.

\bibliography{sn-bibliography}

\end{document}